%% file: ClassicalPaperv3.tex
\documentclass[a4paper,titlepage,12pt]{article}
\pdfoutput=1
\usepackage{graphicx} 
\usepackage[utf8]{inputenc}
\usepackage{geometry}
\usepackage[english]{babel}
\usepackage{amsmath, amsfonts, graphicx}
\usepackage{mathtools}
\usepackage{graphicx}
\usepackage{float}
\usepackage{enumitem}
\usepackage{mciteplus}
\usepackage{bbold}
\usepackage{pstricks}
\usepackage{xcolor}

\usepackage{amsmath,amssymb}
\usepackage[labelfont=bf,font={small}]{caption}
\usepackage{pgfplots}
\usepackage{empheq}
\usepackage{tikz}

\usetikzlibrary{positioning, arrows}
\usetikzlibrary{external}
\tikzset{every picture/.style={line width=0.75pt}}
\usepackage{footmisc}
\usepackage{multirow}

\usepackage{url}
\usepackage{verbatim}

\usepackage[breakable, theorems, skins]{tcolorbox}

\usepackage{enumerate}
\usepackage{physics}
\usepackage{bm}
\usepackage{braket}
\usepackage{epstopdf}
\usepackage{hyperref}
\usepackage[OT2,T1]{fontenc}
\DeclareSymbolFont{cyrletters}{OT2}{wncyr}{m}{n}
\DeclareMathSymbol{\Sha}{\mathalpha}{cyrletters}{"58}
\date{June 2024}

\definecolor{darkblue}{RGB}{0,0,139}

\hypersetup{
    colorlinks,
    citecolor=blue,
    filecolor=blue,
    linkcolor=purple,
    urlcolor=blue
}

	\numberwithin{equation}{section}
    
\begin{document}

\begin{titlepage}

\setcounter{page}{1} \baselineskip=15.5pt \thispagestyle{empty}

\vfil

${}$
\vspace{1cm}

\begin{center}

\def\thefootnote{\fnsymbol{footnote}}
\begin{changemargin}{0.05cm}{0.05cm} 

\begin{center}

{\Large \bf Holographic duals of JT gravity at finite cutoff} \\

\vspace{0.1in}
{\large \bf - the classical case - } \\

\end{center}
\end{changemargin}

\vspace{0.4in}
     {Andreas Belaey$^{1}$, Thomas G. Mertens$^1$, Lorenzo Russo$^2$, Alex Tarana$^3$}

    \vspace{0.4in}
    {$^1$Department of Physics and Astronomy, Ghent University,\\
    Krijgslaan, 299-S9, 9000 Gent, Belgium\\
    $^2$Dipartimento di Fisica, Universit\`a di Firenze and INFN Sezione di Firenze, Via G. Sansone 1, 50019 Sesto Fiorentino, Italy\\
    $^3$Dipartimento SMFI, Universit\`a di Parma and INFN Gruppo Collegato di Parma, Viale G. P. Usberti 7/A, 43100 Parma, Italy}
    \vspace{0.2in}
    
    {\tt andreas.belaey@ugent.be, thomas.mertens@ugent.be, lorenzo.russo@unifi.it, alex.tarana@unipr.it}

\end{center}

\begin{changemargin}{01cm}{1cm} 
{\small  \noindent 
\begin{center} 
\textbf{Abstract}
\end{center} 

}We formulate JT gravity with finite-cutoff boundaries and end-of-the-world branes in the first-order BF formulation. We dynamically reduce the BF model into a one-dimensional system living at the boundaries, generalizing earlier constructions. We obtain in particular a geometrical understanding of the phase space degrees of freedom associated to each of these boundaries. We leverage these results to probe bulk locality and the underlying structure of spacetime. Within our framework, we are able to derive a closed-form expression for the holographic Schwarzian description of JT gravity at finite-cutoff in phase space. The resulting degrees of freedom describe the embedding of the finite-cutoff curve in AdS$_2$. A quantization of these systems is presented in a companion paper.
\end{changemargin}
 \vspace{0.3cm}
\vfil
\begin{flushleft}
\today
\end{flushleft}
\end{titlepage}

{\hypersetup{linkcolor=black}\tableofcontents}

\setcounter{footnote}{0}

\section{Introduction}
Lower-dimensional models in the context of holography have proven to be very powerful in our understanding of low-energy quantum effects of black holes in higher-dimensions. In particular, Jackiw-Teitelboim (JT) gravity \cite{Jackiw:1984je, Teitelboim:1983ux} describes the universal dynamics of dimensionally reduced near-extremal black holes in terms of induced $1+1$d metric and a scalar dilaton field $\Phi$ capturing the transverse area fluctuations of the higher-dimensional black hole above extremality. Near the horizon, these black holes generically develop an infinite AdS$_2$ throat region which is coupled to the asymptotic regime by a strongly coupled gravitational mode that governs fluctuations of the dilaton field in terms of influx of matter. 

When we place the screen separating the throat and the asymptotic regime at infinite AdS$_2$ distance, its holographic description is captured by Schwarzian quantum mechanics \cite{Maldacena:2016upp,Jensen:2016pah,Engelsoy:2016xyb}. 
To resolve the backreaction problem in AdS$_2$ and define a decoupling limit \cite{Maldacena:1998uz}, the dilaton field must blow up near the asymptotic boundary. 

A natural question is therefore how to describe the dynamics when the screen is moved into the throat, and we move away from the asymptotic AdS boundary. In part, this is motivated from trying to get a handle on more local aspects of the bulk gravitational theory, where the moving screen then serves to zoom in on various local regions. This type of procedure can be done in various ways. In this work, we derive an exact holographic equivalence at the level of the action between JT gravity and the gravitational mode at the new boundary. The resulting Schwarzian boundary boundary action governs the embedding of the finite-cutoff boundary curve in AdS$_2$.

Since gravity in $1+1$d has no propagating gravitons, a powerful tool in the quantization of JT gravity is to phrase the model in its first-order formulation in terms of topological BF gauge theory \cite{Fukuyama:1985gg,Isler:1989hq,Chamseddine:1989yz,Jackiw:1992bw}. 
A crucial role now is played by the specific choice of boundary conditions we make at the holographic screen. Boundary conditions have been systematically analyzed in JT gravity in the second-order formalism in \cite{Goel:2020yxl} and in the first-order formalism in \cite{Ferrari:2020yon}. Utilizing the first-order formulation, we systematically develop the classical phase space description of the physical degrees of freedom. Throughout this process, we will find it convenient to analyze the following two types of boundaries in the same framework:
\begin{itemize}
\item \textbf{Finite cut-off boundaries} were motivated from the $T\bar{T}$-deformation in 3d gravity \cite{McGough:2016lol}, and explicitly constructed and analyzed in JT gravity in \cite{Iliesiu:2020zld,Stanford:2020qhm,Griguolo:2021wgy,Ferrari:2024kpz,Chaudhuri:2024yau,Griguolo:2025kpi,Griguolo:2026eoi}.
\item \textbf{End-of-the-world (EOW) branes} were introduced in the JT and SYK context in \cite{Kourkoulou:2017zaj,Penington:2019kki}, and have played a central role as a concrete example of black hole microstates. They were systematically analyzed in \cite{Gao:2021uro} and examined as dynamical object. Further analyses in JT supergravity appeared in \cite{Belaey:2023jtr,Belaey:2024dde}.
\end{itemize}
These two boundary types are naturally cousins of each other, which will become apparent as we develop our framework. We eventually work towards a closed form phase space Lagrangian for the finite-cutoff Schwarzian model, which is an open problem of current interest. \\

In this work, we present the classical treatment. An accompanying work will explore the quantization and applications thereof \cite{toap}. While we were finalizing the manuscript, \cite{Aguilar-Gutierrez:2026qqe} appeared which has some overlap with some parts of this work.

\subsubsection*{Summary of results}
Our main results can be summarized as follows.
\begin{itemize}[leftmargin=*]
\item
We systematically construct at the classical level in the first-order formulation of gravity, the dynamical system governing JT gravity between two finite-cutoff boundaries at dilaton values $\Phi_b^-$ and $\Phi_b^+$ in subsection \ref{s:twophi}. The system is governed by the 2d phase space action
\begin{align}
\int d\tau \,p_L \dot{L} -
    \int_{\partial \text{AdS}_-} \hspace{-0.4cm}d\tau_- \left(\Phi_b^-
-\sqrt{(\Phi_b^-)^2-2\text{Tr}(\mathbf{B}^2)}\right)-
\int_{\partial \text{AdS}_+} \hspace{-0.4cm} d\tau_+\left(\Phi_b^+
-
\sqrt{(\Phi_b^+)^2-2\text{Tr}(\mathbf{B}^2)}\right)
\end{align}
where $
2\text{Tr}(\mathbf{B}^2)=p_L^2+
\frac{2\cosh L\,\Phi_b^-\Phi_b^+-(\Phi_b^+)^2-(\Phi_b^-)^2}
{\sinh^2L}$ \eqref{eq:casimir_length_variables_gjk}, with variables $(L,p_L)$ interpreted as the geodesic length between both boundaries and its conjugate, and with non-trivial support on both boundary curves. The time $\tau$ should be chosen as a function of $\tau_+$ and $\tau_-$, depending on which time evolution one wants to describe. The system between one EOW with tension $\mu$ and one finite-cutoff brane boundary at $\Phi_b$ is constructed in subsection \ref{s:oneeow} and governed by
\begin{align}
\int_{\partial \text{AdS}_+} \hspace{-0.4cm} d\tau_+\left(p_L\dot{L} -\Phi_b
+
\sqrt{(\Phi_b)^2-2\text{Tr}(\mathbf{B}^2)}\right).
\end{align} 
where $2\Tr(\mathbf{B}^2)=p_L^2+\frac{2\mu \Phi_b\sinh L\,+\Phi_b^2-\mu^2}{\cosh^2L}$ \eqref{eq:mixed_casimir_length_general_a}, and with support only on the finite-cutoff boundary.
\item
Using two finite-cutoff boundaries that are very closely together, we probe local features of the gravitational bulk in subsection \ref{s:locality}. In particular we do this to deduce that the reduced algebraic structure that governs gravity instead of gauge theory (argued to be describable in terms of a SL$^+(2,\mathbb{R})$ semigroup structure) emerges locally in spacetime by demanding that at any point on the spatial manifold the infinitesimal area (here length) segment is positive $\sqrt{h}>0$.
\item 
As an application of our technical machinery, we derive in subsection \ref{s:fintesch} a closed form for the finite-cutoff Schwarzian action
\begin{align}
S_{\text{Schwarzian}}
= \int_{\partial} d\tau \left(p_\ell \,\dot{\ell}+ p_f\,\dot{f}-\Phi_b+\sqrt{(e^{-\ell} p_f-\Phi_b)^2-p_\ell^2}\right).
\end{align} 
in terms of four phase space coordinates $(f,p_f,\ell, p_\ell)$ which can be interpreted as the coordinates (and conjugate momenta) of the parametrized boundary curve $(f(\tau), \ell(\tau))$ in terms of proper time $\tau$, in Poincar\'e coordinates $ds^2=\frac{-df^2+dz^2}{z^2}$ where $z\equiv e^{-\ell}$ on the AdS$_2$ manifold.

\end{itemize}

This work is organized as follows. In \textbf{section \ref{sec:boundaryaction}} we present the set-up and analyze and review boundary conditions in JT gravity. We will transfer to the first-order BF formulation and implement the boundary conditions there. \textbf{Section \ref{s:redphase}} collects the phase space treatment of the BF model and applies Hamiltonian reduction of the model when boundary constraints are applied, reducing it to a 2d phase space. The main core of this work is in \textbf{section \ref{s:dsystems}} where we apply the machinery towards finding the Hamiltonian of the two cutoff system and the one cutoff - one EOW brane system. We discuss the classical structure and solutions of the model. \textbf{Section \ref{s:unconstrain}} relaxes the boundary constraint again, and leads to the construction of the phase space Lagrangian of the finite-cutoff Schwarzian as its main application. We provide an outlook in the concluding \textbf{section \ref{s:concl}}, including an application of our formalism to de Sitter spaces. Some additional technical background on the embedding formalism of AdS$_2$, Hamiltonian flow and Poisson algebra, and a phase space analysis directly in the second-order formalism are provided in \textbf{appendices \ref{app:embedding}, \ref{app:sympl}} and \textbf{\ref{app:secondphasespace}}.

\section{Classical JT gravity with boundaries}\label{sec:boundaryaction}
The total Lorentzian JT gravity action (with $8\pi G_N=1$) on a strip topology with a left oriented EOW brane and a right finite cutoff boundary is given by \cite{Gao:2021uro}
\begin{align} 
\label{eq:totalaction}
    S= \frac{1}{2}\int_{\mathcal{M}} d^2x \sqrt{-g} \;\Phi(R+2) +\int_{\partial \text{AdS}} du \sqrt{-h_{uu}}\;\Phi_b (K-1)+\int_{\text{EOW}} dv \sqrt{-h_{vv}}(\Phi K-\mu).
\end{align} 
The bulk fields that make up the action are the scalar dilaton field $\Phi$ and a two-dimensional metric $g_{\mu\nu}$ defined on a Lorentzian manifold $\mathcal{M}$.
The bulk equations of motion obtained by varying w.r.t. the dilaton field and the metric lead respectively to 
\begin{align}\label{eq:secondordereom}
    R+2=0, \qquad \nabla_{\mu }\nabla_{\nu}\Phi -g_{\mu\nu }\Phi=0,
\end{align} 
which fixes the bulk geometry as a patch of AdS$_2$ and controls the shape of the dilaton profile. 
In order to obtain a well-defined variational principle, we must introduce appropriate boundary terms along the one-dimensional boundary curve $\mathcal{M}$, involving the extrinsic curvature \begin{align}
    K=h^{\mu\nu }\nabla_{\mu} n_\nu,
\end{align} 
built from the induced metric $h_{\mu\nu}$ and the outward pointing unit normal vector $n^\mu$ to boundary. In the following we always take the oriented normal vector to be directed outwards.
The one-dimensional boundary $\partial M = \partial \text{AdS}\, \cup\, \text{EOW}$ is made up of a cutoff AdS surface and a geodesic particle with mass $\mu$ whose worldline demarcates where spacetime ends. The latter defines an end-of-the-world (EOW) brane with brane tension $\mu$. 

The allowed boundary conditions along each surface are  
\begin{alignat}{3}
\label{eq: boundary conditions}
  \text{Finite cutoff boundary}
  &:\quad&
  \left.\Phi\right|_{\partial\mathcal M}
  & = \Phi_b,
  &\qquad
  \left.h^{\alpha\beta}\delta g_{\alpha\beta}\right|_{\partial\mathcal M}
  & = 0,
  \\
  \text{EOW brane boundary}
  &:\quad&
  \left.n^\mu\partial_\mu\Phi\right|_{\partial\mathcal M}
  & = \mu,
  &\qquad
  K
  & = 0. \label{eq:eowboundaryconditions}
\end{alignat} 
This amounts to fixing Dirichlet boundary conditions of the dilaton and induced metric along the finite-cutoff surface, and Neumann boundary conditions of the dilaton whose normal derivative equals the brane tension along the $K=0$ brane geodesic. 

For timelike boundaries, the boundary Hamiltonians on either boundary equals
\begin{align}
    T_{\alpha\beta}^{L/R}\equiv -\frac{2}{\sqrt{|h|}}\frac{\delta S}{\delta h^{\alpha\beta}}\Bigg\rvert_{\partial \mathcal{M}_{L/R}} \equiv -H^{L/R}h_{\alpha\beta},
\end{align}
where in the second equality we have defined what we will call the left- and right Hamiltonians in the following (stripping off the induced metric $h_{\alpha\beta}$).
Varying the total action with a finite cutoff boundary in \eqref{eq:totalaction} leads to
\begin{align}
\label{eq:genericADM}
    H_{\partial \text{AdS}}^{L/R}=\left(\Phi-n^\mu \partial_\mu \Phi \right)\rvert_{\partial\mathcal{M}}=\Phi_b-n^\mu \partial_\mu \Phi \rvert_{\partial\mathcal{M}}.
\end{align} 
On the other hand, varying the EOW brane action leads to 
\begin{align}
    H_{\text{EOW}}^{L/R}=\left(\mu-n^\mu \partial_\mu \Phi \right)\rvert_{\partial\mathcal{M}}\equiv 0
\end{align} 
which vanishes identically due to the defining EOW brane boundary condition \eqref{eq:eowboundaryconditions}.

In the well-studied infinite-cutoff regime \cite{Maldacena:2016upp,Jensen:2016pah,Engelsoy:2016xyb}, one takes the finite-cutoff boundary conditions $\Phi_b$ and $g_{\mu\nu}\rvert_{\partial\text{AdS}}$ to be of order $\tfrac{1}{\epsilon}$ resp. $\tfrac{1}{\epsilon^2}$ in the $\epsilon\rightarrow 0$ limit. However, 
in the following it will turn out to be more natural to consider $\Phi_b$ finite as a stepping stone towards the infinite-cutoff limit. 

\subsection{Classical geometry}\label{sec:geometry}
In global coordinates, the metric and dilaton configurations that extremize the action \eqref{eq:totalaction} are given by \cite{Gao:2021uro}
\begin{equation}\label{eq:geometry_on_shell}
    ds^2 = \frac{-\dd\nu^2 + \dd\sigma^2}{\sin^2\sigma}, \qquad \Phi = \Phi_h \frac{\cos\nu}{\sin\sigma}, \qquad
    \begin{cases}
        -\pi/2 \leq \nu \leq +\pi/2, \\
        0 \leq \sigma \leq \pi,
    \end{cases}
\end{equation}
where $\Phi_h$ is the value of the dilaton at the black hole horizon. The global coordinatization has a natural hyperboloid embedding of AdS$_2$ in the coset of SL$(2,\mathbb{R})/\text{SO}(1,1)$ by the local Lorentz group explained in appendix \ref{app:isometries}.   
The physical spacetime is a subregion of the global AdS$_2$ geometry
\eqref{eq:geometry_on_shell}, bounded on the left by the EOW brane and on the
right by the finite cutoff surface. The boundary conditions in \eqref{eq: boundary conditions} determine the locus in spacetime of the associated boundary curves, which in
global coordinates read 
\begin{align}
    \partial\mathcal{M}_{\mathrm{AdS}}
    &:\,\,
    \frac{\cos\nu}{\sin\sigma}
    =
    \frac{\Phi_b}{\Phi_h},
    \label{eq:cutoff_curve_global} \qquad
    \partial\mathcal{M}_{\mathrm{EOW}}
    :\,\,
    \frac{\cos\nu}{\cos\sigma}
    =
    \sqrt{1+\frac{\Phi_h^2}{\mu^2}}.
\end{align}
The first equation follows directly from the Dirichlet condition
\(
    \left.\Phi\right|_{\partial\mathcal{M}_{\mathrm{AdS}}}
    =
    \Phi_b,
\)
whereas the second equation describes the geodesic satisfying
   $ K=0$,
    $\left.n^\mu\partial_\mu\Phi\right|_{\partial\mathcal{M}_{\mathrm{EOW}}}
    =
    \mu.$ We assume $\mu>0$ in which case the brane is in the left half of the Penrose diagram and hence behind the horizon.

The left EOW brane can be parametrized as:\footnote{We identify the isometric embedding of these geodesic solutions in appendix \ref{app:isometricembedding}.}
\begin{align}
\label{eq:branecurve}
    \sigma_{\mathrm{EOW}}(\nu)    =    \text{arccos}\bigl(k\cos\nu\bigr), \qquad k    \equiv    \frac{1}{\sqrt{1+\Phi_h^2/\mu^2}},
    \qquad    0<k<1,
\end{align}
while the finite cutoff condition admits two branches. Taking the finite cutoff surface to be the right boundary of the strip, the appropriate
branch is
\begin{align}\label{eq:finitecutoffcurve}
    \sigma_{\mathrm{AdS}}(\nu)
    =    \pi    -    \text{arcsin}\!\left(        \frac{\Phi_h}{\Phi_b}\cos\nu
    \right).
\end{align} 
Note that the curve is timelike precisely when $\Phi_b> \Phi_h$. The unit normalized outward normal is then determined by the vector components 
\begin{align}
    n^\nu= \frac{\sin \sigma(\nu) \; \sigma'(\nu) }{\sqrt{1-\sigma'(\nu)^2}}, \qquad n^\sigma= \frac{\sin \sigma(\nu)}{\sqrt{1-\sigma'(\nu)^2}},
\end{align}
leading to the normal derivative 
\begin{align}
    n^\mu \partial_\mu \Phi \rvert_\partial= \sqrt{\Phi_b^2-\Phi_h^2 }.
\end{align} 
Inserted in the boundary stress tensor \eqref{eq:genericADM} then leads to an explicit expression of the total boundary energy 
\begin{align}\label{eq:explicitadmenergy}
    E=\Phi_b-\sqrt{\Phi_b^2-\Phi_h^2}.
\end{align}
In the infinite cutoff JT limit $\Phi_b\rightarrow\infty$, this becomes the usual energy dispersion $E=\frac{\Phi_h^2}{2\Phi_b}$

The physical spacetime is the region enclosed between the two boundary curves,
\begin{align}\label{eq:physicalmanifold}
    \mathcal{M}
    =
    \left\{
        (\nu,\sigma)
        \,\middle|\,
        -\frac{\pi}{2}\leq\nu\leq\frac{\pi}{2},
        \quad
        \sigma_{\mathrm{EOW}}(\nu)
        \leq
        \sigma
        \leq
        \sigma_{\mathrm{AdS}}(\nu)
    \right\}.
\end{align}
This physical region is embedded in a geometry with unwrapped global time direction, which we denote as $\widetilde{\text{AdS}}_2$ . On each spacelike slice, the spatial interval begins at the EOW brane and
ends at the finite cutoff boundary. We will use this convention below when
orienting spatial slices and defining the corresponding open holonomies.
\begin{figure}
    \centering
    \input{images/penrose_diagram_EOW.tikz}
    \caption{The Penrose diagram for JT gravity in the presence of an EOW brane boundary (green curve) and a finite-cutoff boundary (red curve). The spacetime is the shaded region between the two boundaries.}
    \label{fig:penrose_diagram_EOW}
\end{figure}
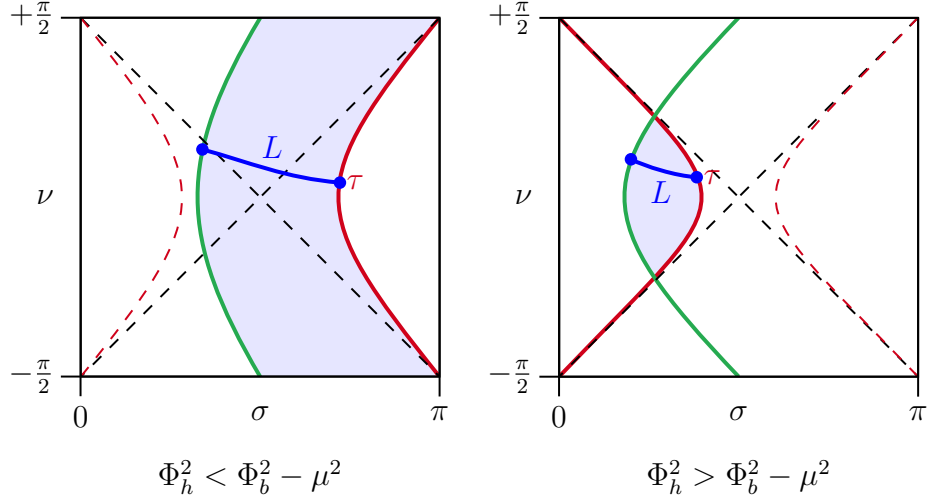
Depending on the value of the solution parameter $\Phi_h$ relative to the boundary data $\Phi_b$, $\mu$, we have two qualitatively different situations, see Figure \ref{fig:penrose_diagram_EOW}. When $\Phi_h^2 < \Phi_b^2 - \mu^2$, the EOW boundary sits on the left of the Penrose diagram, whereas the cutoff boundary sits on the right and the two are separated by a black hole horizon. Instead, when $\Phi_h^2 > \Phi_b^2 - \mu^2$, both boundaries sit on the left of the diagram, and thus there is no black hole horizon separating them and we have a naked brane.

We will reproduce a lot of these features further on from our 2d phase space Hamiltonian systems in subsections \ref{s:claspot} and \ref{s:claspot_EOW}.

\subsection{Review: First-order formulation}\label{sec:firstorderboundaryactions}
It is well understood that the classical bulk JT action can be rewritten as a topological BF gauge theory with an underlying $\mathfrak{sl}(2,\mathbb{R})$ algebra \cite{Fukuyama:1985gg,Isler:1989hq,Chamseddine:1989yz,Jackiw:1992bw}. Here we derive the appropriate first-order analogues of the boundary JT actions of \eqref{eq:totalaction}.

In Lorentzian signature, the bulk manifold $\mathcal{M}$ is spanned by a timelike coordinate $\nu$ and a radial spacelike coordinate $\sigma$ with bulk orientation $d\nu\land d\sigma$. The Lorentzian signature bulk BF action is then given by 
\begin{align}
    S=-2\int_{\mathcal{M}} \text{Tr}(\mathbf{B}\mathbf{F}), 
\end{align} 
where $\mathbf{B}$ is a zero-form field, and $\mathbf{F}=d\mathbf{A}+\mathbf{A}\land \mathbf{A}$ is the two-form field strength build from one-form gauge fields $\mathbf{A}$. They are expanded in terms of the $\mathfrak{so}(2,1)$ generators $P_I$ ($I\in\{0,1,2\}$) satisfying the algebra relations\begin{align}\label{eq:so(2,1)algebra}
    [P_0,P_2]=P_1,\quad [P_1,P_2]=P_0, \quad[P_0,P_1]=-P_2.
\end{align} 
and transform in the adjoint representation $\mathbf{F}\rightarrow h \mathbf{F}h^{-1}$, $\mathbf{B}\rightarrow h\mathbf{B}h^{-1}$. The Cartan-Killing metric  
is\footnote{Raising and lowering indices with this metric, we can write the commutation relations compactly as ($\epsilon_{012}\equiv 1$)\begin{align}\label{eq:algebra}
    [P_{I},P_J]=-\epsilon_{IJK} P^K=-\epsilon_{IJ}^{\;\;\;\;L}P_L.
\end{align}}
\begin{align}\label{eq:CKmertric}
    \text{Tr}(P_IP_J)=\frac{\eta_{IJ}}{2}\qquad \text{with} \qquad\eta_{IJ}=\text{diag}(-1,1,1).
\end{align} 
We can realize these algebra relations using
\begin{align}
    P_0=\frac{1}{2}(E-F), \quad P_1=H, \quad P_2=\frac{1}{2}(E+F)
\end{align} 
in terms of the $\mathfrak{sl}_2$ Cartan-Weyl generators \begin{align}\label{eq:fundamentalgenerators}
    E=\begin{pmatrix}
        0&1\\0&0
    \end{pmatrix}, \quad F=\begin{pmatrix}
        0&0\\1&0
    \end{pmatrix},\quad H=\frac{1}{2}\begin{pmatrix}
        1&0\\0&-1
    \end{pmatrix}.
\end{align}

We expand the gauge field and the auxiliary scalar fields as 
\begin{align}\label{eq:fieldexpansions}
    \mathbf{A}=e^0P_0+e^1P_1+\omega P_2, \qquad \mathbf{B}=-\phi^0P_0-\phi^1P_1+\Phi P_2,
\end{align} 
where $e^a$ are the components of the frame fields and $\omega$ is the spin connection, and where the timelike $e^0$ is aligned with the compact generator $P_0$.  
We can motivate the gravitational gauge field expansion by looking at the on-shell relation with the Cartan-Maurer one-form $\mathbf{A}=g^{-1}dg$ in appendix \ref{app:isometries}. Using the geometrical embedding of AdS$_2=\text{SL}(2,\mathbb{R})/\text{SO}(1,1)$, we identify AdS$_2$ as the homogeneous space generated by the adjoint action $gYg^{-1}$ on the reference point $Y=P_2$ in the algebra. The stabilizer subgroup SO$(1,1)$ generated by non-compact generator $P_2$ can then be identified with the local Lorentz action on the gauge field components.
 
The BF action in components now becomes\footnote{Choosing the timelike $e^0$ along the future oriented direction $d\nu$, we have a positively oriented frame completed by $e^1$ running along the positive radial direction $d\sigma$, leading to the volume form $e^0\land e^1= \sqrt{-g}\; d\nu\land d\sigma$.}
\begin{align}\label{eq:bulkBF}
    S_{bulk}&=-2\int_{\mathcal{M}}\text{Tr}(\mathbf{B}\mathbf{F})=\int_{\mathcal{M}}\phi^a\left(de_a+\epsilon_{ab}\omega\land e^b\right)+\Phi (-d\omega+e^0\land e^1).
\end{align} 
The term involving the dilaton $\Phi$ produces the bulk part of the second-order JT action \eqref{eq:totalaction}: 
\begin{align}\label{eq:firstorderbulkaction}
    S&=\int \phi^a\left(de_a+\epsilon_{ab}\omega\land e^b\right)+\Phi (-d\omega+e^0\land e^1)=\frac{1}{2}\int d^2x \sqrt{-g}\Phi(R+2).
\end{align}
Varying the bulk scalar fields $\phi^a$ leads to flat gauge connections $\mathbf{F}=0$:
\begin{equation}
  \mathrm{d}e^0-\omega\wedge e^1=0,
  \qquad
  \mathrm{d}e^1-\omega\wedge e^0=0,
  \qquad
  \mathrm{d}\omega-e^0\wedge e^1=0.
  \label{eq:flatness_components_gjk}
\end{equation}
The first two implement the no-torsion constraints $T^a=de^a+\epsilon^a_{\;b}\omega\land e^b\equiv 0$.
The last equation geometrically leads to patches of AdS$_2$ spaces. Varying w.r.t. the gauge fields $\mathbf{A}_\mu$ leads to the Gauss law constraint 
\begin{align}\label{eq:gausslawconstraint}
    D_\mu \mathbf{B}\equiv \partial_\mu \mathbf{B}+[\mathbf{A}_\mu,\mathbf{B}]=0,
\end{align} 
which in components gives
\begin{equation}
  \mathrm{d}\phi^0-\omega\phi^1-\Phi e^1=0,
  \qquad
  \mathrm{d}\phi^1-\omega\phi^0-\Phi e^0=0,
  \qquad
  \mathrm{d}\Phi-\phi^0e^1+\phi^1e^0=0,
  \label{eq:DB_components_gjk}
\end{equation} 
which can be combined into the dilaton equation of motion \eqref{eq:secondordereom}. 

\subsection{Boundaries}
 
For a timelike boundary curve $\gamma:\lambda \mapsto X^\mu(\lambda) $, we can introduce the lapse $h_{\lambda\lambda}=g_{\mu\nu}\dot{X}^\mu\dot{X}^\nu=-N^2, \,\, N=\sqrt{-h_{\lambda\lambda}}$ with proper time $d\tau=Nd\lambda$ and unit tangent $t^\mu=\frac{\dot{X}^\mu}{N}$.  We can choose a local Lorentz frame adapted to the curve, and we denote the timelike component $e^{/\!/} \equiv e^0$ and the perpendicular oriented  $e^{\perp} \equiv e^{1}$, pulled back to the boundary curve. Under a local Lorentz transformation \eqref{eq:componenttransformations}, the timelike component $e^{/\!/}$ is still the component of the generator $P_0$. Under any positive local Lorentz transformation, the frame orientation $\sqrt{-g}\;d\nu\land d\sigma=e^{/\!/}\land e^{\perp}$ is preserved.
Therefore, choosing the adapted local Lorentz frame parallel to the tangent of the curve $e_{/\!/}^\mu=t^\mu$, $e_{\perp}$ will be proportional to the normal at any point. The pullback on the curve then satisfies 
\begin{align}\label{eq:lapsepullback}
    \gamma^*e^{/\!/}=d\tau=Nd\lambda, \qquad \gamma^* e^{\perp}=0.
\end{align} 

The variational principle of the bulk action leads to the boundary terms 
\begin{align}
    \delta S_{bulk}&=-2\int_{\partial\mathcal{M}_-} \text{Tr}(\mathbf{B}\delta \mathbf{A}\rvert_{\partial})+2\int_{\partial\mathcal{M}_+} \text{Tr}(\mathbf{B}\delta \mathbf{A}\rvert_{\partial}).
\end{align}
In order to have a well-defined variational principle, we need to impose suitable boundary conditions and/or include additional terms. For the case of producing asymptotic AdS boundary conditions, or to produce general particle on a group boundary actions with the Hamiltonian as the quadratic Casimir, the boundary action and condition was described in \cite{Mertens:2018fds,Gonzalez:2018enk,Gaikwad:2018dfc}. This corresponds to the boundary term one would derive from an $s$-wave dimensional reduction from 3d Chern-Simons theory, and relates this description directly to the Chern-Simons/Wess-Zumino-Witten duality \cite{Elitzur:1989nr}.

However, in order to implement either finite cut-off or EOW brane boundary conditions, we have to generalize beyond this \cite{Ferrari:2020yon}. 

We next describe the specific boundary conditions in the BF first-order language for each of the two brane-like boundaries in turn.

\subsubsection{EOW brane boundary}\label{sec:eowbdryaction}

 Taking the EOW brane to be at the left boundary, we choose to add the following boundary term to the bulk BF action
 \begin{align}
 \label{eq:firstordereowbraneaction}
    S_{\text{EOW}}&=2\int_{\text{EOW}} \text{Tr}(\mathbf{B}\mathbf{A}\rvert_\partial)= \int_{\text{EOW}} \left(-\phi^a\eta_{ab}e^b+\omega \Phi\right).
\end{align}
The variation of the total action $S_{bulk}+S_{\text{EOW}}$ at the EOW brane boundary then reads in components
\begin{align}
    \delta S \equiv \delta S_{bulk}+\delta S_{\text{EOW}} = 2\int_\text{EOW} \text{Tr}(\delta \mathbf{B}\mathbf{A}\rvert_\partial)=-\int_\text{EOW} \left(\delta \phi^{a}\,e_{a}+\delta \Phi \,\omega\right).
\end{align} 
We fix the boundary conditions by taking $\phi^a$ to be constant along the brane $\delta\phi^a\equiv 0$ and fix the spin connection $\omega\equiv 0$ along the brane.

\begin{figure}[h]
    \centering
    \input{images/boundary_frame.tikz}
    \caption{Adapted local frame on each of the boundary brane segments.}
    \label{fig:LLFbranes}
\end{figure}
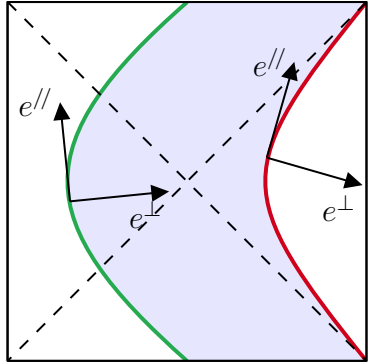

Choosing an adapted local frame field orientation to the EOW brane along the tangent and opposite to the outward normal $e_{/\!/}^\mu=t^\mu$, $e_\perp^\mu=-n^\mu$, the unit normal in components is simply given by $n^a=-\delta^a_1$ (see Fig.~\ref{fig:LLFbranes}).
The extrinsic curvature along the oriented EOW brane geodesic is now given by\footnote{We have used that  $\omega_{\mu\;1}^{0}=-\omega_\mu$ from $\epsilon^{0}_{\;\;1}=-1$ since $\epsilon_{01}=1$ and 0 is timelike.} 
\begin{align}
\label{eq:extrisnsiccruvature}
    K&=e^\mu_a\partial_\mu n^a+e^\mu_a\omega^a_{\mu b}n^b=e^\mu_{/\!/}\omega_\mu.
\end{align}
We can then write the first order boundary action \eqref{eq:firstordereowbraneaction} in second order variables as \begin{align}
    S_{\text{EOW}}=\int_{\text{EOW}}d\lambda \sqrt{-h_{\lambda\lambda}}\left(\Phi K+\phi^{/\!/}\right).
\end{align} 
This boundary action indeed matches with the second-order action defining an EOW brane in \eqref{eq:totalaction}, if we identify $\phi^{/\!/}$ in terms of the constant tension along the brane: \begin{align}\label{eq:firstorderbraneboundarycondition}
    \phi^{/\!/}\equiv -\mu.
\end{align} 
The first-order boundary condition $\omega\equiv0$ corresponds to $K=0$ geodesics, in agreement with the defining EOW brane boundary condition \eqref{eq:eowboundaryconditions}. Furthermore, we can relate the constant $\delta \phi^{/\!/}\equiv 0$ boundary condition in terms of the normal derivative of the dilaton. 
The last Gauss law constraint of \eqref{eq:DB_components_gjk}, $\partial_\mu \Phi+\epsilon_{ab}e_\mu ^a\phi^b=0$, can be inverted by taking the Hodge dual (with $\epsilon_{ab}\epsilon^{ac}=-\delta_b^c$) 
\begin{align}
\label{eq:hodgedual}
    \phi^a=-\epsilon^{ab}e^\mu_b\partial_\mu \Phi,
\end{align} 
which leads to the tangent component of the dilaton field in terms of the normal derivative of the dilaton and vice versa:
\begin{align}
\label{eq:constraints}
    \phi^{/\!/}=e_\perp^\mu \partial_\mu \Phi=-n^\mu \partial_\mu \Phi, \qquad \phi^\perp=-e^\mu_{/\!/}\partial_\mu \Phi=-t^\mu \partial_\mu \Phi,
\end{align}
using the earlier identification of the future directed positively oriented frame fields $e_{/\!/}^\mu=t^\mu$, $e_\perp^\mu=-n^\mu$. Setting $\phi^{/\!/}\equiv -\mu$ \eqref{eq:firstorderbraneboundarycondition} then leads to the defining second order boundary condition of the EOW brane \eqref{eq:eowboundaryconditions}:\footnote{
Note that for an EOW brane on a right oriented boundary, we would set $\phi^{/\!/}=\mu$ and $ n^\mu\partial_\mu\Phi=\mu$.
} 
\begin{align}
    n^\mu\partial_\mu\Phi=\mu.
\end{align}

Having fixed the component of $P_0$ as $-\mu$, we can consider the one-parameter group of gauge transformations $h=e^{\theta P_0}$. The corresponding gauge fields then transform as $\mathbf{A}\mapsto h \mathbf{A}h^{-1}-dhh^{-1}$, leading to \begin{align}
e^{\prime0}&=e^0-d\theta,\\
e^{\prime1}&=\cos\theta\,e^1+\sin\theta\,\omega,\\
\omega'&=-\sin\theta\,e^1+\cos\theta\,\omega.
\end{align}Along the boundary curve, we align the frame at every point such that $\gamma^* e^0=e^{/\!/}$ and $\gamma^*e^\perp=0$ (Fig.~\ref{fig:LLFbranes}).
On a generic curve, the frame field components are no longer aligned with the curve at every point since $e^{\prime \perp}=\sin\theta \omega\neq 0$. On a geodesic with $K=0$, we have \eqref{eq:extrisnsiccruvature}, leading to $\gamma^*\omega=0$. The transformation then preserves the  frame orientation along the curve after parallel transporting the frame 
\begin{align}
    \gamma^*e^{\perp}\rightarrow \gamma^*e^{\prime \perp}=0, \qquad \gamma^*\omega\rightarrow \gamma^*\omega=0.
\end{align} 
Along the boundary curve, the interpretation of the proper time is thus preserved after parallel transport 
\begin{align}
\gamma^*e^{\prime0}=d\tau-d(\theta|_\gamma).
\end{align} 
leading to the boundary time reparametrization $\tau'=\tau-\theta|_\gamma$.

\subsubsection{Finite cutoff boundary}\label{sec:fcbdryaction}
The second order JT action equipped with a finite cutoff boundary  is given by \eqref{eq:totalaction}:
\begin{align}\label{eq:totaladsaction}
   S=\frac{1}{2}\int_{\mathcal{M}} d^2x \sqrt{-g}\Phi(R+2)+\int_{\partial \text{AdS}} du \sqrt{-h}\Phi(K-1).
\end{align}
Taking the finite-cutoff curve to be on the right oriented boundary of the strip, we can make a local Lorentz transformation on each point such that the positive frame field configuration adapted to the cutoff curve points along the tangent and along the outward normal of the curve: $e^\mu_{/\!/}=t^\mu$, $e^{\mu}_{\perp}=n^\mu$. Pulling back the frame fields along the boundary leads to a rewriting of the total action in first order variables:
\begin{align}\label{eq:totalfirstorderfinitecutoffaction}
    \eqref{eq:totaladsaction}=\int_{\mathcal{M}} \left(\phi^a\left(de_a+\epsilon_{ab}\omega\land e^b\right)+\Phi (-d\omega+e^0\land e^1)\right)-\int_{\partial \text{AdS}}\Phi\left( \omega + e^{/\!/}\right).
\end{align}  
Varying this first order action leads to the total boundary term
\begin{align}
    \delta S=\int_{\partial \text{AdS}} \left(\phi^{/\!/}-\Phi\right)\delta e^{/\!/}-\delta \Phi(e^{/\!/}+ \omega).
\end{align} 
Taking Dirichlet boundary conditions for both the tangent frame field $\delta e^{/\!/}\equiv 0$ and the dilaton $\delta\Phi\equiv 0$ along the boundary leads to a well-posed variational principle.
This condition is equivalent to the metric boundary condition \eqref{eq: boundary conditions}. 
Furthermore, setting $\Phi\rvert_{\partial\mathcal{M}}\equiv \Phi_b$ constant leads to the defining finite-cutoff boundary condition. Moreover, since the boundary Hamiltonian is given by $T_{\tau\tau}=-(e^{/\! /})^2\frac{\delta S}{\delta e^{/\!/}}$, the factor multiplying $\delta e^{/\!/}$ is the boundary Hamiltonian:
\begin{align}
\label{eq:Ham}
    H=\Phi-\phi^{/\!/}.
\end{align}
We can then rewrite the AdS boundary action in \eqref{eq:totalfirstorderfinitecutoffaction} as 
\begin{align}
    -\int_{\partial \text{AdS}}\Phi\left( \omega + e^{/\!/}\right)=-\int_{\partial \text{AdS}} (e^{/\!/}\phi^{/\!/}+\omega \Phi) -\int_{\partial \text{AdS}} e^{/\!/}H,
\end{align}
or in the BF formulation:\footnote{On a left oriented boundary,
the Hamiltonian has the same second-order form $H=\Phi+\phi^{/\!/}=\Phi-n^\mu \partial_\mu \Phi$, and the total boundary term looks like 
\begin{align}
\label{eq:canonicalleftfinitecutoff}
    S_{\partial\text{AdS}_-}=2\int_{\partial\mathcal{M}_- } \text{Tr}(\mathbf{B}\mathbf{A}|_\partial)-\int_{\partial\mathcal{M}_-} e^{/\!/}H
\end{align}
} 
\begin{align}\label{eq:canonicalrightfinitecutoff}
    \boxed{S_{\partial\text{AdS}_+}=-2\int_{\partial\mathcal{M}_+} \text{Tr}(\mathbf{B}\mathbf{A}\rvert_{\partial})-\int_{\partial\mathcal{M}_+} e^{/\!/}H}.
\end{align}
This is the action we have to add to the bulk BF action in this case. Note that the EOW boundary action \eqref{eq:firstordereowbraneaction} is just a special case of this one with $H=0$, so both can be discussed simultaneously.

Looking back at the solutions of the Gauss law constraint \eqref{eq:hodgedual}, we can again identify
$\phi^{/\!/}=n^\mu\partial_\mu \Phi$ (with opposite orientation on the right boundary).
Our boundary Hamiltonian \eqref{eq:Ham} is now equal to its second-order counterpart \eqref{eq:genericADM}: 
\begin{align}
\label{eq:Hamnew}
    H=\Phi_b-n^\mu\partial_\mu \Phi\rvert_{\partial\mathcal{M}}. 
\end{align} 
Furthermore, since the dilaton is constant along the brane, the tangent derivative vanishes, such that $\phi^{\perp}=-t^\mu \partial_\mu \Phi=0$. The field $\mathbf{B}$ along the boundary is then generically expanded as $\mathbf{B}_\partial=\Phi_bP_2- \phi^{/\!/} P_{0}$, with $2\text{Tr}(\mathbf{B}^2\rvert_\partial)=\Phi_b^2-(\phi^{/\!/})^2$, such that we can solve for $\phi^{/\!/}$ as 
\begin{align} \label{eq:finitecutoffadm}
    \phi^{/\!/}=\sqrt{\Phi_b^2-2\text{Tr}(\mathbf{B}|_\partial^2)}\equiv n^\mu \partial_\mu \Phi>0
\end{align} 
where we take the positive sign for the root, corresponding to a growing dilaton profile towards to the right boundary.
The Gauss law constraint $D_{\sigma}\mathbf{B}=0$  implies that $\mathbf{B}(\sigma)$ is covariantly constant along the spatial interval. This implies in turn that $\text{Tr}(\mathbf{B}^2)$ is invariant along a spatial slice, since $d\text{Tr}(\mathbf{B}^2)=2\text{Tr}(\mathbf{B}d\mathbf{B})=-2\text{Tr}(\mathbf{B}[\mathbf{A},\mathbf{B}])=0.$ Since the dilaton profile reaches an extremal value in the bulk at the bifurcation point on the horizon, all directional derivatives vanish, by \eqref{eq:constraints} setting $\phi^{a}=0$. The norm of $\mathbf{B}$ then generically equals the value of the dilaton at the horizon
\begin{align}
\label{eq:dilatonhorizon}
    2\text{Tr}(\mathbf{B}^2\rvert_\partial)=2\text{Tr}(\mathbf{B}^2_h)=\Phi_h^2,
\end{align}
and the boundary Hamiltonian observable \eqref{eq:Hamnew} 
\begin{align}
\label{eq:Hamnewest}
    H=\Phi_b-\sqrt{\Phi_b^2-\Phi_h^2} 
\end{align} 
now naturally agrees with the finite-cutoff energy formula \eqref{eq:explicitadmenergy} of the classical solution.

The gauge transformation corresponding to the one-parameter subgroup generated by $P_2$, whose component was fixed as $\Phi_b$, acts as a local Lorentz transformation  $h=e^{\eta P_2}\in \text{SO}(1,1)$. The gauge transformation on the gauge fields $\mathbf{A}\mapsto h \mathbf{A}h^{-1}-dhh^{-1}$ rotate the local Lorentz grid as
\begin{align}\label{eq:componenttransformations}
e'^{0}&=\cosh\eta\,e^0
-\sinh\eta\,e^1,\\
e'^{1}&=-\sinh\eta\,e^0
+\cosh\eta\,e^1,\\\omega'&=\omega-d\eta.
\end{align} The parameter $\eta\in \mathbb{R}$ corresponds physically to the rapidity of the corresponding Lorentz boost.

\section{Reduction to 2d phase space}
\label{s:redphase}
In this section, we will reduce the local BF phase space by its bulk constraints, and implement the gravitational boundary conditions. This first step yields the particle-on-the-group phase space \(T^*G\), with \(G=\widetilde{\mathrm{SL}(2,\mathbb R)}\). It is worth carrying out the reduction explicitly, since the endpoint variables that survive will subsequently become the moment maps through which the gravitational boundary conditions are implemented.

\subsection{Bulk reduction to $T^* G$}
\label{sec:phasespace}
Consider the BF theory on a Lorentzian strip
\(
\mathcal M\simeq\mathbb R_\nu\times I_\sigma
\),
where \(I_\sigma=[\sigma_-,\sigma_+]\) is a spacelike interval, and decompose the connection as $\mathbf A=\mathbf A_\nu d\nu+\mathbf A_\sigma d\sigma$. The field strength can then be written as $\mathbf{F}_{\nu\sigma}
    =
    \partial_\nu\mathbf{A}_\sigma-\partial_\sigma\mathbf{A}_\nu+[\mathbf{A}_\nu,\mathbf{A}_\sigma]
    =
    \dot{\mathbf{A}}_\sigma-D_\sigma\mathbf{A}_\nu$, where the dot denotes a derivative with respect to $\nu$, and $D_\sigma$ is defined as the adjoint covariant derivative along the radial direction.
The full BF action including our boundary terms \eqref{eq:canonicalrightfinitecutoff} can then be written in the canonical form
\begin{align}
S_{\rm bulk}
&=
-2\int d\nu\,d\sigma\,
\Tr\!\left(\mathbf B\,\dot{\mathbf A}_\sigma\right) -
2\int d\nu\,d\sigma\,
\Tr\!\left(D_\sigma\mathbf B \mathbf A_\nu\right) -\int_{\partial\mathcal{M}_-} d\tau_- \;H(\mathcal{C})-\int_{\partial\mathcal{M}_+} d\tau_+ \;H(\mathcal{C}),
\end{align} in terms of the proper time $d \tau_\pm=N_\pm(\nu) d\nu $ defined through the lapse along each boundary. 
The variables \(\mathbf A_\sigma(\sigma)\) and \(\mathbf B(\sigma)\) form a canonical pair with symplectic potential
\begin{align}
\Theta_{\rm unred}
=
-2\int_{\sigma_-}^{\sigma_+}d\sigma\,
\Tr\!\left(\mathbf B\,\delta\mathbf A_\sigma\right).
\label{eq:unred}
\end{align}
The temporal component \(\mathbf A_\nu\) carries no time derivative and acts as a Lagrange multiplier, which enforces the Gauss law constraint in the radial direction along the spatial slice
\begin{align}
D_\sigma\mathbf B=0.
\label{eq:Gauss_constraint}
\end{align}
The field \(\mathbf B\) is thus covariantly constant along the interval, and its value at a single point determines it everywhere by parallel transport along the gauge connection $\mathbf{A}_\sigma$.

Imposing \eqref{eq:Gauss_constraint} is only half of the bulk reduction, since the constraint surface must also be quotiented by the gauge transformations that the constraint generates, which in this case is the usual gauge redundancy of gauge group \(\widetilde{\mathrm{SL}(2,\mathbb R)}\). This is standard procedure in symplectic reduction; we present more explicit details in our set-up in Appendix \ref{app:sympl}. If for a moment we restrict to gauge transformations that are trivial at the endpoints, then after the symplectic quotient the connection \(\mathbf A_\sigma(\sigma)\) retains only the boundary-to-boundary Wilson line defined as
\begin{align}
g
\equiv
\mathcal P\exp\!\left(
-\int_{\sigma_-}^{\sigma_+}\mathbf A_\sigma d\sigma
\right)
\in G .
\label{eq:holonomy}
\end{align}
Under a large gauge transformation, that possibly now has support at the two boundaries, the Wilson line transforms under the endpoint gauge group
\begin{align}
g\longmapsto h_+\,g\,h_-^{-1},
\label{eq:holonomytransf}
\end{align}
The role of such transformations becomes important once the gravitational boundary conditions are imposed.

Therefore in total, before imposing the gravitational boundary conditions, we are left with a finite number of variables: three from the open holonomy $g$ and three from the field $\mathbf B$ evaluated on a point of the interval $[\sigma_-, \sigma_+]$.

What remains now is to determine the symplectic structure carried by these finite-dimensional variables. Denoting the Wilson line starting at the left boundary and ending at a bulk point $\sigma$ by
\begin{align}
U(\sigma) =
\mathcal P\exp\!\left(
-\int_{\sigma_-}^{\sigma}\mathbf A_\sigma d\sigma
\right)
\label{eq:wlineU}
\end{align}
defined by the parallel transport by the gauge connection $\partial_\sigma U(\sigma)
 =
 -\mathbf A_\sigma(\sigma)U(\sigma)$
with  $U(\sigma_-)=\mathbf 1$.
  Using \eqref{eq:Gauss_constraint}, one finds $\partial_\sigma
 \left(
 U^{-1}\mathbf B U
 \right)
 =
 U^{-1}(D_\sigma\mathbf B)U
 =
 0$ and the conjugate field consequently evolves on the Cauchy slice under the adjoint action of the Wilson line $\mathbf{B}(\sigma)=U(\sigma) \mathbf{B}_-U(\sigma)^{-1}$ and contains one independent degree of freedom at the boundary. Between two endpoints, the field configurations are connected by
 \begin{align}
 \mathbf B_+
 =
 g\,\mathbf B_-\,g^{-1},
 \qquad
 \mathbf B_\pm\equiv\mathbf B(\sigma_\pm).
 \label{eq:Btransport}
 \end{align}

Defining
\begin{equation}
K(\sigma)\equiv\delta U(\sigma)U(\sigma)^{-1},
\end{equation}
for which
\(
\delta\mathbf A_\sigma=-D_\sigma K
\),
the potential \eqref{eq:unred} becomes
\begin{align}
\Theta_{\rm unred}
=
2\int_{\sigma_-}^{\sigma_+}d\sigma\,
\Tr\!\left(\mathbf B D_\sigma K\right)
=
2\left.\Tr(\mathbf B K)\right|_{\sigma_-}^{\sigma_+}
-
2\int_{\sigma_-}^{\sigma_+}d\sigma\,
\Tr\!\left((D_\sigma\mathbf B)K\right).
\end{align}
On the Gauss-constraint surface, the bulk contribution vanishes, and since \(U(\sigma_-)=\mathbf 1\) implies \(K(\sigma_-)=0\), while
\(
K(\sigma_+)=\delta g\,g^{-1}
\),
only the right endpoint survives. The reduced symplectic potential is therefore
\begin{align}
\Theta_{\text{red}}
=
2\Tr\!\left(
\mathbf B_+\,\delta g\,g^{-1}
\right)
=
2\Tr\!\left(
\mathbf B_-\,g^{-1}\delta g
\right).
\label{eq:reducedsymplecticpotential}
\end{align}
where in the second equality we expressed it through its left endpoint at $\sigma_-$.
We recognize the components of the $\mathbf{B}$-field as the momenta conjugate to the displacements of the configurational variables measured by the Cartan-Maurer one-forms. We have thus reduced the system from the local fields \((\mathbf A_\sigma(\sigma),\mathbf B(\sigma))\) to a single group element \(g\) together with a single Lie-algebra-valued momentum, i.e. the cotangent bundle $T^*G$.

Viewed as functions on \(T^*G\), the components of $\mathbf{B}$ at either endpoint generate Hamiltonian flows, which follow from the reduced symplectic potential \eqref{eq:reducedsymplecticpotential}. In particular, one finds that \(B_{I}^+\) generates the left action
\begin{align}
g(t)
=
e^{tP_I}g(0),
\qquad
\mathbf B_+(t)
=
e^{tP_I}\mathbf B_+(0)e^{-tP_I},
\label{eq:Bplus_Hamiltonian_flow}
\end{align}
while \(B_{I}^-\) generates the corresponding right action
\begin{align}
g(t)
=
g(0)e^{tP_I},
\qquad
\mathbf B_-(t)
=
e^{-tP_I}\mathbf B_-(0)e^{tP_I}.
\label{eq:Bminus_Hamiltonian_flow}
\end{align}
These Hamiltonian vector fields and their Poisson algebra are derived in detail in Appendix~\ref{app:hamiltonian_vector_fields_TstarG}. Comparing with \eqref{eq:holonomytransf}, the flows are precisely the endpoint transformations.
The moment maps associated with the action
\(
g\mapsto h_+gh_-^{-1}
\)
are
\begin{align}
\mu_{I}^-P^I=-\mathbf B_-,
\qquad
\mu_{I}^+P^I=\mathbf B_+ .
\end{align}

We emphasize that only the bulk gauge constraint has been imposed at this stage. The endpoint transformations \eqref{eq:holonomytransf} have not been quotiented and still act on \(T^*G\) as genuine Hamiltonian symmetries, so that nothing in the reduction so far is specific to JT gravity. The gravitational input enters through the boundary conditions, which act precisely on this residual endpoint structure as we discuss next.

\subsection{Boundary conditions and Hamiltonian reduction}

As we discussed in Section \ref{sec:firstorderboundaryactions}, in the first-order formulation, the gravitational boundary conditions fix particular components of the endpoint fields \(\mathbf B_\pm\). 
A boundary condition imposed on the components of the $\mathbf{B}$-field then takes the form
\begin{align}
B_{I}^\pm
\equiv
2\Tr(\mathbf B_\pm P_I)
= \text{fixed},
\label{eq:constraint}
\end{align}
where \(P_I\) is the Lie-algebra generator selected by the boundary condition.  The boundary conditions \eqref{eq:constraint} fix components of the moment maps, and their implementation amounts to a further Hamiltonian reduction of \(T^*G\).

Once such a component is fixed, only part of the endpoint symmetry remains compatible with it. Since the endpoint fields transform by conjugation under \eqref{eq:Bplus_Hamiltonian_flow} and \eqref{eq:Bminus_Hamiltonian_flow}, the connected transformations preserving the constraint are those generated by elements commuting with \(P_I\), i.e. $\exp(\mathbb R P_I)$.
Which generator \(P_I\) appears, and hence which subgroup survives, depends on the boundary condition. 

At a finite-cutoff boundary, the \(P_2\) component is fixed by the value of the dilaton at the boundary,
\begin{align}\label{eq:Bfinitecutoffconstraint}
    \Phi_b\equiv 2\text{Tr}(\mathbf{B}_\partial \,P_2).
\end{align}
The finite residual gauge transformations compatible with this boundary constraint are therefore the connected local Lorentz transformations,
\begin{align}\label{eq:localLorentzstabiliezr}
H = \exp\bigl(\mathbb R(E+F)\bigr)
\simeq \text{SO}(1,1).
\end{align}
 This identification follows from the coset description of the bulk \(\text{AdS}_2=\text{SL}(2,\mathbb{R})/\text{SO}(1,1)\).
 All points of the homogeneous space stay invariant under the right action of \(\text{SO}(1,1)\). This action therefore preserves the dilaton value \(\Phi_b\), the induced boundary metric and the shape of the physical cutoff curve, whereas other \(\text{SL}(2,\mathbb R)\) transformations generically change the fixed dilaton component or move the cutoff surface.

For an EOW brane boundary, the first-order boundary condition \eqref{eq:firstorderbraneboundarycondition} fixes instead the timelike component $P_0$
\begin{align}\label{eq:ellipticeowboundarycondition}
    -\mu\equiv 2\text{Tr}(\mathbf{B}_-\,P_0).
\end{align}
The corresponding preserved one-parameter subgroup is then
\begin{align}\label{eq:localEOWbranestabilizer}
    H=\exp\bigl(\mathbb R(E-F)\bigr),
\end{align}
namely the universal cover of SO\((2)\) in \(\widetilde{\mathrm{SL}(2,\mathbb R)}\).
Its geometric action, a translation along the EOW geodesic, is derived in Appendix~\ref{app:isometricembedding}.

Fixing the moment-map values does not by itself complete the reduction, since the stabilizers still act on the constraint surface. Whether their orbits are to be retained as symmetries or quotiented as redundancies is settled by the symplectic structure. Let \(X_{B_{I}^\pm}\) denote the Hamiltonian vector field generated by the constrained component \(B_{I}^\pm\), which by \eqref{eq:Bplus_Hamiltonian_flow} and \eqref{eq:Bminus_Hamiltonian_flow} is the infinitesimal action of \(H_I\). Any vector \(V\) tangent to the level set
\(
B_{I}^\pm=b_I^\pm
\)
satisfies
\(
\delta B_{I}^\pm(V)=0
\),
and hence $
\Omega_G\!\left(X_{B_{I}^\pm},V\right)
=
\delta B_{I}^\pm(V)
=
0$. The stabilizer flow is therefore a null direction of the symplectic form pulled back to the constraint surface, and must be quotiented. We see that the transformation that acted as a genuine endpoint symmetry on the partially reduced phase space becomes a redundancy once the associated moment-map value is fixed, as it should be.

How many degrees of freedom survive then follows from a simple count. Each such one-dimensional Hamiltonian reduction removes two dimensions, one by restricting to a fixed moment-map level set and one by quotienting the associated flow. Starting from \(\dim T^*G=6\) and imposing one boundary condition at each endpoint, we are left with
\begin{align}
\dim\mathcal P_{\rm grav}
=
6-2-2
=
2,
\end{align}
so that the physical gravitational phase space must be described by a single canonical pair.

On the holonomy itself, the same reduction identifies configurations related by the stabilizer actions,
\begin{align}
\label{eq:equivclass}
g\sim h_+\,g\,h_-^{-1},
\qquad
h_\pm\in H_\pm ,
\end{align}
and the physical configuration space is therefore the double coset
\begin{align}
\mathcal C_{\rm grav}
=
H_+\backslash G/H_- .
\label{eq:gravitational_double_coset}
\end{align}

\subsection{Geodesic length and the physical canonical pair}\label{sec:gravphasespace}

The Hamiltonian reduction leaves us with a two-dimensional gravitational phase space whose configuration-space projection is the double coset \eqref{eq:gravitational_double_coset}. The reduction by itself, however, does not immediately provide a physical interpretation of the remaining variable. To give it a physical meaning, we choose a representative of the double-coset class adapted to the bulk geometry. Generically, the Gauss constraint mods out interior gauge transformations. In the gravitational theory, these gauge transformations contain both local Lorentz transformations that allow us to choose an adapted frame at every point of the curve, and  the  diffeomorphisms which that freely deform the shape of the curve.  We therefore choose an adapted representative by evaluating the holonomy along a spacelike \emph{geodesic} connecting the two boundaries. As we now show, the corresponding single coordinate is the proper length of this geodesic, as we argued previously in \cite{Belaey:2024dde}.

\begin{figure}[h]
    \centering
    \input{images/local_frame.tikz}
    \caption{Local Lorentz frame adapted to the bulk geodesic curve $\gamma$.}
    \label{fig:LLF}
\end{figure}
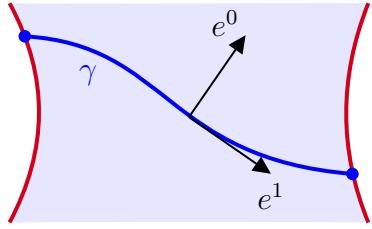

Let \(\gamma\) denote a geodesic connecting the two boundaries. In the gravitational decomposition
\begin{align}
\mathbf A
=
e^0 P_0+e^1 P_1+\omega P_2,
\end{align}
with \(P_1\equiv H\), we choose an orthonormal frame along $\gamma$ such that 
$
\gamma^*e^0=0,
\,
\gamma^*e^1=ds
$,
where \(s\) is proper length along the geodesic, see Fig.~\ref{fig:LLF}. 
In this Lorentz frame, the pullback of the spin connection measures the curvature and since \(\gamma\) is a geodesic \(\gamma^* \omega =0\) by \eqref{eq:extrisnsiccruvature}. Therefore the pullback of $\mathbf A$ reduces to \(\gamma^*\mathbf A=H\,ds\). The Wilson line, hence, admits the representative
\begin{align}
g_\gamma
=
\mathcal P
\exp\!\left(
-\int_\gamma \mathbf A
\right)
=
e^{-LH},
\qquad
L\equiv\int_\gamma ds .
\label{eq:geodesic_representative}
\end{align}

Equation \eqref{eq:geodesic_representative} holds in the gauge adapted to the geodesic, so it does not describe a generic Wilson line \(g\) on \(T^*G\). Residual gauge transformations preserving the boundary conditions, induce physical symmetries at the endpoints \eqref{eq:holonomytransf}. A generic element in the equivalence class \eqref{eq:equivclass} is then written with the inclusion of the boundary stabilizer subgroups 
\begin{align}\label{eq:totalholonomystabilzer}
    g = H_+ e^{-LH}H_-^{-1}.
\end{align} 
The group element $g$ constructed as such does not represent the most general $\widetilde{\text{SL}(2,\mathbb{R})}$ element; a particular patch of the total $\widetilde{\text{SL}(2,\mathbb{R})}$ group manifold is selected that depends on the specific boundary conditions at the two endpoints.

The conjugate momentum paired with this coordinate  follows from the reduced symplectic potential \eqref{eq:reducedsymplecticpotential}. For the representative \(g=e^{-LH}\) one has \(\delta g\,g^{-1}=-H\,\delta L\). Writing the endpoint field as $
\mathbf B_+
=
-\phi^0 P_0-\phi^1 H+\Phi P_2$, 
and using \(2\Tr(H^2)=1\), one obtains
\begin{align}
\Theta_{\text{red}}
&=
2\Tr\!\left(
\mathbf B_+\,\delta g\,g^{-1}
\right)=
\phi^1\,\delta L .
\end{align}
The momentum conjugate to the geodesic length is therefore
\(
p_L=\phi^1,
\label{eq:Pi_L_definition_gjk}
\)
and the gravitational symplectic form takes the canonical form
\begin{align}
\Omega_{\rm red}
=-\delta \Theta_\text{red}=\delta L\land
\delta p_L,
\qquad
\{L,p_L\}=1 ,
\label{eq:gravitational_symplectic_form}
\end{align}
so that the two-dimensional gravitational phase space $\mathcal{P}_{\rm grav}$ is locally parametrized by the canonical pair $\mathcal P_{\rm grav}
\simeq
\{(L,p_L)\}$.\footnote{Since the identification \(p_L=\phi^1\) was obtained with a geodesic representative, it remains to check that the reduced symplectic structure does not depend on this choice. Consider another representative \(g'=h_+\,g\,h_-^{-1}\), with \(h_\pm=e^{\alpha_\pm P_{\pm}}\in H_\pm\) and \(\alpha_\pm\) parametrizing the stabilizer transformations. Because \(b_\pm\) are fixed by the boundary conditions, the change in the symplectic potential on the fixed moment-map level set is exact,
\begin{align}
\Theta_\text{red}'
-
\Theta_\text{red}
=
b_+\,\delta\alpha_+
-
b_-\,\delta\alpha_-
=
\delta\!\left(
b_+\alpha_+-b_-\alpha_-
\right),
\end{align}
and hence \(\delta\Theta_{\text{red}}'=\delta\Theta_{\text{red}}\). Different representatives of the same double-coset class therefore differ only by a canonical reparametrization and define the same reduced symplectic form.}

\section{Dynamical systems between two boundaries}
\label{s:dsystems}
In this section, we apply the machinery developed in the previous sections with the goal of developing closed expressions for the dynamical system describing physics between two boundaries.

\subsection{Two finite-cutoff boundaries}
\label{s:twophi}
We begin with a spatial slice bounded by two finite-cutoff curves and study the general patch decomposition of the group manifold. Consequently, we study the gravitational coset and its corresponding classical phase space.

According to the discussion in section \ref{sec:fcbdryaction}, fixing the dilaton field to $\Phi_b$ imposes a constraint on the $P_2$-component of the $\mathbf{B}$-field. Its conjugate variable in phase space $\eta\in \mathbb{R}$ is the rapidity of the corresponding Lorentz boost $e^{\eta P_2}$ that rotates the local Lorentz grid at every point while keeping its location intact. 

Note that this is a purely first order interpretation. From a second-order gravity perspective, we don't have any a priori geometrical intuition on the conjugate labels of the boundary conditions fixing the dilaton. 

\subsubsection{Gravitational group decomposition and classical symmetries}
Considering the setup with two finite cutoff boundaries, the relevant subgroups that preserve the boundary conditions are the local Lorentz group at each boundary which become physical large local Lorentz symmetries \eqref{eq:localLorentzstabiliezr}. The gravitational group decomposition \eqref{eq:totalholonomystabilzer} for the finite cutoff case then corresponds to the ``hyperbolic Euler angle'' parametrization of an $\widetilde{\text{SL}(2,\mathbb{R})}$ group element:
\begin{align}
\label{eq:finitecutoffdecomp}
        g=e^{\beta(E+F)}e^{-LH}e^{\gamma(E+F)}.
\end{align}
Here, $\beta$ and $\gamma$ are group parameters which are the corresponding rapidities of the large local Lorentz symmetries which preserve both boundaries, and $L>0$ is the bulk geodesic length. Note that the patch only covers the hyperbolic conjugacy classes with $\text{Tr}(g)>2$, which has an important structural implication as we will discuss further on.

In the infinite cutoff asymptotic JT limit, we let $\Phi_b^-=\Phi_b^+\rightarrow\infty$. We should renormalize the bare geodesic length variable $L$ as:
\begin{align}\label{eq:lengthrenormalization}
    L\equiv L_r+2\log \Phi_b,
\end{align}
where the renormalized length $L_r$ takes all real values in $]-\infty, \infty[$. Commuting $e^{-\log \Phi_b H}$ past the ``outer'' subgroups in \eqref{eq:finitecutoffdecomp} leads to the equivalent expression
\begin{align}\label{eq:shiftedhyperbolicEuler}
    g= e^{-\log \Phi_b H}e^{\beta(\Phi_b E+F/\Phi_b) }e^{-L_rH}e^{\gamma(\Phi_bF+E/\Phi_b)}e^{-\log\Phi_b H}.
\end{align} 
In the infinite cutoff limit $\Phi_b\rightarrow\infty$, keeping fixed $\beta_r\equiv \Phi_b\beta$ and $\gamma_r=\Phi_b\gamma$, \eqref{eq:shiftedhyperbolicEuler} reproduces the parabolic Gauss-Euler decomposition of SL$(2,\mathbb{R})$, up to the ``outer'' factors which can be absorbed by a redefinition of the appropriate boundary states on which they act. Hence in this case
\begin{align}\label{eq:gepatch}
    g \simeq e^{\beta \Phi_b E}e^{-L_rH}e^{\gamma \Phi_bF}.
\end{align}

The reduced symplectic potential \eqref{eq:reducedsymplecticpotential} defines a phase space with canonical conjugate momenta $p_\beta,\,p_L,\,p_\gamma$ to the respective group coordinates
\begin{align}\label{eq:canonicaloneformcoordinates}
    \Theta_{\text{red}}
=
2\Tr\!\left(
\mathbf B_+\,\delta g\,g^{-1}
\right)
=
2\Tr\!\left(
\mathbf B_-\,g^{-1}\delta g
\right)=p_\beta\, \delta \beta+p_L\, \delta L+p_\gamma\, \delta \gamma.
\end{align} 
Using the Hamiltonian flow of the moment maps associated to the canonical conjugate momenta, the phase space satisfies the non-vanishing classical Poisson brackets  $\{\gamma, p_\gamma\}=\{L,p_L\}=\{\beta,p_\beta\}=1$. 
The associated right- and left-regular moment maps are derived in Appendix \ref{app:classicalsymmetrygenerators} and given explicitly respectively by
\begin{align}
\mu_{E+F}^-&=-p_\gamma,\label{eq:momentamap1}\\
\mu_{H}^-&=\frac{\sinh2\gamma}{2\sinh L}\,p_\beta
+\cosh2\gamma\,p_L
-\frac12\coth L\,\sinh2\gamma\,p_\gamma,\\
\mu_{E-F}^-&=-\frac{\cosh2\gamma}{\sinh L}\,p_\beta
-2\sinh2\gamma\,p_L
+\coth L\,\cosh2\gamma\,p_\gamma.
\end{align}
and 
\begin{align}
\mu_{E+F}^+&=p_\beta,\label{eq:momentamap2}\\
\mu_{H}^+
&=\frac12\coth L\,\sinh(2\beta)\,p_\beta
-\cosh(2\beta)\,p_L
-\frac{\sinh(2\beta)}{2\sinh L}\,p_\gamma,\\
\mu_{E-F}^+
&=\coth L\,\cosh(2\beta)\,p_\beta
-2\sinh(2\beta)\,p_L
-\frac{\cosh(2\beta)}{\sinh L}\,p_\gamma.
\end{align} 
Using the canonical Poisson brackets on the group manifold, one can check that the left- and right-regular moment maps on both boundaries satisfy the classical $\mathfrak{sl}_2$-symmetry algebra on phase space \begin{align}\label{eq:classicalpoissonalgebra}
    \{\mu_{H}^\pm,\mu_{E+F}^\pm\}=\mu_{E-F}^\pm,\qquad \{\mu_{H}^\pm,\mu_{E-F}^\pm\}=\mu_{E+F}^\pm, \qquad \{\mu_{E-F}^\pm,\mu_{E+F}^\pm\}=4\mu_{H}^\pm.
\end{align} 
Moreover, since the Hamiltonian flow leaves the opposite boundary invariant, the two moment maps also commute $\{\mu_{A}^+,\mu_{B}^-\}
=
0 $, allowing us to simultaneously constrain one independent generator on each boundary.

Inserting the proper moment maps above, we can explicitly evaluate \begin{align}\label{eq:classicalcasimirfc}
2\text{Tr}(\mathbf{B}_\pm^2)=\mu_{H}^{\pm 2}
+\frac14 \mu_{E+F}^{\pm 2}
-\frac14\mu_{E-F}^{\pm 2}=p_L^2+
\frac{2\cosh L\,p_\beta p_\gamma-p_\beta^2-p_\gamma^2}
{4\sinh^2L}.
\end{align}
Using the first equality, one can show explicitly that it Poisson commutes with every other left-and right-regular current: $\{\Tr(\mathbf{B}_\pm^2),\mu_{I}^\pm\}=0$. 

Although the classical Casimir characterizes the bulk solution, it does not generate time evolution of our theory, as that role is played by the boundary Hamiltonian obtained from the boundary action. For a finite-cutoff boundary with fixed dilaton value \(\Phi_b\), the analysis leading to \eqref{eq:finitecutoffadm} gives
\begin{align}\label{eq:finite_cutoff_hamiltonian_casimir}
H_{\partial}
=
\Phi_b
-
\sqrt{\Phi_b^2-2\text{Tr}(\mathbf{B}|_\partial^2)}.
\end{align}

\subsubsection{Reduced gravitational phase space}
Constraining the value of the dilaton at the left- and right boundary as $\Phi_b^-=2\Tr(\mathbf{B}_- P_2)=-\tfrac{1}{2}\mu_{E+F}^-=p_{\gamma}/2$ and $\Phi_b^+=2\Tr(\mathbf{B}_+P_2)=\tfrac{1}{2}\mu_{E+F}^+=p_\beta/2$ \eqref{eq:Bfinitecutoffconstraint},
in \eqref{eq:classicalcasimirfc}
leads to the reduced Hamiltonian 
\begin{align}
\label{eq:casimir_length_variables_gjk}
2\text{Tr}(\mathbf{B}_\pm^2)=p_L^2+
\frac{2\cosh L\,\Phi_b^-\Phi_b^+-(\Phi_b^+)^2-(\Phi_b^-)^2}
{\sinh^2L} \equiv p_L^2 + V_{\text{cutoff}}(L).
\end{align}
This represents a particle in a potential $V_{\text{cutoff}}(L)$, which we will analyze further in subsection \ref{s:claspot} below.

Alternatively, to express the endpoint fields directly in terms of \((L,p_L)\), it is convenient to solve the Gauss constraint along the geodesic itself, where the connection takes the simple form \(\mathbf A=H\,ds\) with \(s\in[0,L]\). Recalling the decomposition of \( \mathbf{B} = -\phi^0 P_0-\phi^1 H+\Phi P_2 \),
the Gauss constraint \(D_s\mathbf B=0\) reduces to
\begin{align}
\frac{d\phi^1}{ds}=0,
\qquad
\phi^0=\frac{d\Phi}{ds},
\qquad
\frac{d\phi^0}{ds}=\Phi .
\label{eq:length_ode_gjk}
\end{align}
The first equation shows that \(\phi^1=p_L\) is constant along the slice, as expected for the momentum conjugate to the total geodesic length, while the other two combine into \(d^2\Phi/ds^2=\Phi\).
The finite-cutoff boundary conditions fix the dilaton at the two ends, \(\Phi(0)=\Phi_b^-\) and \(\Phi(L)=\Phi_b^+\), and the solution of the bulk equations is then
\begin{align}
\Phi(s)
=
\Phi_b^-\cosh s
+
\frac{\Phi_b^+-\Phi_b^-\cosh L}{\sinh L}\,
\sinh s,
\qquad
\phi^0(s)=\frac{d\Phi}{ds}.
\label{eq:phi_solution_length_gauge_gjk}
\end{align}
Substituting \eqref{eq:phi_solution_length_gauge_gjk} in $2\Tr(\mathbf B(s)^2) =\Phi(s)^2+p_L^2-(\phi^0(s))^2$ at either endpoint $s=0$ or $s=L$ leads directly to \eqref{eq:casimir_length_variables_gjk}.

For the special case of equal cutoff values, \(\Phi_b^-=\Phi_b^+\equiv\Phi_b\), this reduces to
\begin{align}
2\Tr(\mathbf{B}^2)
=
p_L^2
+
\frac{\Phi_b^2}{\cosh^2(L/2)} .
\label{eq:equal_cutoff_casimir}
\end{align}
The finite-cutoff boundary conditions are thus encoded in a hyperbolic P\"oschl--Teller potential on the reduced phase space, which decays exponentially at large separation and approaches the finite value \(\Phi_b^2\) as the two boundaries approach one another. This last result was previously obtained by \cite{Griguolo:2026eoi} from a different approach.

Inserting these expressions \eqref{eq:casimir_length_variables_gjk} into the boundary Hamiltonians, we obtain the final phase space action of the system as
\begin{align}
\label{eq:finalfincut}
   \int d\tau\,p_L\dot{L} - \int_{\partial \text{AdS}_-} \hspace{-0.4cm}d\tau_- \left(\Phi_b^-
-\sqrt{(\Phi_b^-)^2-2\text{Tr}(\mathbf{B}_-^2)}\right)-
\int_{\partial \text{AdS}_+} \hspace{-0.4cm} d\tau_+\left(\Phi_b^+
-
\sqrt{(\Phi_b^+)^2-2\text{Tr}(\mathbf{B}_+^2)}\right).
\end{align}
Since the two boundaries have independent proper times $\tau_\pm$
and associated Hamiltonians $H_\pm$, the choice of evolution
parameter $\tau$ is not unique and depends on how the two boundary
clocks are advanced. In the case of equal cutoffs, however, the
two Hamiltonians coincide, $H_-=H_+\equiv H$, and their contributions
to the action combine as
\begin{equation}
    H_-\,d\tau_-+H_+\,d\tau_+
    =H\,d(\tau_-+\tau_+)=H\,d\tau,
\end{equation}
where we have defined $\tau\equiv\tau_-+\tau_+$.
The dynamics is therefore naturally described by a single
Hamiltonian $H$ conjugate to the total boundary time $\tau$.

Taking both finite cutoff boundaries equal, we can define the two-sided time conventionally as \(\tau=\tau_++\tau_-\). We obtain 
 \begin{align}
     S&=\int d\tau\left( p_L\dot{L}  -\Phi_b
 +
 \sqrt{\Phi_b^2-p_L^2
 -
 \frac{\Phi_b^2}{\cosh^2(L/2)}}\right).
 \end{align} 
In the infinite cutoff $\Phi_b \to \infty$ asymptotic JT limit, we should renormalize the bare geodesic length variable $L$ between the two AdS boundaries as  \eqref{eq:lengthrenormalization} 
    $L\equiv L_r+2\log \Phi_b$. Subsequently redefining the proper time $\tau\rightarrow 2\Phi_b \tau$ yields the usual two-sided holographic description of JT gravity in terms of a Liouville action 
\begin{align}
S=\int d\tau \left(p_{L_r} \dot{L}_r-p_{L_r}^2-4e^{-L_r}\right)=\int d\tau \left(\frac{\dot{L}_r^2}{4}-4e^{-L_r}\right).
\end{align}

 A direct second-order derivation of this Hamiltonian sytem, along the lines of \cite{Harlow:2018tqv}, is provided in Appendix \ref{app:ccbc}.

\subsubsection{Classical analysis of the potential}
\label{s:claspot}
\begin{figure}[h]
    \centering
    \includegraphics[width=0.5\linewidth]{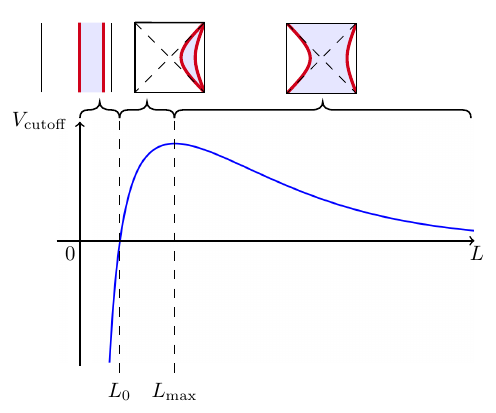}
    \caption{Potential of the dynamical system with two finite-cutoff boundaries as a function of bulk geodesic length $L$, with zero at $L_0$ and maximum at $L_{\text{max}}$.}
    \label{fig:Vfcplot}
\end{figure}

It is now instructive to comment on some features of the potential term appearing in \eqref{eq:casimir_length_variables_gjk}, which is plotted in Fig.~\ref{fig:Vfcplot}. 

We immediately notice that whenever $\Phi_b^{-} \neq \Phi_b^{+}$, the potential exhibits attractive singular behavior $\sim -(\Phi_b^{-} - \Phi_b^{+})^2/L^2$ at the origin $L=0$.\footnote{This constitutes a qualitative difference with respect to the equal-boundary case described in \cite{Griguolo:2026eoi}.}
Without loss of generality, we can take $\Phi_b^- < \Phi_b^+$ from now on.
The potential vanishes at
\begin{equation}
    L_0 = \log\!\left( \frac{\Phi_b^+}{\Phi_b^-} \right),
\end{equation}
whereas its maximum sits at
\begin{equation}
L_\text{max} = \text{arccosh}\!\left( \frac{\Phi_b^{+}}{\Phi_b^-} \right),
\end{equation}
where $V_{\text{cutoff}}(L_\text{max}) = (\Phi_b^{-})^2$. This means that the barrier cannot be overcome as long as we consider configurations with ``energy'' variable $\Phi_h^2$ below this threshold: $\Phi_h^2 < (\Phi_b^{-})^2 < (\Phi_b^{+})^2$.
Our goal is now to provide some geometric interpretation for the different classes of classical solutions. 

For the present discussion, it is convenient to work in Schwarzschild gauge in the bulk:
\begin{equation}\label{eq:metric_SChwarzschild}
    \text{d}s^2 = - (r^2 - \Phi_h^2) \, \text{d}t^2 + \frac{\text{d} r^2}{r^2 - \Phi_h^2}, \qquad \Phi = r.
\end{equation}
Let us first consider the simple case of $p_L = 0$, i.e. we focus on boundary-to-boundary geodesics at constant Schwarzschild time $t$. In Schwarzschild coordinates, it is straightforward to compute the distance between each endpoint and the horizon at $r=\Phi_h$, which reads
\begin{equation}
    L_- = \int_{\Phi_h}^{\Phi_b^-} \frac{\text{d} r}{\sqrt{r^2 - \Phi_h^2}} = \text{arccosh}\left( \frac{\Phi_b^-}{\Phi_h} \right),
    \qquad
    L_+ = \int_{\Phi_h}^{\Phi_b^+} \frac{\text{d} r}{\sqrt{r^2 - \Phi_h^2}} = \text{arccosh}\left( \frac{\Phi_b^+}{\Phi_h} \right).
\end{equation}
Then, since the geodesic we are considering passes through the bifurcate horizon, its length is obtained by summing the above contributions:
\begin{equation}
    L = L_+ + L_- = \text{arccosh}\left( \frac{\Phi_b^+}{\Phi_h} \right) + \text{arccosh}\left( \frac{\Phi_b^-}{\Phi_h} \right) > L_\text{max},
\end{equation}
where the inequality follows from the fact that $\Phi_h < \Phi_b^- < \Phi_b^+$.
Thus, we conclude that every geodesic of this kind is longer than $L_\text{max}$ and therefore lives to the right of the barrier's peak of $V_{\text{cutoff}}(L)$.

Instead, geodesics of shorter lengths are found in the spacetime configuration where both boundaries lie on the same side of the black hole horizon.
In this setup, the distance between two boundary points is given by the \emph{difference}
\begin{equation}
    L = L_+ - L_- = \text{arccosh}\left( \frac{\Phi_b^+}{\Phi_h} \right) - \text{arccosh}\left( \frac{\Phi_b^-}{\Phi_h} \right) < L_\text{max}.
\end{equation}
We also notice that this length increases monotonically from $L_0$ to $L_\text{max}$ as $\Phi_h$ grows from $0$ to $\Phi_b^-$.
Therefore, we conclude that this class of geodesics lie between $L_0$ and $L_{\text{max}}$.

Finally, geodesics with $L < L_0$ can be found by looking at geometries with energy $\Phi_h^2 < 0$. In this case, the metric \eqref{eq:metric_SChwarzschild} does not have any horizon and hence corresponds to empty AdS$_2$. Here, the length of curves of constant $t$ take the form
\begin{equation}
    L = \int_{\Phi_b^-}^{\Phi_b^+} \frac{\text{d} r}{\sqrt{r^2 - \Phi_h^2}} = \text{arcsinh}\!\left( \frac{\Phi_b^+}{\sqrt{-\Phi_h^2}} \right) - \text{arcsinh}\!\left( \frac{\Phi_b^-}{\sqrt{-\Phi_h^2}} \right) < L_0.
\end{equation}
Indeed, as $\Phi_h^2 \to -\infty$, the length vanishes as expected.

So far, the discussion has focused on the special class of curves with $p_L = 0$. Nevertheless, we can show that our conclusions hold in general.
To see this, recall that the derivatives of the dilaton field, w.r.t. the geodesic length $s$, \eqref{eq:phi_solution_length_gauge_gjk} at the endpoints of a geodesic read
\begin{equation}\label{eq:dilaton_derivatives}
    \Phi'(0) = \frac{\Phi_b^+ - \Phi_b^- \, \cosh{L}}{\sinh{L}}, \qquad \Phi'(L) = \frac{\Phi_b^+ \, \cosh{L} - \Phi_b^-}{\sinh{L}}.
\end{equation}
Since $\Phi_b^- < \Phi_b^+$, the right derivative is always positive, whereas the left one changes sign precisely at $L = L_\text{max}$, which is consistent with our earlier results.

Since the dilaton is constant along each boundary, its gradient points in the normal direction, i.e. $\nabla^\mu \Phi = (n \cdot \nabla \Phi) n^\mu$. 
Instead, the tangent vector along the geodesic points inwards on the left endpoint and outwards on the right one. As a result, the sign of the derivative of the dilaton at each endpoint can be related to the sign of its outward derivative as
   $ \text{sgn}(n \cdot \nabla \Phi) = - \text{sgn}(\Phi'(0))$ and $\text{sgn}(n \cdot \nabla \Phi) = \text{sgn} (\Phi'(L))$,
on the left and on the right boundary, respectively.
By virtue of the discussion after Equation \eqref{eq:dilaton_derivatives}, we conclude that the value of the length $L$ relative to the maximum of the potential $L_\text{max}$ is directly related to the sign of the normal derivatives of the dilaton as
\begin{align}
        L > L_\text{max} &: \quad n \cdot \nabla \Phi \lvert_{\partial\mathcal{M}_\text{L}} > 0, \qquad n \cdot \nabla \Phi \lvert_{\partial\mathcal{M}_\text{R}} > 0, \\
        L < L_\text{max} &: \quad n \cdot \nabla \Phi \lvert_{\partial\mathcal{M}_\text{L}} < 0, \qquad n \cdot \nabla \Phi \lvert_{\partial\mathcal{M}_\text{R}} > 0.
\end{align}
We thus conclude that whenever the distance between boundary points satisfies $L > L_\text{max}$, they are connected across the horizon, whereas for $L_0 < L < L_\text{max}$, they both sit to the right of the black hole. Finally, $0 < L < L_0$ means that we are considering global AdS.

\subsubsection{Application: positivity constraints in the gravity phase space}
\label{s:locality}
\begin{figure}[h]
    \centering
    \begin{minipage}{0.4\textwidth}
    \includegraphics[width=\textwidth]{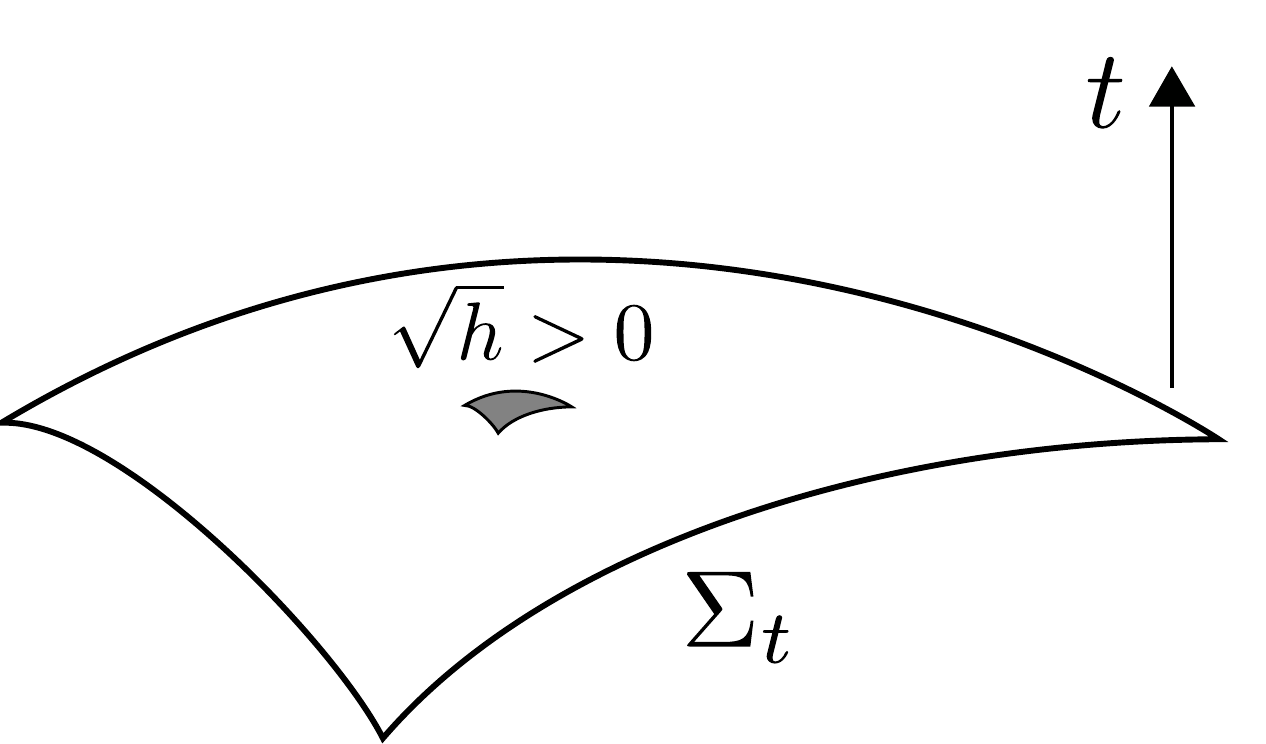}
    \caption{Infinitesimal spatial areas on a slice $\Sigma_t$ have to be positive in any gravitational phase space.}
    \label{fig:posgen}
    \end{minipage}
    \hspace{0.05\textwidth}
    \begin{minipage}{0.5\textwidth}
    \centering
    \input{images/bulklocality.tikz}
    \caption{Spacetime between two nearby finite cutoff boundaries (here in Poincar\'e coordinates) allows us to probe emergence of bulk locality. The bulk geodesic length between both adjacent boundaries is positive $L>0$ for any choice of $\Phi_b^-$ and $\Phi_b^+$.}
    \label{fig:bulklocality}
    \end{minipage}
\end{figure}

In the canonical approach of gravity where one works with the gravitational variables in the ADM decomposition
\begin{equation}
\label{eq:adm}
ds^2 = -N^2 dt^2 + h_{ij} (dx^i + N^i dt)(dx^j + N^j dt),
\end{equation}
one should impose positivity of the pullback metric determinant on each Cauchy slice $\Sigma_t$ (Fig.~\ref{fig:posgen}) \cite{Loll:1997mi,Donnelly:2016auv}:
\begin{equation}
\label{eq:posco}
\sqrt{h} \geq0,
\end{equation}
such that each small patch of the spatial submanifold has positive infinitesimal area. 

We can see the appearance of such inequalities here in our BF formulation as well. 
Firstly, since the dilaton field plays the role of the transverse area when obtaining JT gravity in an $s$-wave reduction from higher dimensions (see e.g. \cite{Mertens:2022irh}), it should be a positive phase space observable $\Phi>0$ everywhere in the bulk in the canonical framework of JT gravity. 
This means in particular the component of $\mathbf{B}$ along $P_2=(E+F)/2$ is positive $\Phi>0$. 
Additionally for the specific case of an EOW brane, the component along the timelike generator $P_0=(E-F)/2$ is given by the positive brane tension $\mu$, which is positive as well $\mu >0$. This constraint corresponds to the fact that the EOW brane must be classically shielded behind a black hole horizon, a sort of cosmic censorship for branes. This is in particular required in order for these branes to play any role in a microstate interpretation \cite{Gao:2021uro}. 
Combining these facts, we are led to propose that \emph{all components of the $\mathbf{B}$-field itself are constrained to be positive observables in the phase space of gravity}:
\begin{equation}
B_I \geq 0,
\end{equation}
and as operators in the canonical framework upon quantization.

In $1+1$d, the constraint \eqref{eq:posco} comes down to demanding positivity of the (geodesic) length between any two infinitesimally separated spatial points. 
We can exactly probe this constraint using our set-up of two finite-cutoff boundaries (Fig.~\ref{fig:bulklocality}), which when we vary $\Phi_b^-$ and $\Phi_b^+$ allows us to isolate any spatial bulk segment we want, and in particular \emph{locally} access this gravitational constraint.

The positivity of the geodesic length $L$ is implemented by the choice of reduced patch of the $\widetilde{\text{SL}(2,\mathbb{R}})$ manifold, compatible with the finite-cutoff boundary conditions. The group parameterization for two finite-cutoff boundaries in the fundamental representation \eqref{eq:fundamentalgenerators} is given by \eqref{eq:finitecutoffdecomp}: 
\begin{align}
    g&\mapsto e^{\beta(E+F)}e^{-LH}e^{\gamma(E+F)}\\&=
\begin{pmatrix}
e^{-L/2}\cosh\beta\cosh\gamma+e^{L/2}\sinh\beta\sinh\gamma
&
e^{-L/2}\cosh\beta\sinh\gamma+e^{L/2}\sinh\beta\cosh\gamma
\\[2mm]
e^{-L/2}\sinh\beta\cosh\gamma+e^{L/2}\cosh\beta\sinh\gamma
&
e^{-L/2}\sinh\beta\sinh\gamma+e^{L/2}\cosh\beta\cosh\gamma
\end{pmatrix},\nonumber
\end{align} 
with $L>0$. We observe now that there exists a conjugation matrix 
\begin{align}
    C=\frac1{\sqrt2}
\begin{pmatrix}
1&1\\
-1&1
\end{pmatrix}
\in \text{SL}(2,\mathbb R),
\end{align} 
which maps $g$ into an equivalent hyperbolic form 
\begin{align}
g\rightarrow CgC^{-1}&=e^{2\beta H}e^{\frac{L}{2}(E+F)}e^{2\gamma H}=\begin{pmatrix}
e^{\beta+\gamma}\cosh\frac L2
&
e^{\beta-\gamma}\sinh\frac L2
\\[2mm]
e^{-\beta+\gamma}\sinh\frac L2
&
e^{-\beta-\gamma}\cosh\frac L2
\end{pmatrix}\in \text{SL}^+(2,\mathbb{R}),
\end{align}  
where all four matrix entries are manifestly positive if $L>0$, making it an element of $\text{SL}^+(2,\mathbb{R})$. The patch covered by the finite-cutoff group element with $L>0$ is hence equivalent to the positive $\text{SL}^+(2,\mathbb{R})$ patch.\footnote{Positivity of all four entries of an $\text{SL}(2,\mathbb{R})$ matrix defines a subsemigroup. Changing basis by conjugating this entire subsemigroup leads to an equivalent semigroup whose entries are not all positive, but which is algebraically equivalent.} This is consistent with the $\text{SL}^+(2,\mathbb{R})$ semigroup proposal of the more precise \emph{local} bulk algebraic structure governing JT gravity amplitudes \cite{Blommaert:2018oro, Blommaert:2018iqz, Fan:2021wsb, Belaey:2024dde}. This was argued to realize the structural reduction of gauge theory to gravity, corresponding to removing singular Euclidean geometries in a Euclidean gravitational path integral framework.\footnote{For 2d JT and 3d pure gravity, which are topological, this mathematically corresponds to reducing from the moduli space of flat SL$(2,\mathbb{R})$ connections to the Teichm\"uller component, see e.g. \cite{Witten:2007kt}.}
    
\subsection{EOW brane and finite-cutoff boundary}
\label{s:oneeow}
We now replace the left finite-cutoff boundary by a timelike geodesic EOW brane, keeping a finite-cutoff boundary at the right oriented boundary.

According to the analysis of section \ref{sec:eowbdryaction}, an EOW brane boundary fixes the tangent $P_0$-component of $\mathbf{B}$-field in terms of the tension $\mu$ along the brane, whose corresponding gauge transformations are $h=e^{\theta P_0}$. The variable $\theta$ is then the conjugate variable to the tension $\mu$, and has a geometric interpretation as a proper time translation along the brane. 
To fix the mass tension along the brane, we have to choose a preferred orientation of the frame field tangent to the boundary curve at every point. Since the brane is geodesic, its tangent is parallel transported along the curve. 
Looking back at the action of the EOW brane \eqref{eq:totalaction} in second-order form, we can immediately identify the conjugate parameter in terms of the boundary time $d\tau=dv \sqrt{-h_{vv}}$. 
Indeed, on geodesics ($K=0$) the action is already of the $\int pdq$ form:
\begin{align}
    \int_{\text{EOW}} dv \sqrt{-h_{vv}}(\Phi K-\mu)=- \int_{\text{geodesic}} \mu \,d\tau.  
\end{align}

\subsubsection{Gravitational group decomposition and classical symmetries}
Along a local frame of the timelike EOW brane geodesic, the boundary condition of the EOW brane is preserved by the one-parameter subgroup \eqref{eq:localEOWbranestabilizer} $H=\exp\bigl(\mathbb R(E-F)\bigr)\simeq \widetilde{\text{SO}(2)}$.
The compatible group element then covers a subpatch of the SL$(2,\mathbb{R})$ manifold described by:
\begin{align}\label{eq:finitecutoffeowpatch}
        g = e^{\beta(E+F)}e^{-LH}e^{\theta(E-F)}.
    \end{align}
Inserting the patch decomposition in the canonical symplectic potential on the group manifold \eqref{eq:canonicaloneformcoordinates}, one can in principle repeat the analysis done in appendix \ref{app:classicalsymmetrygenerators} to derive the classical symmetry generators on phase space. Appendix \ref{sec:eowbranesymmetrygenerators} however uses a different perspective by conjugating the EOW brane patch \eqref{eq:finitecutoffeowpatch} to an analytically continued finite cutoff patch \eqref{eq:finitecutoffdecomp} in terms of the phase space coordinates $ L\rightarrow L'=L+\frac{i\pi}{2},\; \gamma'\rightarrow i\theta,\;p_\gamma'\rightarrow -ip_\theta$. Doing these substitutions in the above classical symmetry generators \eqref{eq:momentamap1}, \eqref{eq:momentamap2} yields directly the classical symmetry generators on the EOW brane patch, \eqref{eq:eowclassicalregulargenrators}:
\begin{align}
\mu_{E-F}^-&=-p_\theta,\\
\mu_{H}^-&=\frac{\sin(2\theta)}{2\cosh L}\,p_\beta
+\cos(2\theta)\,p_L
-\frac12\tanh L\sin(2\theta)\,p_\theta,\\
\mu_{E+F}^-&=-\frac{\cos(2\theta)}{\cosh L}\,p_\beta
+2\sin(2\theta)\,p_L
+\tanh L\cos(2\theta)\,p_\theta,
\end{align}
and
\begin{align}
\mu_{E+F}^+
&=p_\beta,\\
\mu_{H}^+
&=\frac12\tanh L\sinh(2\beta)\,p_\beta
-\cosh(2\beta)\,p_L
+\frac{\sinh(2\beta)}{2\cosh L}\,p_\theta,\\
\mu_{E-F}^+
&=\tanh L\cosh(2\beta)\,p_\beta
-2\sinh(2\beta)\,p_L
+\frac{\cosh(2\beta)}{\cosh L}\,p_\theta.
\end{align}
Using the non-vanishing canonical Poisson brackets on the group manifold $\{\beta,p_\beta\}=\{L,p_L\}=\{\theta, p_\theta\}=1$, one can check that these are again a realization of the classical $\mathfrak{sl}_2$-symmetry algebra \eqref{eq:classicalpoissonalgebra}. 
Inserted in the quadratic norm yields the classical Casimir 
\begin{align}
\label{eq:wzwham}
    2\text{Tr}(\mathbf{B}_\pm^2)=\mu_{H}^{\pm 2}
+\frac14\mu_{E+F}^{\pm 2}
-\frac14\mu_{E-F}^{\pm 2}=p_L^2-
\frac{2\sinh L\,p_\beta p_\theta-p_\beta^2+p_\theta^2}
{4\cosh^2L},
\end{align}
which equivalently can be obtained directly from the classical Casimir of the finite cutoff realization \eqref{eq:classicalcasimirfc}, doing the phase space substitutions \eqref{eq:coordinatesubstitutions} $L\rightarrow L+\frac{i\pi}{2}$, $\mu_{E+F}^-\rightarrow-i\mu_{E-F}^-$.

\subsubsection{Reduced EOW brane phase space}

We fix the symmetry charges at the left EOW brane boundary according to the boundary condition \eqref{eq:ellipticeowboundarycondition} $-\mu=2\Tr(\mathbf{B}_-P_0)=-\frac{1}{2}\mu_{E-F}^-$ and the usual finite cutoff boundary condition \eqref{eq:Bfinitecutoffconstraint} at the right boundary $\Phi_b=2\Tr(\mathbf{B}_+P_2)=\frac{1}{2}\mu_{E+F}^+$, leading to the reduced Hamiltonian 
\begin{align}
\label{eq:mixed_casimir_length_general_a}
2\Tr(\mathbf{B}^2)=p_L^2+\frac{2\mu \Phi_b\sinh L\,+\Phi_b^2-\mu^2}{\cosh^2L} \equiv p_L^2 + V_{\text{EOW}}(L).
\end{align} 
The EOW brane tension therefore enters in the reduced dynamics entirely through the potential term in the Casimir.
This classical potential  \eqref{eq:mixed_casimir_length_general_a} will be discussed in subsection \ref{s:claspot_EOW} below.

In the asymptotic limit, we renormalize the geodesic length that reaches the AdS$_2$ boundary on one side as $L\equiv L_r+\log \Phi_b$, leading to the known Morse potential of asymptotic JT with EOW branes \cite{Gao:2021uro} 
\begin{align}
    2\text{Tr}(\mathbf{B}^2)=p_{L_r}^2+4\mu e^{-L_r}+4e^{-2L_r}.
\end{align}

Alternatively, in the geodesic representative, the connection is still \(\mathbf A=H\,ds\) with \(s\in[0,L]\), so the bulk equations \eqref{eq:length_ode_gjk} remain unchanged and the new configuration differs from the previous one only in its endpoint data. The EOW brane obeys the Neumann boundary condition \(n^\mu\partial_\mu\Phi=\mu\). To write it in terms of \(s\), we orient the spatial coordinate from the EOW brane at \(s=0\) toward the finite-cutoff boundary at \(s=L\), so that the outward-pointing normal at the left endpoint is opposite to the direction of increasing \(s\). The condition then becomes
\begin{align}
\phi^0(0)
=
\left.\frac{d\Phi}{ds}\right|_{s=0}
=
-\mu ,
\label{eq:eow_length_boundary_condition}
\end{align}
while the right endpoint obeys the Dirichlet condition \(\Phi(L)=\Phi_b\). Solving \(d^2\Phi/ds^2=\Phi\) with these boundary conditions gives
\begin{align}
\Phi(s)
=
\frac{\Phi_b+\mu\sinh L}{\cosh L}\,
\cosh s
-
\mu\sinh s ,
\qquad
\phi^0(s)=\frac{d\Phi}{ds}.
\label{eq:mixed_phi_solution}
\end{align}

The quantity $2\Tr(\mathbf B^2)$ is again constant along the slice, so it can be evaluated at the EOW endpoint, where the boundary condition fixes \(\phi^0(0)=-\mu\) and \eqref{eq:mixed_phi_solution} gives \(\Phi(0)=(\Phi_b+\mu\sinh L)/\cosh L\). Using \(p_L=\phi^1\), one recovers \eqref{eq:mixed_casimir_length_general_a}.

In the two-cutoff case each boundary carried its own proper time and contributed its own Hamiltonian. At the EOW brane side however, the corresponding boundary Hamiltonian is zero. This also means that the final case of two EOW brane boundaries is not interesting, since the Hamiltonian vanishes entirely.

Again we can derive the same Hamiltonian system from a direct second-order analysis as we present in Appendix \ref{app:cebc}.

\subsubsection{Classical analysis of the potential}\label{s:claspot_EOW}
\begin{figure}[h]
    \centering
    \includegraphics[width=\linewidth]{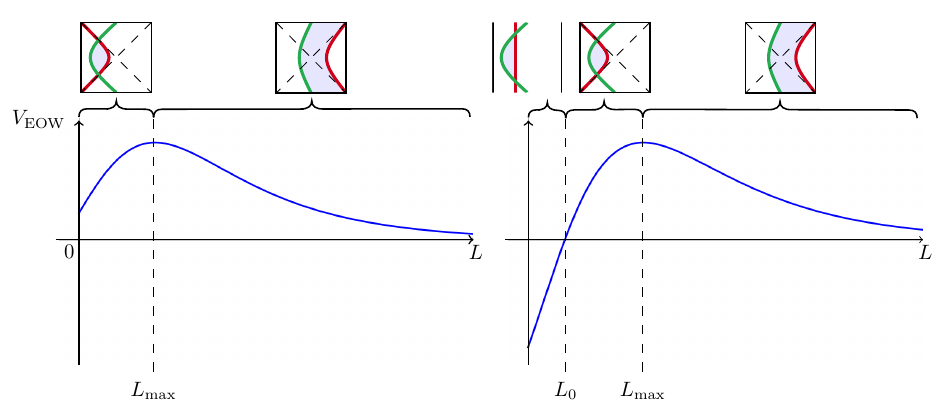}
    \caption{Potential $V_{\text{EOW}}(L)$ for the two cases: $\Phi_b^2-\mu^2>0$ (left) and $\Phi_b^2-\mu^2<0$ (right). At the top we denote the various EOW brane (green) and finite-cutoff boundary (red) configurations depending on $L$.}
    \label{fig:VEOW}
\end{figure}

We now discuss some features of the brane-cutoff potential \eqref{eq:mixed_casimir_length_general_a} and provide a geometric interpretation for the different regimes that it exhibits (Fig.~\ref{fig:VEOW}).

The potential vanishes at large distance, $L \to \infty$, and it is regular at the origin,
\begin{equation}
    V_\text{EOW}(0) = \Phi_b^2 - \mu^2,
\end{equation}
meaning that there is no attractive $-1/L^2$ singularity as opposed to the two finite-cutoff boundaries case from subsection \ref{s:claspot}.
Furthermore, it has a maximum at
\begin{equation}
    L_\text{max} = \text{arcsinh}\!\left( \frac{\mu}{\Phi_b} \right),
\end{equation}
where $V_\text{EOW}(L_\text{max}) = \Phi_b^2$.
This means that requiring the finite-cutoff boundary to be timelike, $\Phi_h^2 < \Phi_b^2$, is the same as considering trajectories that do not cross the peak of the potential.
It is clear that some features of the potential depend on the hierarchy between the boundary data $\mu$ and $\Phi_b$, suggesting to consider two separate regimes:
\begin{itemize}
    \item[$\bullet$] $\Phi_b > \mu$ (Figure \ref{fig:VEOW}, left): the potential is positive on the entire physical range $L \geq 0$ and grows from $V_\text{EOW}(0) =  \Phi_b^2 - \mu^2$ to $V_\text{EOW}(L_\text{max}) = \Phi_b^2$ before decaying,
    \item[$\bullet$] $\Phi_b < \mu$ (Figure \ref{fig:VEOW}, right): the potential is negative near the origin and vanishes at $L_0 = \log\!\left( \frac{\mu}{\Phi_b} \right) < L_\text{max}$.
\end{itemize}
We now provide some geometric interpretation for each range $0 < L <L_0$, $L_0 < L < L_\text{max}$ and $L_\text{max} < L < \infty$ of the potential \eqref{eq:mixed_casimir_length_general_a}
for both the regimes of the boundary data.

For simplicity, let us start by considering $p_L = 0$ first.
This condition characterizes turning points in the potential \eqref{eq:mixed_casimir_length_general_a}, which can be obtained by imposing $V_\text{EOW}(L_\text{T}) = \Phi_h^2$, yielding
\begin{equation}\label{eq:turning_points}
    \sinh{L^\pm_\text{T}} = \frac{\mu \Phi_b \pm \sqrt{(\Phi_b^2 - \Phi_h^2) (\mu^2 + \Phi_h^2)}}{\Phi_h^2}.
\end{equation}
The two turning points satisfy $L^-_\text{T} < L_\text{max} < L^+_\text{T}$ as long as $0 < \Phi_h^2 < \Phi_b^2$.
In this range there are always scattering trajectories with $L \geq L^+_\text{T}$. Trajectories confined to the left of the peak, $0 \leq L \leq L^-_\text{T}$, exist only when $\Phi_h^2 > V_\text{EOW}(0) = \Phi_b^2 - \mu^2$, which for $\mu > \Phi_b$ includes negative values of $\Phi_h^2$.

On the geometric side, instead, $p_L = 0$ means that $\nabla\Phi$ is tangent to the geodesic. Since the geodesic reaches the brane orthogonally, at its endpoint on the brane the dilaton gradient is purely normal, $\nabla\Phi = \mu \, n$. In the Schwarzschild coordinates \eqref{eq:metric_SChwarzschild}, this selects the slice $t=0$ and the value of the dilaton at this endpoint is given by \eqref{eq:cutoff_curve_global}:
\begin{equation}
    \Phi_\text{EOW} = \sqrt{\Phi_h^2 + \mu^2}.
\end{equation}
Moving from the brane into the bulk the dilaton decreases since $\Phi'(0) = -\mu < 0$, which leaves three possibilities.

First, let us consider the case where the slice starts at the brane in the left exterior, passes through the bifurcation point and ends on the finite-cutoff boundary in the right exterior. Its length is given by
\begin{equation}
    L = \int_{\Phi_h}^{\Phi_\text{EOW}} \frac{\text{d}r}{\sqrt{r^2 - \Phi_h^2}} + \int_{\Phi_h}^{\Phi_b} \frac{\text{d} r}{\sqrt{r^2 - \Phi_h^2}}
    = \text{arccosh}\!\left(\frac{\Phi_b}{\Phi_h}\right) + \text{arcsinh}\!\left(\frac{\mu}{\Phi_h}\right),
\end{equation}
which is precisely $L^+_\text{T}$ in \eqref{eq:turning_points}, and decreases monotonically from $+\infty$ to $L_\text{max}$ as $\Phi_h$ grows from $0$ to $\Phi_b$. These geodesics therefore live to the right of the peak of Fig.~\ref{fig:VEOW}.

Second, we consider the case where the brane and the cutoff boundary both lie on the same side of the horizon. There, we have $\Phi_\text{EOW} > \Phi_b$ and the distance between them reads
\begin{equation}
    L = \int_{\Phi_b}^{\Phi_\text{EOW}} \frac{\text{d}r}{\sqrt{r^2 - \Phi_h^2}} = \text{arcsinh}\!\left(\frac{\mu}{\Phi_h}\right) - \text{arccosh}\!\left(\frac{\Phi_b}{\Phi_h}\right),
\end{equation}
which coincides with $L^-_\text{T}$ in \eqref{eq:turning_points}.
This length is positive precisely when $\Phi_\text{EOW} > \Phi_b$, i.e.\ $\Phi_h^2 + \mu^2 > \Phi_b^2$, and it increases with $\Phi_h$, reaching $L_\text{max}$ as $\Phi_h \to \Phi_b$. The lower end of its range depends on the regime:
\begin{itemize}
    \item[$\bullet$] $\Phi_b > \mu$: the distance vanishes as $\Phi_h^2 \to \Phi_b^2 - \mu^2$, where $\Phi_b = \Phi_\text{EOW}$ and the brane intersects the finite-cutoff boundary,
    \item[$\bullet$] $\Phi_b < \mu$: the length approaches $L_0$ as $\Phi_h \to 0$.
\end{itemize}

Finally, we address the case where $\Phi_h^2 < 0$ and the geometry is global AdS$_2$.
Since $\Phi_h^2 = V_\text{EOW}(L) < 0$, this is only possible for $\mu > \Phi_b$ and $L < L_0$.
The dilaton is again monotonic along the slice, and
\begin{equation}
    L = \int_{\Phi_b}^{\Phi_\text{EOW}} \frac{\text{d}r}{\sqrt{r^2 - \Phi_h^2}}
    = \text{arcsinh}\!\left(\frac{\sqrt{\mu^2 + \Phi_h^2}}{\sqrt{-\Phi_h^2}}\right) - \text{arcsinh}\!\left(\frac{\Phi_b}{\sqrt{-\Phi_h^2}}\right).
\end{equation}
This expression matches the turning point $L^-_\text{T}$, which is the only physical one in this case.
Moreover, it interpolates between $L_0$ as $\Phi_h^2 \to 0$ and $0$ as $\Phi_h^2 \to \Phi_b^2 - \mu^2 < 0$.

To conclude, we underline that the conclusions above do not rely on $p_L = 0$ and hold in general. To see this, recall that the endpoint derivatives of \eqref{eq:mixed_phi_solution} are
\begin{equation}\label{eq:dilaton_derivatives_DN}
    \Phi'(0) = -\mu, \qquad \Phi'(L) = \frac{\Phi_b \sinh L - \mu}{\cosh L},
\end{equation}
and depend on $L$ only.
The left derivative is always negative, while the right one changes sign precisely at $L = L_\text{max}$. Using the same argument as in the cutoff-cutoff case, $\text{sgn}(n\cdot\nabla\Phi) = \text{sgn}\,\Phi'(L)$ on the finite-cutoff boundary, so that
\begin{align}
    L > L_\text{max} &: \quad n\cdot\nabla\Phi\big\rvert_\text{EOW} = \mu > 0, \qquad n\cdot\nabla\Phi\big\rvert_{\partial\mathcal{M}_\text{R}} > 0, \\
    L < L_\text{max} &: \quad n\cdot\nabla\Phi\big\rvert_\text{EOW} = \mu > 0, \qquad n\cdot\nabla\Phi\big\rvert_{\partial\mathcal{M}_\text{R}} < 0.
\end{align}
Hence, for $L > L_\text{max}$ the outward normal of the Dirichlet boundary points towards increasing dilaton, and the brane and the boundary are connected across it. For $L < L_\text{max}$ it points towards decreasing dilaton, and both lie in the same exterior region, or in a horizonless geometry, which requires $\mu > \Phi_b$ and $L \leq L_0$.

\section{Finite-cutoff holography and unconstrained boundaries}
\label{s:unconstrain}
So far, we have studied the reduction of a two-sided setup with two boundary constraints. Fixing boundary conditions on each side in the first-order formalism reduces the six-dimensional phase space to the two-sided gravitational phase space in terms of the geodesic length and its conjugate according to section
 \ref{sec:gravphasespace}.

There are various situations where one might be interested in relaxing some of these constraints. The main interest in this is to describe intrinsically one-sided systems, where one endpoint would now be ``free''. This is in particular what is needed to describe the Schwarzian phase space itself. This is also needed when introducing an entanglement surface in the system with a complete set of degrees of freedom on it, so-called edge modes, explored in the infinite cutoff JT case in \cite{Blommaert:2018iqz,Mertens:2022ujr}. We will denote such systems as \emph{one-sided} since they only have a single constrained boundary.

It must be noted that such a procedure strongly relies on the first-order framework, and it is unclear how a direct second-order framework, as in Appendix \ref{app:secondphasespace}, is directly feasible. This is even more so because of the following reason.

Setting one of the endpoints free, adds exactly one degree of freedom and its conjugate. These have direct physical and geometrical interpretations. We have described earlier the relevant one-parameter subgroups that leave fixed our chosen boundary conditions. This was:\footnote{On AdS$_2$, the range of $\theta$ is a priori restricted to $[0,2\pi]$, leading to discrete conjugate brane tension. On the universal cover $\widetilde{\text{AdS}}_2$, the time direction is unwrapped to $\theta\in \mathbb{R}$, allowing for continuous brane tension. } 
\begin{equation}
\partial \mathcal{M}_{\text{AdS}}: \{e^{\eta P_2},\eta \in \mathbb{R}\}, \quad\quad \partial \mathcal{M}_{\text{EOW}}:\{e^{\theta P_0},\theta \in \mathbb{R}\}.
\end{equation}
The parameters $\eta$ and $\theta$ play the role of the canonically conjugate variables to the cutoff value $\Phi_b$ and brane tension $\mu$ respectively: 
\begin{itemize}
\item A finite-cutoff description of the free boundary would introduce additionally the conjugate pair $(\Phi_b, \eta)$ where $\eta$ is the rapidity of the local Lorentz transformation at the cutoff brane, conjugate to $\Phi_b$. 
\item An EOW brane description of the free boundary introduces the conjugate pair $(\mu,\theta)$ where $\theta$ is a proper time shift along the brane, conjugate to the brane tension $\mu$. 
\end{itemize}
Note that whereas the conjugate variable for the EOW brane has a direct second-order interpretation, this is not true for the conjugate variable of a finite-cutoff boundary. This is perhaps related to the fact that EOW brane particles have naturally appeared to cut open the bulk geometry in the related DSSYK model / sine dilaton gravity \cite{Lin:2023trc,Cui:2025sgy,vanderHeijden:2025zkr}, whereas finite-cutoff boundaries have not.

\subsection{Schwarzian actions at finite-cutoff}
\label{s:fintesch}
As an application of this generalization of our techniques, we will consider the scenario of a one-sided setup with a single finite cutoff boundary. Since there is only a single constraint on phase space, the phase space becomes 4-dimensional. Considering a finite cutoff constraint at the right boundary $2\Tr(\mathbf{B}_+P_2)=\Phi_b$, the associated redundancy is now with respect to a left local Lorentz boost $g\sim e^{\eta P_2} g$ with $g\in \widetilde{\text{SL}(2,\mathbb{R})}$. The corresponding physical configuration space can be geometrically interpreted in this case in terms of two coordinates in an emergent AdS geometry:  
\begin{align}
\mathcal{C}_{\text{phys}}=H_+\backslash \widetilde{\text{SL}(2,\mathbb{R}})=\widetilde{\text{AdS}_2}.
\end{align} 
This part of the phase space must correspond to the geometrical configuration of the bulk, since the configurational degrees of freedom are Wilson lines composed of the gauge fields describing the geometry in the first order action. In this sense, the quotient by the finite cutoff curves under the local Lorentz group preserves the boundary points of the geometry, and the Wilson line identifies the physical spacetime locally and isometrically with the coset.
In this identification, we have to make a choice on precisely which coordinate patch of the AdS$_2$ manifold we wish to describe. 

\subsubsection{Zero-temperature Schwarzian}
We first restrict to the Poincar\'e frame in terms of a temporal coordinate $f$ and a radial coordinate $z\equiv e^{-\ell}>0$ 
\begin{align}
\label{eq:poincarémeteric}
    ds^2=\frac{-df^2+dz^2}{z^2}=d\ell^2-e^{2\ell}df^2.
\end{align} 
The asymptotic boundary is located at $\ell\rightarrow+\infty$ in terms of the proper length $\ell$ measured from left to right. We will briefly come back to other coordinate frames further on.
The corresponding frame fields and spin connection can be chosen according to \eqref{eq:flatness_components_gjk}
\begin{align}
    e^0=e^{\ell}df,\qquad e^1=d\ell, \qquad \omega=-e^\ell df,
    \label{eq:poincarégaugefields}
\end{align}
where a positively oriented radial points towards decreasing radial $\ell$. These fields make up the components of the gauge field $\mathbf{A}=e^0 P_0+e^1P_1+\omega P_2$ as
\begin{align}
\label{eq:poicnarégaugefields}
    \mathbf{A}= d\ell H-e^\ell F=-\frac{dz}{z}H-\frac{df}{z}F.
\end{align} 
Since this decomposition must descend from a homogeneous space generated by the left coset under the local Lorentz group, we can find a representative $c$ in the coset that implements $\mathbf{A}=-dc \,c^{-1}$ \eqref{eq:gaugeidentification}.\footnote{Note that in appendix , the conventions are fixed in terms of a right coset under the Lorentz group. Simply setting $g\rightarrow c^{-1}$ switches the coset in the construction.} By integrating the gauge connection from a bulk reference point  towards a point at the boundary labeled by $(f,z)$, the coset element takes the form of a group element parametrized as in \eqref{eq:gepatch}:
\begin{align}
\label{eq:poincarérep}    
c(\ell,f)=e^{ -\ell H}e^{fF}=\begin{pmatrix}
        \sqrt{z}&0\\f/\sqrt{z} &1/\sqrt{z}
    \end{pmatrix}
\end{align}  
The shape of the cutoff curve can be found from the solution of the Gauss law constraint $D\mathbf{B}=0$ in components \eqref{eq:DB_components_gjk}. Inserting the Poincaré frame gauge field components  \eqref{eq:poincarégaugefields} yields the known classical solution of the dilaton \cite{Almheiri:2014cka}
\begin{align}
    \Phi(f,z)=\frac{a+bf+c(f^2-z^2)}{z}.
\end{align} 
At infinite cutoff $\Phi_b\rightarrow\infty$, the dilaton locus with $\Phi(f,z)=\Phi_b$ becomes the asymptotic boundary curve as $z\sim 1/\Phi_b$.

To determine the geometrical phase space corresponding to the one-sided Schwarzian at finite cutoff, we relate the classical symmetry generators with the canonical conjugate momenta on AdS. Since the finite-cutoff surface is at the right boundary, the constraint on the dilaton $\Phi_b\equiv 2\Tr(\mathbf{B}_+P_2)$ is most transparent 
by considering $\Theta_{\text{red}}=2\Tr(\mathbf{B}_+ \delta gg^{-1})$, in terms of $\theta_R=\delta gg^{-1}$ on the full patch decomposition $g=e^{\eta P_2}c(\ell,f)$:
\begin{align}
    \delta g g^{-1}=P_2\delta \eta+e^{\eta P_2}\delta cc^{-1}e^{-\eta P_2}. 
\end{align}
Inserted in the potential directly leads to 
\begin{align}\label{eq:schwarzianprepotential}
    \Theta_{\text{red}}&=2\Tr(\mathbf{B}_+P_2)\delta \eta +2 \Tr(\mathbf{B}_+e^{\eta P_2}\delta cc^{-1}e^{-\eta P_2})=\Phi_b \delta \eta +2\Tr(\widetilde{\mathbf{B}}_+ \delta cc^{-1})\\&\equiv p_\eta \delta \eta +p_\ell \delta \ell +p_f \delta f
\end{align}
in terms of the conjugate fields on the group manifold with $\widetilde{\mathbf{B}}_+\equiv e^{-\eta P_2}\mathbf{B}_+e^{\eta P_2}$, given by 
\begin{align}
    \widetilde{\mathbf{B}}_+= e^{-\eta P_2}\left(-\phi_+^0P_0-\phi_{+}^1P_1+\Phi_b P_2\right)e^{\eta P_2}\equiv -\widetilde{\phi}_{+}^0P_0-\widetilde{\phi}_{+}^1P_1+\Phi_b P_2.
\end{align}
Inserting \eqref{eq:poicnarégaugefields} in the potential \eqref{eq:schwarzianprepotential} leads to an identification with the conjugate momenta: 
\begin{align}
    \Phi_b=p_\eta, \qquad \widetilde{\phi}_{+}^1=p_\ell, \qquad \widetilde{\phi}_{+}^0=-e^{-\ell}p_f+\Phi_b.
\end{align}
By fixing $p_\eta=\Phi_b$, the symplectic potential on the group manifold reduces to $\Theta_{\text{red}}=p_\ell \delta \ell+p_f\delta f$, up to an exact term $\Phi_b \delta \eta$ that doesn't contribute to the symplectic form. The classical Casimir on the constrained group manifold is now finally
\begin{align}
2\Tr(\mathbf{B}_+^2)=2\Tr(\widetilde{\mathbf{B}}_+^2) &= p_\ell^2+2\Phi_b e^{-\ell} p_f-e^{-2\ell}p_f^2 \\
&= p_\ell^2+\Phi_b^2-(\Phi_b-e^{-\ell}p_f)^2,
\label{eq:morseschwarzian}
\end{align}
leading to the physical boundary Hamiltonian 
\begin{align}
\label{eq:Hbdyschw}
    H_\partial= \Phi_b-\sqrt{\Phi_b^2-2\Tr(\mathbf{B}\rvert_\partial^2)}=\Phi_b-\sqrt{(e^{-\ell} p_f-\Phi_b)^2-p_\ell^2}.
\end{align} 
This expression is subject to two inequalities in phase space as follows. 

Firstly, note that from \eqref{eq:finitecutoffadm}, we have $\phi_{+}^{/\!/}=n^\mu \partial_\mu \Phi=\sqrt{\Phi_b^2-2\Tr(\mathbf{B}_+^2)}>0$ from the orientation of the outward normal along a growing dilaton profile. Because the dilaton is constant along the boundary, its tangential derivative vanishes. In that frame, $\mathbf{B}_+=-\phi_{+}^{/\!/} P_0+\Phi_bP_2$, leading to \begin{align}
\widetilde{\mathbf{B}}_+=-\phi_{+}^{/\!/}\cosh\eta\,P_0-\phi_{+}^{/\!/}\sinh\eta\,P_1+\Phi_bP_2
\end{align}
for a suitable rapidity angle that aligns it with the fixed global frame. We can then identify \begin{align}
\Phi_b-e^{-\ell} p_f=\phi_{+}^{/\!/}\cosh\eta,
\qquad
p_\ell=\phi_{+}^{/\!/}\sinh\eta.
\end{align}
Since $\phi_{+}^{/\!/}=n^\mu\partial_\mu \Phi>0$ at the right boundary and $\cosh\eta>|\sinh\eta|$, this becomes 
\begin{align}\label{eq:positive}
    \Phi_b-e^{-\ell} p_f>|p_\ell|.
\end{align} 
Therefore, the argument under the square root in \eqref{eq:Hbdyschw} is always positive. This corresponds to our restriction of the finite-cutoff boundary as a timelike curve. Allowing it be spacelike would correspond to complex energies and a bulk black hole that is larger than $\Phi_b$, see \cite{AliAhmad:2025kki} for some recent ideas in this direction.

Secondly, we impose
\begin{equation}
\label{eq:positive2}
p_f >0.
\end{equation}
Under both \eqref{eq:positive} and \eqref{eq:positive2}
the classical Casimir is always positive, making the boundary Hamiltonian \eqref{eq:Hbdyschw} bounded from below.\footnote{Explicitly, when $p_f>0$, the quantity on the left in \eqref{eq:positive} ranges in $[|p_\ell|,\Phi_b]$ assuming $|p_\ell|<\Phi_b$. This makes \eqref{eq:morseschwarzian} greater than $p_\ell^2$. If we would allow any $p_f<0$ on the other hand, we can make the quantity on the left in \eqref{eq:positive} as large as we want by taking $\ell \to -\infty$. This would make the classical Casimir \eqref{eq:morseschwarzian} arbitrarily negative, and the Hamiltonian \eqref{eq:Hbdyschw} unbounded from below.}

With \eqref{eq:Hbdyschw}, the final one-sided bulk phase space action is
\begin{align}
\label{eq:final}
\boxed{S_{\text{Schwarzian}}
= \int_{\partial} d\tau \left(p_\ell \,\dot{\ell}+ p_f\,\dot{f}-\Phi_b+\sqrt{(e^{-\ell} p_f-\Phi_b)^2-p_\ell^2}\right)}.
\end{align} 
From the Hamilton equations 
\begin{align}
\label{eq:poincaréhamiltoneq}
    \dot{\ell}=\frac{p_\ell}{\sqrt{(e^{-\ell} p_f-\Phi_b)^2-p_\ell^2}}, \qquad \dot{f}=\frac{e^{-\ell} \Phi_b-e^{-2\ell} p_f}{\sqrt{(e^{-\ell} p_f-\Phi_b)^2-p_\ell^2}},
\end{align}
we find 
\begin{align}
\label{eq:eom}
    e^{2\ell}\dot{f}^2-\dot{\ell}^2=1.
\end{align}  
Evaluating the Poincar\'e metric \eqref{eq:poincarémeteric} along the finite-cutoff curve leads to a precise identification of $\tau$ as proper time on the boundary curve $ds^2_\partial=-d\tau^2$.
Since $f$ is a cyclic coordinate in the Lagrangian, its conjugate momentum is a constant of motion $\dot{p}_f=0$. The last Hamilton equation is $\dot{p}_\ell = p_f \dot{f}$. It is possible to solve these classical equations of motion systematically, but we refrain from analyzing the solution here.\footnote{We immediately obtain $p_\ell(t)=p_f f(t)+C_1$ and $\dot{\ell}(t)= \frac{p_\ell(t)}{\Phi_b-E}$, where $E=\Phi_b-\sqrt{(e^{-\ell} p_f-\Phi_b)^2-p_\ell^2}$ is a constant on the equations of motion. Plugging this in \eqref{eq:eom} leads to $e^{2\ell} = \frac{1}{\dot{f}^2}\left(1+\frac{p_\ell^2(t)}{(\Phi_b-E)^2}\right)$, which can then be plugged into the right equation of \eqref{eq:poincaréhamiltoneq} and is directly integrable to obtain the solution $f(t)$.}

Furthermore, from \eqref{eq:positive}, we have that $\dot{f}>0$ for a classical solution \eqref{eq:poincaréhamiltoneq}.

When moving the finite cutoff surface towards the asymptotic boundary, the radial coordinate scales as $z\equiv z_r/\Phi_b$ 
We then introduce the renormalized proper length coordinate $\ell_r=-\log z_r=\ell-\log \Phi_b$. 
The boundary Hamiltonian reduces to the standard Schwarzian at infinite cutoff $\Phi_b\rightarrow\infty$ with respect to the renormalized boundary time $\tau=\Phi_b u$ 
\begin{align}
\label{eq:Hinfcut}
    H_u\equiv \Phi_b H_\partial\xrightarrow{\Phi_b\rightarrow\infty}\Tr(\mathbf{B}_\partial^2)=\frac{p_{\ell_r}^2}{2}+p_fe^{-\ell_r},
\end{align}
which is positive provided $p_f>0$ on phase space. This is the usual formulation of the Schwarzian at infinite cutoff, leading to the phase space action \cite{Engelsoy:2016xyb,Mertens:2017mtv}
\begin{align}
S_\partial=\int du\left[p_{\ell_r}\ell_r'+p_ff' - p_fe^{-\ell_r}-\frac{p_{\ell_r}^2}{2}
\right],
\end{align}
where now primes imply derivatives with respect to the renormalized boundary time $u$.
The transition to the usual Schwarzian action is well-known: the canonical momenta implement the off-shell constraints $f'=e^{-\ell_r}$, $\ell'_r=p_{\ell_r}$,
corresponding to fixing $z=f'/\Phi_b$, and the renormalized boundary Poincar\'e asymptotics $ds^2=-\frac{d\tau^2}{\Phi_b^2}=-du^2$. 
Plugging this in leads to
\begin{align}
    S_\partial=\frac{1}{2}\int_\partial du \left(\frac{f''}{f'}\right)^2=-\int_\partial du \,\{ f,u\},\quad \{f,u\}\equiv\frac{f'''}{f'}-\frac{3}{2}\left(\frac{f''}{f'}\right)^2.
\end{align}

\subsubsection{Other coordinate frames}
The finite-temperature generalization of this can be found immediately by instead starting from the AdS$_2$ representative 
\begin{align}
 c(\ell,\gamma)=e^{-\ell H}e^{\gamma(E+F)},
\end{align} 
for which the metric on the AdS$_2$ coset \eqref{eq:metriccoset} yields the Rindler patch (with temperature $\beta=\pi$) \begin{align}
\label{eq:rindlermetric}
    ds^2=2\Tr(dcc^{-1}|_{P_2=0})^2=d\ell^2-4\sinh^2\ell \;d\gamma^2.
\end{align}
The boundary Hamiltonian follows directly from  \eqref{eq:classicalcasimirfc} (with the same orientation of $L$ and $\ell$ and one boundary constraint $p_\beta = 2\Phi_b$):
\begin{equation}
\label{eq:Hbdyschw2}
H = \Phi_b- \sqrt{\Phi_b^2-p_\ell^2-
\frac{4\cosh \ell\,\Phi_b p_\gamma-4\Phi_b^2-p_\gamma^2}
{4\sinh^2 \ell}}.
\end{equation}
In terms of the renormalized length $\ell_r = \ell - \log \Phi_b$, we analogously obtain
\begin{equation}
\label{eq:finitetschw}
\Phi_b H \xrightarrow{\Phi_b\rightarrow\infty} \frac{p_{\ell_r}^2}{2}+e^{-\ell_r} p_\gamma-2e^{-2\ell_r},
\end{equation}
which can be identified with the finite-temperature phase space description of the Schwarzian \cite{Mertens:2017mtv}, coming from a Schwarzian action of the form 
\begin{align}
    S_\partial=-\int_\partial du\,\left(\{\gamma,u\}-2(\gamma')^2\right) =-\int_\partial du\;\{\tanh\gamma(u),u\}.
\end{align}
This is exactly the asymptotic transformation needed to transfer from the Poincar\'e patch metric \eqref{eq:poincarémeteric} to the Rindler metric \eqref{eq:rindlermetric} when $\ell\rightarrow \infty$:\footnote{The finite coordinate transformations are explicitly: 
\begin{align}
    e^\ell\rightarrow \cosh\ell+\sinh\ell\cosh(2\gamma),\qquad f\rightarrow\frac{\sinh\ell\sinh(2\gamma)}
{\cosh\ell+\sinh\ell\cosh(2\gamma)}.
\end{align} 
The Rindler exterior (with $\ell>0$ and $\gamma\in \mathbb{R}$) then only covers a section of the Poincar\'e patch bounded by $|f|+e^{-\ell}<1$, or equivalently $\ell>0$ and $-1+e^{-\ell}<f<1-e^{-\ell}$. } \begin{align}
    \ell\rightarrow \ell +2\log \cosh\gamma+\mathcal{O}(e^{-2\ell}), \qquad f\rightarrow \tanh(\gamma)+\mathcal{O}(e^{-2\ell}).
\end{align}
At the asymptotic boundary, the accessible interval of the Poincar\'e time then becomes limit to $-1<f<1$, which induces a temperature.

Finally, when considering an unconstrained family of EOW branes as interior states, the AdS$_2$ representative is given by $c=e^{-\ell H}e^{\gamma(E-F})$, used in appendix \ref{app:isometries}. Inserted in the symplectic potential $\Theta_{\text{red}}=2\Tr(\mathbf{B}_+ \delta gg^{-1})$ with $g=e^{\eta P_2}c$, and imposing the finite cutoff constraint $\Phi_b=p_\eta$  yields \begin{align}
p_\eta=\Phi_b,\qquad\widetilde\phi^1_+=p_\ell,\qquad
\widetilde\phi^0_+=\frac{p_\gamma+2\Phi_b\sinh\ell}{2\cosh\ell}
\end{align}leading to the classical Casimir \begin{align}
2\operatorname{Tr}(B_+^2)=p_\ell^2+\Phi_b^2-(\widetilde\phi^0_+)^2=p_\ell^2+
\frac{4\Phi_b^2-4\Phi_bp_\gamma\sinh\ell-p_\gamma^2}
{4\cosh^2\ell}.
\end{align} 
and corresponding Hamiltonian 
\begin{align}
\label{eq:Hbdyschw3}
H_\partial=\Phi_b-\sqrt{\left(\frac{p_\gamma+2\Phi_b\sinh\ell}{2\cosh\ell}\right)^2
-p_\ell^2}.
\end{align}
After renormalizing $\ell_r=\ell-\log \Phi_b$, leads to the infinite cutoff Hamiltonian
\begin{equation}
\Phi_b H \xrightarrow{\Phi_b\rightarrow\infty} \frac{p_{\ell_r}^2}{2}-e^{-\ell_r} p_\gamma+2e^{-2\ell_r},
\end{equation}
which upon setting $p_\gamma \to - p_\gamma$ (corresponding to flipping the clock $\gamma$),
is the same expression as \eqref{eq:finitetschw} except for the sign of the last term \cite{Jafferis:2019wkd}. This now matches with the global frame Schwarzian model, with action
\begin{align}
S_\partial=-\int_\partial du \,\left(\{ \gamma,u\}+2\gamma'^2\right) =-\int_\partial du\;\{\tan\gamma(u),u\}.
\end{align}
We can summarize our observations as follows, relating the type of free boundary with the coordinate patch for which the finite-cutoff Schwarzian model is formulated:
\begin{center}
\begin{tabular}{|c|c|}
\hline 
type of free boundary & coordinate patch \\
\hline\hline
infinite cutoff \eqref{eq:Hbdyschw} & Poincar\'e frame \\
finite cutoff \eqref{eq:Hbdyschw2} & thermal (Rindler) frame \\
EOW \eqref{eq:Hbdyschw3} & global frame \\
\hline
\end{tabular}
\end{center}

\subsubsection{Symmetries}

All four-dimensional phase space models above have a single set of $\mathfrak{sl}(2,\mathbb{R})$ symmetry generators (down from the double set of left- and right-symmetry generators on the fully unconstrained group theoretic models we started with) that act as isometries on the coordinate patch of AdS$_2$ under consideration.

The right-regular symmetry generators obtained by right multiplying $g\rightarrow gh $ act as isometry generators, leaving the Cartan-Maurer form $dgg^{-1}$ invariant. Using the full group decomposition $g=e^{\eta P_2}e^{-\ell H}e^{fF}$, the corresponding Cartan-Maurer form can be worked out as 
\begin{align}
    g^{-1}\delta g=H\left(e^\ell f\delta \eta-\delta \ell\right)+E \frac{e^\ell}{2}\delta \eta+F\left(f\delta \ell +\delta f+\left(e^{-\ell}-e^\ell f^2\right)\frac{\delta \eta}{2}\right).
\end{align}
Plugged in the expression of the right-regular symplectic potential on the group manifold $\Theta_{\text{red}}=2 \Tr\left(\mathbf{B}_- g^{-1}\delta g\right)=p_\eta \delta \eta +p_\ell \delta \ell +p_f\delta f$, we determine the right-regular symmetry generators as the components of $\mu_{I}^-=-2\Tr(\mathbf{B}^-P_I)$: 
\begin{align}
\label{eq:symgen}
    \mu^-_{F}=-p_f, \qquad \mu_{H}^-=p_\ell -fp_f, \qquad \mu_{E}^-=-2e^{-\ell }p_\eta-2p_\ell f+p_f\left(f^{2}+{\color{darkblue}e^{-2\ell}}\right).
\end{align}
The last term in $\mu_{E}^-$ (colored in darkblue) is the only term that deviates from the known infinite-cutoff symmetry generators. Since these symmetry generators do not depend on $\eta$, they Poisson commute with the left-regular constraint $p_\eta=\mu_{P_2}^+\equiv \Phi_b$. Using the canonical Poisson brackets on phase space, they satisfy the classical symmetry algebra 
\begin{align}
\{\mu_{H}^-,\mu_{E}^-\}=\mu_{E}^-,\qquad
\{\mu_{H}^-,\mu_{F}^-\}=-\mu_{F}^-,\qquad
\{\mu_{E}^-,\mu_{F}^-\}=2\mu_{H}^-.
\end{align} 
Their classical Casimir $2\Tr(\mathbf{B}^2_-)=(\mu_{H}^-)^2+\mu_{E}^-\mu_{F}^-$ yields back the Schwarzian expression \eqref{eq:morseschwarzian}. After imposing the above constraint, we therefore preserve a full right-regular $\mathfrak{sl}_2$ symmetry algebra that commutes with the Hamiltonian $\{\mu_{I}^-,H_\partial\}=0$ and acts as isometries on the AdS$_2$ manifold. In particular, an observable $F(c)$ on AdS$_2$ transforms 
under the symmetry generators $B_{I}^-$ as \eqref{eq:app_Bminus_action_on_functions} 
$\delta_I F\equiv \{F(c),B_{I}^-\}=R_I F(c)$. 
Projected onto the physical coordinates $f$ and $z=e^{-\ell}$ leads explicitly to the isometry generators 
\begin{align}
\zeta_F=\partial_f,\qquad\zeta_H=f\partial_f+z\partial_z,\qquad\zeta_E=-(f^2+z^2)\partial_f-2fz\partial_z.
\end{align}
Infinitesimally, these generate the coordinate transformations 
\begin{align}
f\rightarrow f+\epsilon\left[c+af-b(f^2+z^2)\right]+\mathcal{O}(\epsilon^2),\qquad z\rightarrow z+\epsilon\left[az-2bfz\right]+\mathcal{O}(\epsilon^2).
\end{align}

\section{Concluding remarks}
\label{s:concl}
In this work, we have provided a first-principles derivation of finite-cutoff and end-of-the-world boundaries and their dynamical systems in JT gravity in the first-order formulation. We leveraged those technical results to probe the local structure of the gravitational bulk in subsection \ref{s:locality}, and to derive a closed-form action in phase space for the finite-cutoff Schwarzian model in section \ref{s:unconstrain}. In our companion work \cite{toap}, we will present results on the quantization of these models, and in particular discuss the role of these boundaries in factorization of the gravitational Hilbert space and edge state decompositions. We will end here with a couple of interesting open questions. 

\paragraph{Supersymmetrization.} There are several direct structural extensions that are interesting to consider. The first is supersymmetrization of our construction, to $\mathcal{N}=1,2,4$ JT supergravity \cite{Fan:2021wsb,Lin:2022zxd,Belaey:2024dde,Lin:2026sli,Forste:2017kwy}, and in particular construct the analogous finite-cutoff Schwarzian action. 

\paragraph{Comparison with other finite-cutoff descriptions.}
It would be interesting to compare the phase space action \eqref{eq:final} with other classical descriptions of JT gravity at finite cutoff. In \cite{Iliesiu:2020zld}, the boundary action of the Euclidean theory was obtained perturbatively in the cutoff as a functional of a single reparametrization $f(u)$, and takes the schematic form
\begin{equation}
\label{eq:lit}
    S[f] = -\int \frac{du}{\epsilon^2}\left(\sqrt{1+2\epsilon^2\{f,u\}}-1+\text{derivatives of }\{f,u\}\right).
\end{equation}
The leading term is the one-dimensional $T\bar{T}$-deformation of the Schwarzian \cite{Gross:2019ach}, in which the deformed energy is a square root of the undeformed one. Our Hamiltonian \eqref{eq:Hbdyschw} also contains a square root, but of a different quantity. Setting $\epsilon\equiv 1/\Phi_b$ and using the renormalized variables of \eqref{eq:Hinfcut}, we write
\begin{equation}
\label{eq:comparison}
    \Phi_b H_\partial = \frac{1}{\epsilon^2}\left(1-\sqrt{1-\epsilon^2\left(p_{\ell_r}^2+2p_f e^{-\ell_r}\right)+\epsilon^4\, e^{-2\ell_r}p_f^2}\right).
\end{equation}
What appears under the square root is the classical Casimir \eqref{eq:morseschwarzian},
which contains the term $-\epsilon^2 e^{-2\ell_r}p_f^2$ that comes from the finite-cutoff correction to the generator $\mu_E^-$ in \eqref{eq:symgen}. The same happens in the two-sided system. For equal cutoffs $\Phi_b^-=\Phi_b^+\equiv\Phi_b$ and $L=L_r+2\log\Phi_b$, the Casimir \eqref{eq:equal_cutoff_casimir} becomes
\begin{equation}
    2\text{Tr}(\mathbf{B}^2) = p_{L_r}^2+\frac{4e^{-L_r}}{\left(1+\epsilon^2 e^{-L_r}\right)^2},
\end{equation}
so the square root now acts on the P\"oschl--Teller Hamiltonian of \cite{Griguolo:2026eoi} instead of the Liouville Hamiltonian $p_{L_r}^2+4e^{-L_r}$ of the asymptotic theory \cite{Harlow:2018tqv}.

At least naively, our Hamiltonian is therefore not a square-root deformation of the Schwarzian or Liouville Hamiltonian. This difference is invisible on classical solutions, where the Casimir takes the value $\Phi_h^2$ and the finite-cutoff energy \eqref{eq:explicitadmenergy} is recovered, so that the two descriptions agree on-shell while being realized differently on phase space. To make the comparison precise, one could expand \eqref{eq:comparison} in powers of $\epsilon$ and integrate out the momenta order by order, and compare the resulting derivative expansion with \eqref{eq:lit}. A complementary approach would be to start from the exact Riccati equation for the extrinsic curvature of the boundary curve derived in \cite{Griguolo:2025kpi} and relate it to the Hamilton equations \eqref{eq:poincaréhamiltoneq}. We cannot in any case integrate out $p_f$ to produce a closed-form coordinate space path integral.\footnote{This is just like in the $q$-deformed case (the $q$-Schwarzian) \cite{Blommaert:2023opb,Blommaert:2023wad} where we are also stuck with the more fundamental phase space description.}

Finally, because our description keeps track of both coordinates $(f(\tau),\ell(\tau))$ of the boundary curve together with their conjugate momenta, it would be suitable for a direct comparison with the Euclidean geometric approaches of \cite{Stanford:2020qhm,Ferrari:2024kpz}, in which the finite-cutoff boundary is treated as a curve in the hyperbolic plane. In particular, it would be interesting to understand whether the requirement that such a curve bounds a regular bulk region is reflected in the inequalities \eqref{eq:positive} and \eqref{eq:positive2} or in more restrictive ones, and how it connects to the positivity structure discussed in subsection \ref{s:locality}.

\paragraph{Towards quantization of the finite-cutoff Schwarzian.}
To analyze the finite-cutoff system \eqref{eq:final} at the quantum level, 
within canonical quantization, the system is constrained by the inequalities \eqref{eq:positive} and \eqref{eq:positive2} as operator relations. 
These inequalities in phase space correspond to a positive classical Casimir expression. Upon quantization, this implies a positive quantum Casimir for $\mathfrak{sl}(2,\mathbb{R})$ which pinpoints the principal series representations as the only (tempered) unitary irreducible representation that are allowed to appear, but with an additional upper bound on the Casimir eigenvalue. These are precisely the gravitational constraints again that are implemented by restricting to the positive semigroup SL$^+(2,\mathbb{R})$ \cite{Blommaert:2018oro, Blommaert:2018iqz, Fan:2021wsb, Belaey:2024dde}.

Alternatively, one could contemplate the Euclidean phase space path integral
\begin{align}
\label{eq:Z}
Z \overset{?}{=} \int_{\mathcal{I}} \mathcal{D}\ell \mathcal{D}p_\ell \mathcal{D}f \mathcal{D}p_f \, e^{\int_{\partial} d\tau \left(ip_\ell \,\dot{\ell}+ ip_f\,\dot{f}-\Phi_b+\sqrt{(e^{-\ell} p_f-\Phi_b)^2-p_\ell^2}\right)},
\end{align}
where the integration domain $\mathcal{I}$ needs to be suitably defined, and with the flat path integral measure in phase space. 

In the infinite-cutoff regime \eqref{eq:Hinfcut}, it is known that the constraint $p_f>0$ is replaced by $p_f \in i\mathbb{R}$ in the Euclidean gravitational path integral \cite{toap2}, playing the role of a Lagrange multiplier, and very similar to the usual discussion on the choice of the lapse contour in canonical gravity formulations \cite{Halliwell:1989dy,Marolf:1996gb,Banihashemi:2024aal,Liu:2023jvm}. The resulting infinite-cutoff path integral localizes on $f'=e^{-\ell_r} >0$ and hence only paths that monotonically reparametrize the background time $f$ contribute to the path integral. 

\begin{figure}[h]
    \centering    \includegraphics[width=0.5\textwidth]{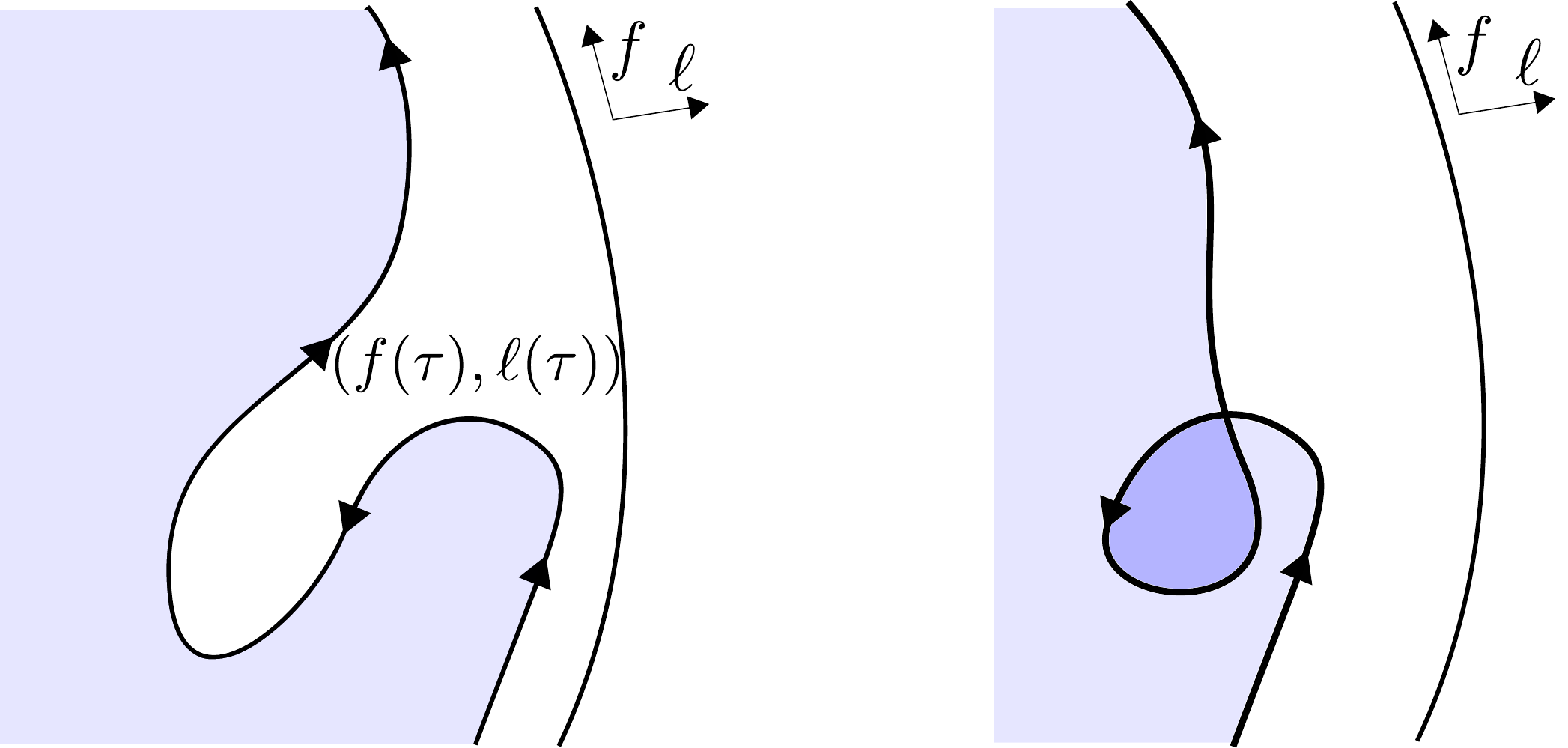}
    \caption{Left: example of Euclidean boundary curve $(f(\tau),\ell(\tau))$ parametrized in proper time $\tau$, that is not monotonic in $f$. Right: example of a boundary curve that self-intersects.}
    \label{fig:selfint}
\end{figure}

It is unclear how the other constraint \eqref{eq:positive} would need to be imposed for the finite-cutoff case in the Euclidean path integral, but we note that the Hamiltonian is already bounded from below by just letting $p_f \in i\mathbb{R}$ (although it is complex-valued), so the Euclidean path integral is expected to converge. Regardless, the finite-cutoff path integral \eqref{eq:Z} does not localize on $\dot{f}>0$, and hence the boundary curve can run backwards in Poincar\'e time, and even locally self-intersect (see Fig.~\ref{fig:selfint} for a sketch). Self-intersections are then in principle allowed in the specific case of self-overlapping regions \cite{Ferrari:2024ndr}, and it would be interesting to understand this better.

\paragraph{$T\bar{T}$-flow of symmetry generators.} The finite-cutoff system has a set of $\mathfrak{sl}(2,\mathbb{R})$ symmetry generators \eqref{eq:symgen}, where only the last one is different from the infinite cutoff generators. Can we determine directly the flow of these symmetry generators by integrating the $T\bar{T}$-flow equations \cite{Guica:2022gts}?

\paragraph{Baby universe from finite cutoff and EOW brane?} In subsection \ref{sec:geometry} we derived that the bulk patch between an EOW brane and a finite-cutoff boundary in the regime $\Phi_h^2 > \Phi_b^2 - \mu^2$ (see Fig.~\ref{fig:penrose_diagram_EOW}) exhibits an open universe big bang big crunch topology in Lorentzian signature. It would be interesting to understand the significance of this.

\begin{figure}[h]
    \centering
    \input{images/raddress.tikz}
    \caption{Diff-invariant construction of the dilaton field using two-sided gravitational information ($L$) and the geodesic length $s$ along this geodesic.}
    \label{fig:raddress}
\end{figure}
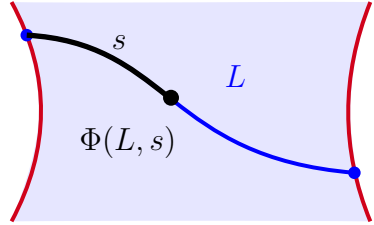

\paragraph{Gravitational dressing in a two-sided setup.}
Using the Gauss law constraint $D_\sigma \mathbf{B}=0$ \eqref{eq:gausslawconstraint}, 
we can determine the radial profile of the dilaton between two boundaries in the geodesic gauge using only the boundary data as \eqref{eq:phi_solution_length_gauge_gjk}.
This defines in particular the dilaton field as an observable in terms of the gravitational state labeled by $L$, and the geodesic distance $s$ along the slice: $\Phi(L,s)$. This is geometrically invariant information, and this hence gives an operational definition to the dilaton field on a Cauchy slice. It is also intrinsically two-sided since we used the boundary-to-boundary geodesic length $L$ (Fig.~\ref{fig:raddress}).
Adding additional matter fields and relating it to the construction of \cite{Harlow:2021dfp} would be interesting. The relation with one-sided dressing constructions as in \cite{Blommaert:2019hjr,Mertens:2025rpa,Franken:2026bff} would also be interesting to understand.

The dilaton field usually plays an inverse role as follows. The value of any scalar field, such as $\Phi$, can be used to label bulk points in terms of the value of the field itself \cite{Giddings:2005id}. Taking the dilaton field $\Phi$ on a fixed Cauchy slice then naturally defines a radial location in the bulk.

\paragraph{Timelike boundaries in de Sitter JT gravity.}
Unlike AdS$_2$, de Sitter space only has spacelike conformal boundaries, so
that JT gravity with positive cosmological constant
\cite{MaldacenaTuriaciYang2019,CotlerJensenMaloney2019} lacks an asymptotic
timelike boundary whose proper time could serve as a clock. Such clocks can
be introduced by restricting to a region bounded by timelike curves of
constant dilaton, whose proper times then describe the evolution of the
enclosed subsystem. These boundaries were used in \cite{SveskoEtAl2022} to
define quasi-local energies in dS$_2$, while timelike boundaries motivated by
the de Sitter static patch have been studied in higher dimensions in
\cite{AnninosGalanteManeerat2023,AnninosGalanteManeerat2024}. In the
following, we briefly sketch how the reduction developed in this work could
be adapted to this setting.

In the first-order language, the de Sitter theory is described by the same
BF action \eqref{eq:bulkBF} with $\mathfrak{so}(2,1)$ gauge algebra
\cite{Fukuyama:1985gg,Isler:1989hq,Chamseddine:1989yz,Jackiw:1992bw,CotlerJensenMaloney2019,CotlerJensen2024} with
only the identification of the metric changing with respect to the AdS case. Locally, both geometries are
described by the adjoint orbit of $P_2$ of appendix~\ref{app:isometries}, now
equipped with the opposite metric
\begin{equation}
    ds^2_{\text{dS}} = 2\,\text{Tr}(dY^2) = -ds^2_{\text{AdS}} = (e^0)^2 - (e^1)^2 .
\end{equation}
The frame field $e^0$, aligned with the compact generator $P_0$, is now
spacelike while $e^1$ becomes timelike.
The bulk equations $\mathbf{F}=0$ and $D\mathbf{B}=0$ are therefore unchanged,
and only the causal interpretation of the variables, and the
gravitational patch selected in the reduction differ.

Consider then two such boundaries at dilaton values $\Phi_b^-$ and
$\Phi_b^+$, connected by a spacelike geodesic $\gamma$ of length $L$. In a
frame adapted to $\gamma$, with $\gamma^* e^0 = ds$ and
$\gamma^* e^1 = \gamma^*\omega = 0$, the connection pulls back to
$\gamma^*\mathbf{A} = P_0\, ds$, so that the Wilson line representative
\eqref{eq:geodesic_representative} is replaced by $g_\gamma = e^{-LP_0}$.
Along the geodesic, the Gauss law constraint $D_s\mathbf{B}=0$ reduces to
$d\phi^0/ds = 0$, $\phi^1 = -d\Phi/ds$ and $d\phi^1/ds = \Phi$. Compared with
\eqref{eq:length_ode_gjk}, the dilaton now obeys $d^2\Phi/ds^2 = -\Phi$, and
the constant component $p_L = -\phi^0$ plays the role of the momentum
conjugate to $L$. Repeating the reduction of subsection~\ref{s:twophi} then
leads to
\begin{equation}
\label{eq:casds}
    2\,\text{Tr}(\mathbf{B}^2) = -p_L^2
    + \frac{(\Phi_b^-)^2 + (\Phi_b^+)^2 - 2\Phi_b^-\Phi_b^+\cos L}{\sin^2 L},
    \qquad 0 < L < \pi ,
\end{equation}
which is the trigonometric counterpart of \eqref{eq:casimir_length_variables_gjk}, with
a kinetic term of opposite sign. As in \eqref{eq:dilatonhorizon}, this Casimir
equals $\Phi_h^2$, with $\Phi_h$ now the value of the dilaton at the
bifurcation point of the cosmological horizon, and the boundaries are
timelike precisely when $\Phi_b^- , \Phi_b^+< \Phi_h$, opposite to the AdS condition
$\Phi_b^- , \Phi_b^+ > \Phi_h$.

The boundary Hamiltonians depend on where the two boundaries sit relative
to the horizon. When they lie in opposite static patches, with the region
between them containing the cosmological horizon, the dilaton decreases
towards each boundary and the outward normal derivative is negative on
both sides. Inserting the Casimir \eqref{eq:casds} into the boundary Hamiltonians as in
\eqref{eq:finalfincut}, and parametrizing the evolution by a common
parameter $\lambda$, with boundary proper times $\tau_{\pm}(\lambda)$,
gives the reduced phase space action
\begin{equation}
\begin{aligned}
    S_{\mathrm{red}}
    &= \int d\lambda \left[
        p_L \frac{dL}{d\lambda}
        - H_\partial^- \frac{d\tau_-}{d\lambda}
        - H_\partial^+ \frac{d\tau_+}{d\lambda}
    \right], \\[2mm]
    H_\partial^{\pm}
    &= E_0(\Phi_b^{\pm})
    + \sqrt{2\,\operatorname{Tr}(\mathbf{B}^2)
        - (\Phi_b^{\pm})^2},
\end{aligned}
\end{equation}
with $E_0(\Phi_b^{\pm})$ a constant set by the boundary counterterm
at fixed dilaton values.

As we discussed in the AdS case of subsection~\ref{s:claspot}, the two boundaries
can also lie within a single static
patch, and this configuration is the natural
setting to repeat the local probe of the bulk structure of subsection~\ref{s:locality},
by sending the separation between the two boundaries to zero. Including the
boundary stabilizers as in \eqref{eq:finitecutoffdecomp}, and using the same conjugation
matrix $C$ of subsection~\ref{s:locality}, which commutes with $P_0$, maps
the geodesic patch $g = e^{\beta(E+F)} e^{-LP_0} e^{\gamma(E+F)}$ to\footnote{The entries satisfy, for $0 < L < \pi$,
\begin{equation}
    \hat{g}_{11}, \hat{g}_{22} > 0, \qquad
    \hat{g}_{12} < 0 < \hat{g}_{21}, \qquad
    \hat{g}_{12}\,\hat{g}_{21} = -\sin^2\frac{L}{2}.
\end{equation}
This is to be contrasted with the finite-cutoff patch of
subsection~\ref{s:locality}, where all entries are positive and the patch is
the semigroup $\text{SL}^+(2,\mathbb{R})$.}
\begin{equation}
    \hat{g} \equiv CgC^{-1} = e^{2\beta H} e^{-LP_0} e^{2\gamma H} =
    \begin{pmatrix}
        e^{\beta+\gamma}\cos\frac{L}{2} & -e^{\beta-\gamma}\sin\frac{L}{2} \\[2mm]
        e^{-\beta+\gamma}\sin\frac{L}{2} & e^{-\beta-\gamma}\cos\frac{L}{2}
    \end{pmatrix},
\end{equation}
The de Sitter domain is not
closed under multiplication, since composing two geodesic segments with total
length exceeding $\pi$ yields $\hat{g}_{11}<0$. A related global limitation is that
spacelike geodesics connecting the two boundaries can cease to exist at
finite boundary time
\cite{ChapmanGalanteKramer2021,AalsmaEtAl2022,ChapmanEtAl2023}, so that the
geodesic description used above does not cover all configurations. It would be
interesting to understand the origin of this different algebraic structure,
and whether a suitable replacement of $\text{SL}^+(2,\mathbb{R})$ governs the
gluing of gravitational regions in dS JT gravity.

\paragraph{Finite cutoff $q$-Schwarzian and sine dilaton gravity.}
A remarkable one-parameter generalization of JT gravity is \textit{sine dilaton gravity}, which is a model characterized by the periodic potential
\begin{equation}
V(\Phi)=\frac{\sin(2 \abs{\log{q}} \Phi)}{\abs{\log{q}}},
\end{equation}
and reduces to the JT potential $V(\Phi)=2\Phi$ of \eqref{eq:totalaction} as $q \to 1$. Its holographic relation with double-scaled SYK \cite{Blommaert:2024ymv} makes it a candidate all-energy generalization of JT gravity in which exact quantum results remain accessible.
In the same way JT gravity is a BF theory, sine dilaton gravity can be recast as a Poisson sigma model \cite{Schaller:1994es,Ikeda:1993fh,Cattaneo:2001bp,Fan:2021bwt,Blommaert:2023opb,Blommaert:2023wad}, i.e. a generalization of the former where the $\mathfrak{sl}(2,\mathbb{R})$ Poisson brackets of the components of $\mathbf{B}$ are replaced by a non-linear Poisson structure (the so-called Semenov-Tian-Shansky bracket \cite{Semenov-Tian-Shansky:1985mgd}), which is a classical limit of the $\mathcal{U}_q(\mathfrak{sl}_2)$ quantum algebra governing DSSYK \cite{Berkooz:2022mfk,Blommaert:2023opb,Belaey:2025ijg,Schouten:2025tvn}. Within that framework, the bulk reduction of section \ref{s:redphase} again leaves a six-dimensional phase space. Imposing a single gravitational boundary constraint
reduces it to a four-dimensional phase characterizing the $q$-Schwarzian, which is the $q$-analogue of the phase space $(f,p_f,\ell,p_\ell)$ of subsection \ref{s:fintesch}.
The strategy here would be to formulate the boundary conditions directly at the level of symplectic reduction of the classical Poisson-Lie non-linear symmetries of the Poisson-sigma model.

It would be interesting to study how the finite cutoff techniques developed in this work extend to the Poisson sigma model formulation.
For a general dilaton potential with prepotential $W'(\Phi)=V(\Phi)$, the finite cutoff boundary Hamiltonian is readily found to be
\begin{equation}\label{eq:SD_hamiltonian}
H = \sqrt{W(\Phi_b)}-\sqrt{W(\Phi_b)-\mathcal{C}},\qquad W(\Phi) = -\frac{\cos(2 \abs{\log{q}} \Phi)}{2 \log^2{q}}.
\end{equation}
Here $\mathcal{C}$ is the Casimir function of the Poisson structure, which reduces to $2\Tr(\mathbf{B}^2)$ as $q\to1$ and equals $W(\Phi_h)$ on shell. The first term in \eqref{eq:SD_hamiltonian} is a counterterm needed that is required as we expect the finite cutoff sine dilaton boundary conditions to be some deformation of those proposed in \cite{Blommaert:2024ymv}
\begin{equation}
    2 \abs{\log{q}} \Phi_b = \frac{\pi}{2} + i \infty, \qquad \sqrt{h} \, e^{i \log{q} \Phi_b} = \frac{i}{2 \log{q}}.
\end{equation}
Moreover, in our analysis the finite cutoff boundary condition constrains the hyperbolic generator $E+F$. By analogy, we expect the finite cutoff boundary conditions in sine dilaton gravity to translate into constraints on hyperbolic elements of $\mathcal{U}_q(\mathfrak{sl}_2)$.
This is consistent with the findings for Liouville gravity at finite cutoff \cite{toap3}, which exhibits an analogous structure.

Finally, it would be interesting to relate the resulting finite cutoff theory to $T\bar{T}$-inspired deformations of the DSSYK model, which have already been discussed in the literature \cite{Aguilar-Gutierrez:2024oea,Aguilar-Gutierrez:2026ogo}.

\section*{Acknowledgments}
We thank Andreas Blommaert, Luca Griguolo, Domenico Seminara, Thomas Tappeiner and Qi-Feng Wu. We especially thank Jacopo Papalini for early collaboration, discussions and comments on the draft. AB acknowledges the UGent Special Research Fund (BOF) for financial support. LR and AT have been supported in part by the Italian Ministero
dell’Università e della Ricerca (MIUR), and by Istituto Nazionale di Fisica Nucleare (INFN)
through the “Gauge and String Theory” (GAST) research project. TM acknowledges financial support from the European Research Council (grant BHHQG-101040024). Funded by the European Union.
Views and opinions expressed are however those of the author(s) only and do not necessarily reflect those of the European Union or the European Research Council. Neither the European Union nor the granting authority can be held responsible for them. We acknowledge fruitful discussions during the workshop ``Observers, wormholes, and complex saddles in cosmology”, organized at the Bernoulli Center for Fundamental Studies (EPFL, Lausanne) from 18--22 May 2026.

\appendix
\section{Isometries of AdS$_2$}
\label{app:embedding}
In this appendix, we review the geometry of AdS$_2$ in the embedding formalism.

\subsection{AdS$_2$ as a homogeneous space}\label{app:isometries}
The $\mathfrak{so}(2,1)$-algebra defined in \eqref{eq:so(2,1)algebra} 
\begin{align}
\label{eq:algebraappendix}
    [P_0,P_2]=P_1,\quad [P_1,P_2]=P_0, \quad[P_0,P_1]=-P_2
\end{align}
can be equipped with an invariant Cartan-Killing form \eqref{eq:CKmertric} $\text{Tr}(P_IP_J)=\frac{\eta_{IJ}}{2}$ with $\eta_{IJ}=\text{diag}(-1,1,1)$. 
Expanding a general element as  $Y=Y_{0}P_0+Y_1P_1+Y_2P_2$, we can define a metric in $\mathbb{R}^{2,1}$ as 
\begin{align}
    ds^2\equiv -2\text{Tr}(dY^2)=-dY_1^2-dY_2^2+dY_0^2.
\end{align}
One can realize AdS$_2$ as the homogeneous space generated by the right coset under the local Lorentz group SO$(1,1)$ \begin{align}
    \text{AdS}_2=\frac{\text{SO}(2,1)}{\text{SO}(1,1)}.
\end{align} Starting from the reference point $Y=P_2$, the homogeneous space is covered by the orbit of the adjoint action $Y\equiv gP_2g^{-1}$. Its stabilizer is generated by the group action generated by $P_2$. Consequently,
\begin{align}
\label{eq:stabilizertransf}
    g\longrightarrow ge^{-\eta P_2}
\end{align}
leaves the spacetime point $Y$ unchanged. This is the freedom to choose a local Lorentz frame at the same point. The remaining generators move the point according to the algebra \eqref{eq:algebraappendix}. Since $\Tr(Y^2)=1/2$, the squared norm $2\text{Tr}(Y^2)=Y^2_{1}+Y^2_{2}-Y^2_{0}$ defines the isometric embedding of AdS$_2$ in $\mathbb{R}^{2|1}$ as the hyperboloid 
\begin{align}
\label{eq:hyperboloid}
    -Y_{1}^2-Y_2^2+Y_0^2=-1.
\end{align}
Parameterizing the compact direction with a periodic global time coordinate $-\pi/2<\nu<\pi/2$,\footnote{Taking \(\nu\sim\nu+2\pi\) and \(\rho\in\mathbb R\), this representative covers the ordinary \(\mathrm{AdS}_2\) hyperboloid. The physical spacetime region for which the dilaton profile is positive, is obtained by restricting to \(-\pi/2\leq\nu\leq\pi/2\). On \(\widetilde{\mathrm{AdS}}_2\), the angular time coordinate is unwrapped to \(\nu\in\mathbb R\).} and a radial global coordinate $\rho$ which interpolates between the left boundary at $\rho\rightarrow -\infty$ and a right boundary at $\rho \rightarrow \infty$, the transformation is determined by $g=e^{\nu P_0}e^{\rho P_1}$, leading to the adjoint orbit\begin{align}
    Y=\cosh\rho \cos\nu P_2+\cosh\rho \sin\nu P_1+\sinh\rho P_0
\end{align}
which lives on the locus described by the hyperboloid \eqref{eq:hyperboloid}. The pullback of the metric on $\mathbb{R}^{2|1}$ naturally lives on the coset space \begin{align}
dY&=g[g^{-1}dg,P_2]g^{-1}=g[g^{-1}dg\rvert_{P_2=0},P_2]g^{-1}
\end{align} leading to the identification with the Cartan-Maurer metric of AdS$_2$ in global coordinates \begin{align}\label{eq:metriccoset}
    ds^2\equiv  -2\text{Tr}(dY^2)=2\text{Tr}(g^{-1}dg\rvert_{P_2=0})^2.
\end{align} Decomposing the Cartan-Maurer in terms of the metric and spin connections \begin{align}\label{eq:gaugeidentification}
    g^{-1}dg=e^0P_0+e^1P_1+\omega P_2,
\end{align} the embedding the metric is given in terms of local co-frames \begin{align}
ds^2=-2\text{Tr}(dY^2)
=-(e^0)^2+(e^1)^2.
\end{align} In particular, using the previous seed coordinate $g=e^{\nu P_0}e^{\rho P_1}$, one explicitly finds \begin{align}
    g^{-1}dg
=\cosh\rho\,d\nu\,P_0
+d\rho\,P_1
-\sinh\rho\,d\nu\,P_2,
\end{align}leading to the metric and the frame identification 
\begin{align}
    ds^2=d\rho^2-\cosh^2\rho \,d\nu^2, \qquad e^0=\cosh\rho\,d\nu,\quad
e^1=d\rho,\quad
\omega=-\sinh\rho\,d\nu.
\end{align} 
The transformation by the stabilizer $h=e^{\eta P_2}$ in the coset $g\rightarrow gh^{-1}$ \eqref{eq:stabilizertransf}, leads to:
\begin{align}
    g^{-1}dg\longrightarrow
h(g^{-1}dg)h^{-1}-dh\,h^{-1}.
\end{align}
Its components then indeed transform correctly as the local gauge fields \eqref{eq:componenttransformations}
\begin{align}
    \begin{pmatrix}e^0\\ e^1\end{pmatrix}
\longrightarrow
\begin{pmatrix}
\cosh\eta&-\sinh\eta\\
-\sinh\eta&\cosh\eta
\end{pmatrix}
\begin{pmatrix}e^0\\ e^1\end{pmatrix},
\qquad
\omega\longrightarrow\omega-d\eta.
\end{align} 
On shell, this geometrically identifies the gauge field with the Cartan-Maurer one-form $\mathbf{A}=g^{-1}dg$ in the expansion of a general gravitational gauge field \eqref{eq:fieldexpansions}. 

Redefining $\cosh\rho=\csc \sigma$, $\sinh\rho=-\cot \sigma$, we can further relate this with the AdS$_2$ metric in global coordinates \eqref{eq:geometry_on_shell} 
\begin{align}
    ds^2 = \frac{-d\nu^2 + d\sigma^2}{\sin^2\sigma}
\end{align}
for which the hyperboloid embedding is then given by 
\begin{align}
     Y=\csc\sigma \cos\nu P_2+\csc\sigma \sin\nu P_1-\cot\sigma P_0
\end{align}

Since the hyperboloid is given by $\text{Tr}(Y^2)=1/2$, isometries are realized by the adjoint action on the algebra $Y\rightarrow gYg^{-1}$. The infinitesimal adjoint action of the generators is then given as $\delta_X Y=[X,Y]$, leading to the Killing vector fields $\delta_X Y=-\zeta^\nu_X \partial_\nu Y-\zeta^\sigma_X \partial_\sigma Y$. In particular, using the algebra
\begin{align}
    [P_0,Y]=-Y_{1}P_2+Y_2P_1, \quad[P_1,Y]=Y_0P_2+Y_2P_0,\quad [P_2,Y]=-Y_{0}P_1-Y_1P_0,
\end{align}
    leads to the $\mathfrak{sl}_2$-realization of the Killing vectors\begin{align}
    \zeta_{P_0}=-\partial_\nu,\quad \zeta_{P_1}&=\sin\nu \tanh \rho \;\partial_\nu-\cos\nu\; \partial_\rho,\quad \zeta_{P_2}= \cos\nu \tanh \rho \;\partial_\nu+\sin\nu \;\partial_\rho.
\end{align}
Equivalently, in terms of the global coordinates used in section \ref{sec:geometry}, the Killing vectors read \begin{align}
    \zeta_{P_0}=-\partial_\nu,\quad \zeta_{P_1}=-\sin\nu \cos\sigma \;\partial_\nu-\cos\nu\sin\sigma\; \partial_\sigma,\quad \zeta_{P_2}= -\cos\nu \cos \sigma \;\partial_\nu+\sin\nu\sin\sigma \;\partial_\sigma.
\end{align}
Importantly, we see that the Lorentzian time generator corresponds to the action of the compact elliptic generator $P_0=\frac{1}{2}(E-F)$. Therefore, the corresponding coordinate $\nu\sim \nu+2\pi$ is periodic on the global patch of AdS$_2$. Taking the universal cover $\widetilde{\text{AdS}}_2=\widetilde{\text{SO}(2,1)/}\text{SO}(1,1)$, leads to an unwrapped time coordinate $\nu\in \mathbb{R}$.

\subsection{Isometric embedding of the boundary geodesics}\label{app:isometricembedding}

Starting from the global AdS\(_2\) metric and frame \eqref{eq:geometry_on_shell}
\begin{align}
ds^2=\frac{-d\nu^2+d\sigma^2}{\sin^2\sigma},
\qquad
e^0=\frac{d\nu}{\sin\sigma},
\qquad
e^1=\frac{d\sigma}{\sin\sigma}
\end{align} 
the pullback along the curve defined by  $\sigma=\pi/2$ describes a geodesic $\gamma_0$. Starting from this reference geodesic, we can reach the general solution \eqref{eq:branecurve} by performing a radial deformation.  As a homogeneous space, AdS$_2$ is realized by the adjoint orbit of the hyperbolic reference element $P_2$, whose stabilizer is the $\text{SO}(1,1)$ local Lorentz subgroup. Following appendix \ref{app:isometries}, the pullback of the hyperboloid embedding of AdS$_2$ on the reference geodesic is derived from the adjoint orbit \begin{align}
    Y(\nu)
=
e^{\nu P_0}P_2e^{-\nu P_0}
=
\cos\nu\,P_2+\sin\nu\,P_1.
\end{align} In other words, a circle on the intersection of the hyperboloid $-Y_1^2-Y_2^2+Y_0^2=-1$ with the $Y_0=0$ plane. The infinitesimal variation under the compact elliptic generator $\delta_{P_0} Y=[P_0,Y]=\frac{dY}{d\nu}$ leads to the Killing vector $\zeta_{P_0}=-\partial_\nu$, which remains tangent to $\sigma=\pi/2$. Alternatively, $\partial_\nu=Y_2\partial_{Y_1}-Y_1\partial_{Y_2}$ is the rotation matrix leaving the circular intersection of the $Y_0$-plane with the hyperboloid invariant.

Now let $g_\gamma\in \text{SL}(2,\mathbb R)$ map the reference geodesic to another timelike geodesic $Y_\gamma(\nu)=g_\gamma Y(\nu)g_\gamma^{-1}$. The Killing transformation tangent to the curve is then generated by $P_\gamma=g_\gamma P_0g_\gamma^{-1}$, whose finite action gives \begin{align}
e^{sP_\gamma}Y_\gamma(\nu)e^{-sP_\gamma}=g_\gamma e^{sP_0}\bigl(e^{\nu P_0}P_2e^{-\nu P_0}\bigr)e^{-sP_0}g_\gamma^{-1}=Y_\gamma(\nu+s)
\end{align} 
Taking a particular deformation of the reference brane in the radial direction with $g_\gamma=e^{-\eta P_1},$ with $k=\tanh\eta,$,  the tangent generator to the reference plane transforms as
\begin{align}
    g_\eta P_0g_\eta^{-1}=\cosh\eta\,P_0-\sinh\eta\,P_2.
\end{align} Consequently, the transformed plane is determined by \begin{align}
    2\text{Tr}
\left[
\left(
\cosh\eta P_0-\sinh\eta P_2
\right)Y
\right]=0,
\end{align} which becomes $Y_0+kY_2=0$. Using the hyperboloid embedding coordinates $Y_0=-\cot \sigma$, $Y_2=\csc\sigma \cos\nu$, we obtain precisely the brane geodesic \eqref{eq:branecurve} \begin{align}
    \cos\sigma=k\cos\nu.
\end{align}

\section{Endpoint symmetries and Hamiltonian flows on $T^*G$}
\label{app:sympl}
In this appendix, we collect, review and apply Hamiltonian and symplectic geometry techniques to the BF model with boundaries. This appendix serves to make the treatment in the main text in section \ref{s:redphase} self-contained.

\subsection{Gauge moment map and endpoint charges}
\label{app:gauge_moment_map_endpoint_charges}
Our conventions on the canonical phase with the prepotential defined from 
\begin{align}
    S=\int dt [p_i\dot{q}^i-H], \qquad \Theta=p_i\delta q^i,
\end{align} 
fix the symplectic two-form as 
\begin{align}
    \Omega=-\delta \Theta=dq^i\land dp_i.
\end{align}
Symplectomorphisms along $\mathcal{L}_{X_H}\Omega=0$ are generated by the moment maps $H$ defined from 
\begin{align}
\label{eq:app_hamiltonian_vector_field_convention}
    \iota_{X_H}\Omega=\delta H=\frac{\partial H}{\partial q^i}\delta q^i+\frac{\partial H}{\partial p_i}\delta p_i,
\end{align}
leading to 
\begin{align}
    X_H=\frac{\partial H}{\partial p_i}\partial_{q^i}-\frac{\partial H}{\partial q^i}\partial_{p_i}.
\end{align}
The variation along the vector field is then defined in terms of the Poisson bracket 
\begin{align}\label{eq:variation}
    \dot{F}=\mathcal{L}_{X_H}F\equiv -\{H,F\}.
\end{align}

On the BF theory, the starting point is the symplectic potential \eqref{eq:unred},
\begin{align}
\Theta_{\rm unred}
=
-2\int_{\sigma_-}^{\sigma_+}d\sigma\,
\Tr\!\left(
\mathbf B\,\delta\mathbf A_\sigma
\right),
\end{align}
whose exterior derivative on field space is the symplectic two-form
\begin{align}
\Omega_{\rm unred}
=-
\delta\Theta_{\rm unred}
=-
2\int_{\sigma_-}^{\sigma_+}d\sigma\,
\Tr\!\left(
\delta\mathbf A_\sigma\wedge\delta\mathbf B
\right).
\label{eq:app_unreduced_symplectic_form}
\end{align}

Under an infinitesimal gauge transformation \(h=1+\epsilon\), with a field-independent parameter \(\epsilon(\sigma)\in\mathfrak{sl}(2,\mathbb R)\), the fields transform as
\begin{align}
\delta_\epsilon\mathbf A_\sigma
=
-D_\sigma\epsilon,
\qquad
\delta_\epsilon\mathbf B
=
[\epsilon,\mathbf B],
\label{eq:app_infinitesimal_gauge_transformations}
\end{align}
so that the associated vector field on field space is
\(
X_\epsilon
=
\left(
-D_\sigma\epsilon,\,
[\epsilon,\mathbf B]
\right)
\).
Contracting it with $\Omega_{\rm unred}$ \eqref{eq:app_unreduced_symplectic_form} gives
\begin{align}
\iota_{X_\epsilon}\Omega_{\rm unred}
=
2\int_{\sigma_-}^{\sigma_+}d\sigma\,
\left[
\Tr\!\left(
D_\sigma\epsilon\,\delta\mathbf B
\right)
+
\Tr\!\left(
\delta\mathbf A_\sigma[\epsilon,\mathbf B]
\right)
\right],
\label{eq:app_gauge_contraction}
\end{align}
and by \eqref{eq:app_hamiltonian_vector_field_convention} the generator we are looking for is the function whose variation equals this expression.

Consider the Gauss constraint, smeared with the same parameter,
\begin{align}
\mathcal G_{\rm bulk}[\epsilon]
=-
2\int_{\sigma_-}^{\sigma_+}d\sigma\,
\Tr\!\left(
\epsilon\,D_\sigma\mathbf B
\right).
\end{align}
Using \(\delta(D_\sigma\mathbf B)=D_\sigma\delta\mathbf B+[\delta\mathbf A_\sigma,\mathbf B]\) and integrating the first term by parts, its variation is
\begin{align}
\delta\mathcal G_{\rm bulk}[\epsilon]
&=-
2\int_{\sigma_-}^{\sigma_+}d\sigma\,
\Tr\!\left[
\epsilon
\left(
D_\sigma\delta\mathbf B
+
[\delta\mathbf A_\sigma,\mathbf B]
\right)
\right]
\nonumber\\
&=
2\int_{\sigma_-}^{\sigma_+}d\sigma\,
\Tr\!\left(
D_\sigma\epsilon\,\delta\mathbf B
\right)
+
2\int_{\sigma_-}^{\sigma_+}d\sigma\,
\Tr\!\left(
\delta\mathbf A_\sigma[\epsilon,\mathbf B]
\right)
\nonumber\\
&\hspace{1cm}
-
2\Tr\!\left(
\epsilon_+\,\delta\mathbf B_+
\right)
+
2\Tr\!\left(
\epsilon_-\,\delta\mathbf B_-
\right),
\label{eq:app_bulk_gauss_variation}
\end{align}
where \(\epsilon_\pm\equiv\epsilon(\sigma_\pm)\). The first two terms reproduce \eqref{eq:app_gauge_contraction} exactly, so \(\mathcal G_{\rm bulk}\) does generate the gauge transformation, but only up to the endpoint terms in the last line.

For a parameter with arbitrary endpoint values, the endpoint terms must therefore be compensated, and the generator is
\begin{align}
\mathcal G[\epsilon]
=-
2\int_{\sigma_-}^{\sigma_+}d\sigma\,
\Tr\!\left(
\epsilon\,D_\sigma\mathbf B
\right)
-
2\Tr\!\left(
\epsilon_-\,\mathbf B_-
\right)
+
2\Tr\!\left(
\epsilon_+\,\mathbf B_+
\right),
\label{eq:app_full_gauge_generator}
\end{align}
satisfying
$\delta\mathcal G[\epsilon]
=
\iota_{X_\epsilon}\Omega_{\rm unred}$.

Although \eqref{eq:app_full_gauge_generator} generates the gauge action for any \(\epsilon\), its value on the Gauss-constraint surface depends only on the endpoint data, and the two cases are physically different. Writing \(\approx\) for equality on the constraint surface, a parameter with
\(
\epsilon_-=\epsilon_+=0
\)
gives
\begin{align}
\mathcal G[\epsilon]
=-
2\int_{\sigma_-}^{\sigma_+}d\sigma\,
\Tr\!\left(
\epsilon\,D_\sigma\mathbf B
\right)
\approx0 ,
\end{align}
so that for any vector \(V\) tangent to the constraint surface
\begin{align}
\Omega_{\rm unred}(X_\epsilon,V)
=
\delta\mathcal G[\epsilon](V)
=
0.
\end{align}
Gauge transformations trivial at the endpoints thus span null directions of the symplectic form pulled back to the Gauss-constraint surface, and are precisely the redundancies quotiented in the bulk reduction. This is the same argument that later identifies the stabilizer flows as null directions once the moment-map values are fixed.

When the endpoint values are allowed to be nonzero, the generator instead retains the finite value
\begin{align}
\mathcal G[\epsilon]
\approx-
2\Tr\!\left(
\epsilon_-\,\mathbf B_-
\right)
+
2\Tr\!\left(
\epsilon_+\,\mathbf B_+
\right),
\end{align}
so the transformation carries a nonvanishing Hamiltonian charge. Because \(\epsilon_+\) and \(\epsilon_-\) can be varied independently, this splits into two separate charges,
\begin{align}
Q_-[\epsilon_-]
=-
2\Tr\!\left(
\epsilon_-\,\mathbf B_-
\right),
\qquad
Q_+[\epsilon_+]
=
2\Tr\!\left(
\epsilon_+\,\mathbf B_+
\right),
\label{eq:app_endpoint_charges}
\end{align}
whose relative sign is fixed by the orientation of the interval through the boundary terms in \eqref{eq:app_bulk_gauss_variation}. Writing \(Q_\pm[\epsilon_\pm]=2\Tr(\epsilon_\pm\,\mu^\pm)\) gives the endpoint moment maps used in the main text,
\begin{align}
\mathbf{B}_\pm=\pm\left[
\mu_{H}^\pm H
+\frac14\mu_{E+F}^\pm(E+F)
-\frac14\mu_{E-F}^\pm(E-F)
\right].
\end{align}

The distinction between bulk gauge redundancy and endpoint symmetry is therefore already visible before passing to the finite-dimensional description: transformations vanishing at the endpoints are generated entirely by the Gauss constraint and are removed in the quotient, whereas transformations with nonzero endpoint values survive on the bulk-reduced phase space with the charges \eqref{eq:app_endpoint_charges}. Consistently with \eqref{eq:holonomytransf}, the open holonomy transforms infinitesimally as
\begin{align}
\delta_\epsilon g
=
\epsilon_+\,g
-
g\,\epsilon_-,
\label{eq:app_infinitesimal_holonomy_action}
\end{align}
so that the right endpoint acts by left multiplication and the left endpoint by right multiplication. In the next subsection we recover these actions as Hamiltonian flows of the endpoint momentum components on \(T^*G\) and derive their Poisson algebra.

\subsection{Hamiltonian vector fields on \(T^*G\)}
\label{app:hamiltonian_vector_fields_TstarG}

We now derive, directly on the bulk-reduced phase space, the Hamiltonian flows generated by the endpoint components of the $\mathbf{B}$-field. 

The reduced symplectic potential \eqref{eq:reducedsymplecticpotential} can be written in either right- or left-trivialized variables,
\begin{align}
\Theta_{\text{red}}
=
2\Tr\!\left(\mathbf B_+\,\theta_R\right)
=
2\Tr\!\left(\mathbf B_-\,\theta_L\right),
\qquad
\theta_R\equiv\delta g\,g^{-1},
\qquad
\theta_L\equiv g^{-1}\delta g ,
\label{eq:app_two_trivializations}
\end{align}
the two momenta being related by parallel transport, \(\mathbf B_+=g\,\mathbf B_-\,g^{-1}\). Each trivialization is adapted to one endpoint and we now treat the two cases in turn. Throughout, wedge products of one-forms are evaluated as \((\alpha\wedge\beta)(V_1,V_2)=\alpha(V_1)\beta(V_2)-\alpha(V_2)\beta(V_1)\), and a dot denotes the derivative along the flow whose generator is under discussion.

In the right-trivialized description \((g,\mathbf B_+)\), the Maurer--Cartan equation \(\delta\theta_R=\theta_R\wedge\theta_R\) gives:\footnote{Using the right Maurer-Cartan equation $\delta\theta_R=\theta_R\wedge\theta_R,$
and the algebra \eqref{eq:algebra} $[P_I,P_J]=-\epsilon_{IJ}^{\;\;\;\;K}P_K$,
we can write in component form
\begin{align}\label{eq:rightsymplectictwoform}
    \Omega_{\text{red}}
    =-
    \delta B_I\wedge \theta^I
    +
    \frac{1}{2}B_I\epsilon_{JK}^{\;\;\;\;\;I}\theta^J\wedge\theta^K.
\end{align}
The first term is the ordinary canonical contribution, Since the configuration space is the group manifold SL\((2,\mathbb{R})\), the second term is the non-abelian correction.}
\begin{align}
\Omega_{\text{red}}
=-
\delta\Theta_{\text{red}}
=-
2\Tr\!\left(
\delta\mathbf B_+\wedge\theta_R
\right)
-
2\Tr\!\left(
\mathbf B_+\,\theta_R\wedge\theta_R
\right).
\label{eq:app_right_trivialized_symplectic_form}
\end{align}
Representing a tangent vector to \(T^*G\) at \((g,\mathbf B_+)\) as
\begin{align}
V=(\xi,\beta),
\qquad
\xi\equiv\dot g\,g^{-1},
\qquad
\beta\equiv\dot{\mathbf B}_+,
\end{align}
so that \(\theta_R(V)=\xi\) and \(\delta\mathbf B_+(V)=\beta\), the symplectic form evaluates on two such vectors to
\begin{align}
\Omega_{\text{red}}(V_1,V_2)
=-
2\Tr\!\left(
\beta_1\xi_2
-
\beta_2\xi_1
+
\mathbf B_+[\xi_1,\xi_2]
\right).
\label{eq:app_right_symplectic_on_vectors}
\end{align}

Consider now the charge associated with a fixed \(X\in\mathfrak{sl}(2,\mathbb R)\),
\begin{align}
B_{X,+}
\equiv
2\Tr(\mathbf B_+X),
\end{align}
and let \(X_{B_{X,+}}=(\xi_X,\beta_X)\) be its Hamiltonian vector field, defined as in \eqref{eq:app_hamiltonian_vector_field_convention} by
\begin{align}
\iota_{X_{B_{X,+}}}\Omega_{\text{red}}
=
\delta B_{X,+}.
\label{eq:app_Hamiltonian_vector_field_convention}
\end{align}
Evaluating both sides on a generic tangent vector \(V=(\xi,\beta)\) gives
\begin{align}
-2\Tr\!\left(
\beta_X\xi
-
\beta\xi_X
+
\mathbf B_+[\xi_X,\xi]
\right)
=
2\Tr(\beta X),
\end{align}
where \(\xi\) and \(\beta\) may be varied independently. Matching the coefficient of \(\beta\) fixes \(\xi_X=X\), and the remaining terms then give \(\beta_X=-[\mathbf B_+,X]\). The Hamiltonian equations generated by \(B_{X,+}\) are therefore
\begin{align}
\dot g\,g^{-1}
=
X,
\qquad
\dot{\mathbf B}_+
=
[X,\mathbf B_+],
\label{eq:app_Bplus_Hamiltonian_equations}
\end{align}
and integrate to
\begin{align}
g(t)
=
e^{tX}g(0),
\qquad
\mathbf B_+(t)
=
e^{tX}\mathbf B_+(0)e^{-tX}.
\label{eq:app_Bplus_Hamiltonian_flow}
\end{align}
Thus \(B_{X,+}\) generates left multiplication of the open holonomy, under which the momentum written in the opposite endpoint frame stays fixed,
\begin{align}
\mathbf B_-(t)
=
g(t)^{-1}\mathbf B_+(t)g(t)
=
\mathbf B_-(0).
\label{eq:app_Bminus_constant_under_plus_flow}
\end{align}

The left endpoint is treated in the same way, now in the left-trivialized variables \((g,\mathbf B_-)\), in which \(B_{X,-}\) is linear in the momentum. From \(\delta\theta_L=-\theta_L\wedge\theta_L\),
\begin{align}
\Omega_{\text{red}}
=-
2\Tr\!\left(
\delta\mathbf B_-\wedge\theta_L
\right)
+
2\Tr\!\left(
\mathbf B_-\,\theta_L\wedge\theta_L
\right),
\label{eq:app_left_trivialized_symplectic_form}
\end{align}
and writing
\begin{align}
V=(\xi,\beta),
\qquad
\xi\equiv g^{-1}\dot g,
\qquad
\beta\equiv\dot{\mathbf B}_-,
\end{align}
one obtains
\begin{align}
\Omega_{\text{red}}(V_1,V_2)
=-
2\Tr\!\left(
\beta_1\xi_2
-
\beta_2\xi_1
-
\mathbf B_-[\xi_1,\xi_2]
\right).
\label{eq:app_left_symplectic_on_vectors}
\end{align}
For
\(
B_{X,-}\equiv2\Tr(\mathbf B_-X)
\),
the defining relation \(\iota_{X_{B_{X,-}}}\Omega_G=\delta B_{X,-}\) again fixes \(\xi_X=X\) from the coefficient of \(\beta\), the sign of the commutator term in \eqref{eq:app_left_symplectic_on_vectors} now giving
\begin{align}
g^{-1}\dot g
=
X,
\qquad
\dot{\mathbf B}_-
=
[\mathbf B_-,X],
\label{eq:app_Bminus_Hamiltonian_equations}
\end{align}
with finite flow
\begin{align}
g(t)
=
g(0)e^{tX},
\qquad
\mathbf B_-(t)
=
e^{-tX}\mathbf B_-(0)e^{tX}.
\label{eq:app_Bminus_Hamiltonian_flow}
\end{align}
Hence \(B_{X,-}\) generates right multiplication, leaving the momentum in the right-endpoint frame invariant,
\begin{align}
\mathbf B_+(t)
=
g(t)\mathbf B_-(t)g(t)^{-1}
=
\mathbf B_+(0).
\label{eq:app_Bplus_constant_under_minus_flow}
\end{align}

That the endpoint label and the side on which the group acts appear crossed is a consequence of the orientation chosen for the Wilson line in \eqref{eq:holonomy}. Under a finite endpoint gauge transformation
\begin{align}
g
\longmapsto
h_+\,g\,h_-^{-1},
\label{eq:app_endpoint_action_on_holonomy}
\end{align}
a transformation at the right endpoint acts from the left and one at the left endpoint from the right. Comparing this with the flows just derived also fixes the signs of the endpoint charges. For \(h_-=e^{t\epsilon_-}\) the left-endpoint action is \(g\mapsto g\,e^{-t\epsilon_-}\), whereas \eqref{eq:app_Bminus_Hamiltonian_flow} generates $g\mapsto g\,e^{tX}$ with \(X=\epsilon_-\). The gauge transformation is therefore not generated by \(B_{\epsilon_-}^-\) but by its opposite, so that the gauge generator is \(-B_{\epsilon_-}^-\). At the right endpoint, by contrast, \(h_+=e^{t\epsilon_+}\) gives \(g\mapsto e^{t\epsilon_+}g\), which is \eqref{eq:app_Bplus_Hamiltonian_flow}. The two endpoint charges are then indeed $Q_+[\epsilon_+]
= B_{\epsilon_+}^+,
\;
Q_-[\epsilon_-]
=- B_{\epsilon_-}^-,$
in agreement with \eqref{eq:app_endpoint_charges}. This is the origin of the relative sign between the two moment maps,
\begin{align}
\mu_{A}^+P^A=\mathbf B_+,
\qquad
\mu_{A}^-P^A= -\mathbf{B}_- .
\end{align}

Equations \eqref{eq:app_Bminus_constant_under_plus_flow} and \eqref{eq:app_Bplus_constant_under_minus_flow} show that each endpoint flow leaves  the momentum expressed in the frame of the other endpoint invariant, and on the group element the two act by multiplication from opposite sides. The endpoint transformations thus define independent left and right Hamiltonian actions on \(T^*G\), a statement we make precise in the next subsection by deriving their Poisson algebra.

\subsection{Poisson algebra and left/right regular actions}
\label{app:poisson_algebra_regular_actions}

The flows derived above determine both the Poisson algebra of the endpoint momenta and their action on functions of the holonomy. Since every sign below depends on it, we fix the bracket convention first. With the Hamiltonian vector field defined as in \eqref{eq:app_hamiltonian_vector_field_convention} \(\iota_{X_H}\Omega_G=\delta H\), we define the Poisson bracket as
\begin{align}
\{H,F\}
\equiv
-\mathcal L_{X_H}F,
\label{eq:app_poisson_bracket_convention}
\end{align}
so that an observable evolves as \(\dot F=\{F,H\}\). We also fix the normalization of the structure constants in
\begin{align}
[P_A,P_B]
=
-\epsilon_{AB}{}^{C}P_C ,
\label{eq:app_sl2_structure_constants}
\end{align}
taking \(\epsilon_{ABC}\) totally antisymmetric with \(\epsilon_{012}=1\) and indices raised with \(\eta^{AB}\), where \(\eta_{AB}=2\Tr(P_AP_B)=\mathrm{diag}(-1,1,1)\). This reproduces the algebra \eqref{eq:algebraappendix} of the basis used throughout.

The action of the endpoint momenta on functions of the group element is read off directly from the corresponding flows. Introducing the left- and right-regular differential operators
\begin{align}
L_XF(g)
&\equiv
\left.
\frac{d}{dt}
F\!\left(e^{-tX}g\right)
\right|_{t=0},
\label{eq:app_left_regular_definition}
\\[1ex]
R_XF(g)
&\equiv
\left.
\frac{d}{dt}
F\!\left(ge^{tX}\right)
\right|_{t=0},
\label{eq:app_right_regular_definition}
\end{align}
whose opposite exponents are adapted to the endpoint action \(g\mapsto h_+gh_-^{-1}\), the right-endpoint flow \eqref{eq:app_Bplus_Hamiltonian_flow}, \(g(t)=e^{tX}g\), gives
\begin{align}
\{B_{X}^+,F(g)\}
=
L_XF(g),
\label{eq:app_Bplus_action_on_functions}
\end{align}
while the left-endpoint flow \eqref{eq:app_Bminus_Hamiltonian_flow}, \(g(t)=ge^{tX}\), gives
\begin{align}
\{B_{X}^-,F(g)\}
=-
R_XF(g).
\label{eq:app_Bminus_action_on_functions}
\end{align}

The brackets among the momentum components follow from a similar logic. Writing \(B_{A}^\pm\equiv2\Tr(\mathbf B_\pm P_A)\), the right-endpoint flow gives \(\dot{\mathbf B}_+=[P_A,\mathbf B_+]\), so that with \eqref{eq:app_poisson_bracket_convention}
\begin{align}
\{B_{A}^+,B_{B}^+\}
&=
2\Tr\!\left(
[\mathbf B_+,P_A]P_B
\right)
\nonumber
=
2\Tr\!\left(
\mathbf B_+[P_A,P_B]
\right)
\nonumber\\
&=-
\epsilon_{AB}{}^{C}B_{C}^+.
\label{eq:app_Bplus_poisson_algebra}
\end{align}
At the left endpoint the flow instead gives \(\dot{\mathbf B}_-=[\mathbf B_-,P_A]\), and cyclicity of the trace now produces the opposite sign in the second step,
\begin{align}
\{B_{A}^-,B_{B}^-\}
&=
2\Tr\!\left(
[P_A,\mathbf B_-]P_B
\right)
\nonumber
=
-2\Tr\!\left(
\mathbf B_-[P_A,P_B]
\right)
\nonumber\\
&=
\epsilon_{AB}{}^{C}B_{C}^-.
\label{eq:app_Bminus_poisson_algebra}
\end{align}
The opposite signs in \eqref{eq:app_Bplus_poisson_algebra} and \eqref{eq:app_Bminus_poisson_algebra} is the familiar distinction between the left- and right-trivialized momentum maps on \(T^*G\).
The two endpoint algebras commute, as the flows already indicate: \(B_{X}^+\) leaves \(\mathbf B_-\) invariant by \eqref{eq:app_Bminus_constant_under_plus_flow}, and \(B_{Y}^-\) leaves \(\mathbf B_+\) invariant by \eqref{eq:app_Bplus_constant_under_minus_flow}. Hence
\begin{align}
\{B_{A}^+,B_{B}^-\}=0,
\label{eq:app_cross_endpoint_poisson}
\end{align}
or equivalently, on functions of the group element, \([L_X,R_Y]=0\).

These components should be distinguished from the generators of the endpoint gauge transformations, which by Appendix~\ref{app:gauge_moment_map_endpoint_charges} are
\begin{align}
\mu_{A}^-=-B_{A}^-,
\qquad
\mu_{A}^+=B_{A}^+.
\label{eq:app_endpoint_charge_components}
\end{align}
In terms of these variables, we obtain the Poisson algebra relations:
\begin{align}
\{\mu_{A}^+,\mu_{B}^+\}
&=
-\epsilon_{AB}{}^{C}\mu_{C}^+,
\label{eq:app_Qplus_algebra}
\\
\{\mu_{A}^-,\mu_{B}^-\}
&=
-\epsilon_{AB}{}^{C}\mu_{C}^-,
\label{eq:app_Qminus_algebra}
\\
\{\mu_{A}^+,\mu_{B}^-\}
&=
0 .
\label{eq:app_Qcross_algebra}
\end{align}
Thus, both endpoint charges generate \(\mathfrak{sl}(2,\mathbb R)\) with the structure constants \eqref{eq:app_sl2_structure_constants}, acting independently at the two ends of the interval. This structure is well-known, and is known to persist even in the non-linear deformation of BF theory \cite{Cattaneo:2001bp}.

\subsubsection{Classical finite cutoff symmetry generators}\label{app:classicalsymmetrygenerators}
As an explicit example, consider the geometrical setup of two timelike separated finite-cutoff boundary curves. The relevant hyperbolic Euler patch of the group manifold is given by the group element \eqref{eq:finitecutoffdecomp}: 
\begin{align}
\label{eq:gfc}
    g= e^{\beta(E+F)}e^{-LH}e^{\gamma(E+F)}.
\end{align} 
Evaluating the Cartan-Maurer form of this patch decomposition:
\begin{align}
\theta_L=g^{-1}\delta g&=
H\left(2\sinh L\,\sinh2\gamma\,\delta\beta
-\cosh2\gamma\,\delta L\right)+(E+F)\left(\cosh L\,\delta\beta+\delta\gamma\right)\nonumber\\
&+(E-F)\left(\sinh L\,\cosh2\gamma\,\delta\beta
-\frac12\sinh2\gamma\,\delta L\right).
\end{align}
Substituting this into the symplectic potential $\Theta_{\text{red}}=2\text{Tr}(\mathbf{B}_-\theta_L)\equiv p_\beta \,\delta\beta+p_L\,\delta L+p_\gamma\,\delta \gamma$ leads to 
\begin{align}
p_\beta&=-2\sinh L\,\sinh2\gamma\,\mu_{-,H}
-\cosh L\,\mu_{-,E+F}-\sinh L\,\cosh2\gamma\,\mu_{-,E-F},\\
p_L&=\cosh2\gamma\,\mu_{-,H}+\frac12\sinh2\gamma\,\mu_{-,E-F},\\
p_\gamma&=-\mu_{-,E+F},
\end{align}
where $p_\beta$, $p_L$, $p_\gamma$ are the Darboux conjugate momenta to the respective group coordinates, satisfying
$\{\beta, p_\beta\}=\{L,p_L\}=\{\gamma,p_\gamma\}=1$. Inverting these relations leads to 
\begin{align}\label{eq:rightfcclassicalsymmetrygenerators}
\mu_{E+F}^-&=-p_\gamma,\\
\mu_{H}^-&=\frac{\sinh2\gamma}{2\sinh L}\,p_\beta
+\cosh2\gamma\,p_L
-\frac12\coth L\,\sinh2\gamma\,p_\gamma,\\
\mu_{E-F}^-&=-\frac{\cosh2\gamma}{\sinh L}\,p_\beta
-2\sinh2\gamma\,p_L
+\coth L\,\cosh2\gamma\,p_\gamma.
\end{align}
One can explicitly check that these are right regular symmetry generators on phase space: 
\begin{align}\label{eq:classicalrightsymmetry}
    \{\mu_{H}^-,\mu_{E+F}^-\}=\mu_{E-F}^-,\qquad \{\mu_{H}^-,\mu_{E-F}^-\}=\mu_{E+F}^-, \qquad \{\mu_{E-F^-},\mu_{E+F}^-\}=4\mu_{H}^-.
\end{align} 

Analogously, evaluating the left regular Cartan-Maurer one-form on the hyperbolic Euler patch leads to 
\begin{align}
\theta_R=\delta g\,g^{-1}
&=H\left(2\sinh L\,\sinh(2\beta)\,\delta\gamma
-\cosh(2\beta)\,\delta L\right)+(E+F)\left(\delta\beta+\cosh L\,\delta\gamma\right) \nonumber\\
&+(E-F)\left(\frac12\sinh(2\beta)\,\delta L
-\sinh L\,\cosh(2\beta)\,\delta\gamma\right).
\end{align} 
Using $\Theta_{\text{red}}=2\text{Tr}(\mathbf{B}_+\theta_R)\equiv p_\beta \,\delta\beta+p_L\,\delta L+p_\gamma\,\delta \gamma$, we can identify the conjugate momenta in terms of the left-regular symmetry currents $\mathbf{\mu}_{A}^+P^A=\mathbf{B}_+$: 
\begin{align}
p_\beta&=\mu_{E+F}^+,\\
p_L&=-\cosh(2\beta)\mu_{H}^+
+\frac12\sinh(2\beta)\mu_{E-F}^+,\\
p_\gamma&=2\sinh L\,\sinh(2\beta)\mu_{H}^+
+\cosh L\,\mu_{E+F}^+
-\sinh L\,\cosh(2\beta)\mu_{E-F}^+.
\end{align}
Inverting these relations yields the classical left regular symmetry generators on phase space \begin{align}\label{eq:leftfcclassicalsymmetrygenerators}
\mu_{E+F}^+&=p_\beta,\\
\mu_{H}^+
&=\frac12\coth L\,\sinh(2\beta)\,p_\beta
-\cosh(2\beta)\,p_L
-\frac{\sinh(2\beta)}{2\sinh L}\,p_\gamma,\\
\mu_{E-F}^+
&=\coth L\,\cosh(2\beta)\,p_\beta
-2\sinh(2\beta)\,p_L
-\frac{\cosh(2\beta)}{\sinh L}\,p_\gamma.
\end{align}
One can again check that these classical currents satisfy the correct symmetry algebra 
\begin{align}\label{classicalleftsymetry}
    \{\mu_{H}^+,\mu_{E+F}^+\}=\mu_{E-F}^+,\qquad \{\mu_{H}^+,\mu_{E-F}^+\}=\mu_{E+F}^+, \qquad \{\mu_{E-F}^+,\mu_{E+F}^+\}=4\mu_{H}^+.
\end{align} 

\subsubsection{Classical EOW brane symmetry generators}\label{sec:eowbranesymmetrygenerators}
If we replace the left finite cutoff boundary by a timelike geodesic EOW brane, the boundary condition of the EOW brane is preserved by the one-parameter subgroup \eqref{eq:localEOWbranestabilizer} $H=\exp\bigl(\mathbb R(E-F)\bigr)$.
The group decomposition of the EOW brane is then given by the subpatch \eqref{eq:finitecutoffeowpatch}: 
\begin{align}
 g=e^{\beta(E+F)}e^{-LH}e^{\theta(E-F)}.
\end{align}
Inserting this patch decomposition in the canonical symplectic potential \eqref{eq:canonicaloneformcoordinates}, one can in principle repeat the analysis of the previous subsection  \ref{app:classicalsymmetrygenerators} to derive the classical symmetry generators on phase space. We will, however, use a trick to relate them directly with the classical symmetry generators of the finite cutoff patch by conjugation within SL$(2,\mathbb{C})$ by an SU$(1,1)$ matrix. In particular, we have the relation $e^{\theta(E-F)}=C^{-1}e^{i\theta(E+F)}C$ with 
\begin{align}\label{eq:conjugationmatrix}
    C=\begin{pmatrix}
        e^{i\pi/4}&0\\0&e^{-i\pi/4}
    \end{pmatrix}=e^{i\pi H/2}\in \text{SU}(1,1).
\end{align} 
This manifold parametrization then looks like an analytically continued version of \eqref{eq:gfc}:
\begin{align}\label{eq:patchconjugation}
    g=e^{\beta(E+F)}e^{-LH}e^{\theta(E-F)} = \underbrace{
e^{\beta(E+F)} e^{-L'H} e^{i\theta(E+F)}
}_{g'\equiv \, \eqref{eq:gfc}}
\, C,
\end{align} 
with we identify the parameters of \eqref{eq:gfc} as
\begin{align}
\label{eq:coordinatesubstitutions}
 L=L'-\frac{i\pi}{2},\qquad \gamma = i\theta. 
\end{align} 
If the symplectic potential on the group manifold \eqref{eq:reducedsymplecticpotential} is preserved under this substitution $\Theta_G=p_\beta \delta \beta+p_L\delta L+p_\theta \delta \theta=p_{\beta'}\delta \beta'+p_{L'}\delta L'+p_{\gamma'}\delta \gamma'$, the conjugate momenta are related to the finite cutoff patch as 
\begin{align}
    p_{\beta'}=p_\beta, \qquad p_{L'}=p_L, \qquad p_\gamma=-ip_\theta.
\end{align} 
These redefinitions consequently preserve the canonical Poisson brackets on phase space. The Cartan-Maurer forms transform as 
\begin{align}
    \theta_L\equiv g^{-1}\,\delta g = C^{-1}{g'}^{-1}\delta g'C, \qquad \theta_R\equiv \delta g\,g^{-1} = \delta g'\,{g'}^{-1}.
\end{align}
Preserving the symplectic potential $\Theta_{\text{red}}=2\Tr(\mathbf{B}_-\theta_L)=2\Tr(\mathbf{B}_+\theta_R)$ consequently relates the EOW brane variable $\mathbf{B}$ to the finite cutoff patch variable $\mathbf{B}'$ as
\begin{align}
    \mathbf{B}'_- = C\mathbf{B}_-C^{-1}, \qquad \mathbf{B}_+'=\mathbf{B}_+.
\end{align}
This relates the components of the moment map expansion 
\begin{align}
\mathbf{B}_\pm=\pm\left[
\mu_{H}^\pm H
+\frac14\mu_{E+F}^\pm(E+F)
-\frac14\mu_{E-F}^\pm(E-F)
\right]
\end{align}
as:\footnote{One can also see this directly from the group actions:
\begin{align}
    ge^{tX}= g'\,e^{t(CXC^{-1})}C,\qquad
e^{-tX}g=(e^{-tX} g')C.
\end{align} 
} 
\begin{align}
    \begin{aligned}
\mu_{H}^-
&= {\mu'}_{H}^-, 
&\qquad
\mu_{E+F}^-
&= i{\mu'}_{E-F}^-, 
&\qquad
\mu_{E-F}^-
&= i{\mu'}_{E+F}^-, \\
\mu_{H}^+
&= {\mu'}_{H}^+, 
&\qquad
\mu_{E+F}^+
&= {\mu'}_{E+F}^+, 
&\qquad
\mu_{E-F}^+
&= {\mu'}_{E-F}^+ .
\end{aligned}
\end{align} 
Explicitly, performing the coordinate substitutions \eqref{eq:coordinatesubstitutions} in \eqref{eq:rightfcclassicalsymmetrygenerators}, \eqref{eq:leftfcclassicalsymmetrygenerators}, we obtain directly the classical right- and left-regular symmetry generators 
\begin{align}\label{eq:eowclassicalregulargenrators}
\mu_{E-F}^-&=-p_\theta,\\
\mu_{H}^-&=\frac{\sin(2\theta)}{2\cosh L}\,p_\beta
+\cos(2\theta)\,p_L
-\frac12\tanh L\sin(2\theta)\,p_\theta,\\
\mu_{E+F}^-&=-\frac{\cos(2\theta)}{\cosh L}\,p_\beta
+2\sin(2\theta)\,p_L
+\tanh L\cos(2\theta)\,p_\theta.
\end{align}
and 
\begin{align}
\mu_{E+F}^+
&=p_\beta,\\
\mu_{H}^+
&=\frac12\tanh L\sinh(2\beta)\,p_\beta
-\cosh(2\beta)\,p_L
+\frac{\sinh(2\beta)}{2\cosh L}\,p_\theta,\\
\mu_{E-F}^+
&=\tanh L\cosh(2\beta)\,p_\beta
-2\sinh(2\beta)\,p_L
+\frac{\cosh(2\beta)}{\cosh L}\,p_\theta.
\end{align} 
satisfying the $\mathfrak{sl}_2$-symmetry algebra relations
\begin{align}
    \{\mu_{H}^\pm,\mu_{E+F}^\pm\}=\mu_{E-F}^\pm,\qquad \{\mu_{H}^\pm,\mu_{E-F}^\pm\}=\mu_{E+F}^\pm, \qquad \{\mu_{E-F}^\pm,\mu_{E+F}^\pm\}=4\mu_{H}^\pm.
\end{align}

\section{Hamiltonian reduction in the second-order formalism}
\label{app:secondphasespace}
In this appendix, we show how the reduced Hamiltonians \eqref{eq:casimir_length_variables_gjk} and \eqref{eq:mixed_casimir_length_general_a} can be recovered by choosing an explicit foliation of the bulk geometry, resulting in a “dynamical” generalization of the considerations presented in Section \ref{s:claspot} and \ref{s:claspot_EOW}.
What we learn from this construction is not only how the Hamiltonians emerge from the geometry, but also how classical trajectories of those Hamiltonians are mapped into the dynamics of Cauchy slices.
This approach is inspired by the one adopted in \cite{Harlow:2018tqv} for the case of two asymptotic Dirichlet boundaries, as well as in \cite{Gao:2021uro} for the case of one asymptotic Dirichlet boundary and an EOW brane.
Our goal in this section is to generalize those constructions to the case of finite cutoff boundary conditions.

First, we compute the Hamiltonian corresponding to each type of boundary. To do so, we first need the Brown-York stress-energy tensor
\begin{equation}
    T_{\alpha\beta}\bigg\lvert_{\partial\mathcal{M}}
    \equiv
    -\frac{2}{\sqrt{\abs{h}}}
    \frac{\delta S}{\delta h^{\alpha\beta}}\bigg\lvert_{\partial\mathcal{M}}.
\end{equation}
By differentiating \eqref{eq:totalaction}, we find that on a cutoff boundary
\begin{equation}\label{eq:brown_york_cutoff}
    T_{\alpha\beta}\bigg\lvert_{\partial\mathcal{M}_\text{cutoff}}
    =
    (\partial_n \Phi - \Phi)\bigg\lvert_{\partial\mathcal{M}_\text{cutoff}} h_{\alpha \beta}
    =
    - \left( \Phi_b  - \epsilon \sqrt{\Phi_b^2 - \Phi_h^2} \right) h_{\alpha\beta},
\end{equation}
where $\epsilon = +1$ if the outward normal is in the direction of increasing dilaton, and $\epsilon = -1$ otherwise.
Contracting \eqref{eq:brown_york_cutoff} with the vector field $\partial_\tau$ twice (here $\tau$ is the proper time along $\partial\mathcal{M}_\text{cutoff}$) we obtain
\begin{equation}\label{eq:ADM_energy}
    H = \Phi_b - \epsilon \sqrt{\Phi_b^2 - \Phi_h^2}. 
\end{equation}
An analogous procedure on $\partial\mathcal{M}_\text{EOW}$ leads to
\begin{equation}
    T_{\alpha\beta}\bigg\lvert_{\partial\mathcal{M}_\text{EOW}}
    =
    (\partial_n \Phi - \mu) h_{\alpha \beta} \bigg\lvert_{\partial\mathcal{M}_\text{EOW}} = 0,
\end{equation}
where we used the boundary condition \eqref{eq: boundary conditions}. Hence, we conclude that EOW boundaries carry no energy and do not contribute to the total Hamiltonian.

Our strategy consists of foliating the geometry \eqref{eq:geometry_on_shell} with spacelike boundary-to-boundary Cauchy slices on which we specify the state of the system. Since the two-sided phase space of JT gravity is two-dimensional, states at a given time are fully described by the value of an observable of our choice on the slice and the corresponding conjugate momentum.
Since the foliation procedure is qualitatively different depending on the chosen boundary conditions, we will consider each case separately in the following.

\subsection{Cutoff-cutoff boundary conditions}
\label{app:ccbc}
First, we consider the case of Dirichlet boundary conditions \eqref{eq: boundary conditions} on both the left and the right boundaries:
\begin{equation}
    \Phi \big\lvert_{\partial \mathcal{M}_{-}} = \Phi_b^-,
    \qquad
    \Phi \big\lvert_{\partial \mathcal{M}_{+}} = \Phi_b^+.
\end{equation}
Without loss of generality, we can choose $\Phi_b^- \leq \Phi_b^+$.
The special case $\Phi_b^- = \Phi_b^+$ was extensively studied in \cite{Griguolo:2026eoi}. Here, we briefly review that construction and generalize it to the case of unequal values of the boundary dilaton.

\begin{figure}
    \centering
    \input{images/penrose_diagram_cutoff_cutoff.tikz}
    \caption{The on-shell geometry for JT gravity with two different Dirichlet boundaries, with $\Phi_b^-$ and $\Phi_b^+$ corresponding to the left and right red curves, respectively. On the left panel, the boundaries lie on different sides of the black hole. On the right panel, the boundary curves both lie to the right of the horizon.}
    \label{fig:penrose_diagram_cutoff_cutoff}
\end{figure}
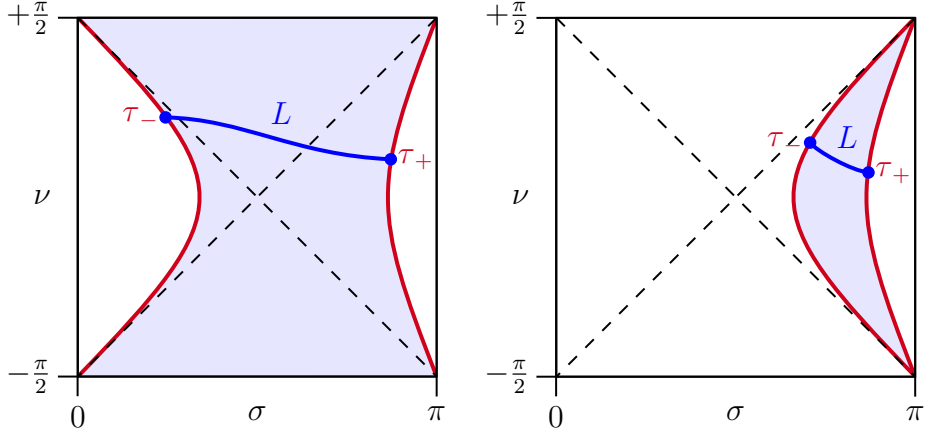

The resulting geometries are depicted in Figure \ref{fig:penrose_diagram_cutoff_cutoff}. We immediately notice that when $\Phi_b^- \neq \Phi_b^+$, we must deal with two qualitatively different cases. The two boundaries can lie on two causally disconnected regions on opposite ends of the black hole, or alternatively they can both lie on the same spacetime wedge. Remarkably, we will see that both these classes of classical geometries are encoded in the same Hamiltonian description, in the sense that they share the same Casimir $\Phi_h^2(L,P)$ defined below.

Let us start by considering the regular case of two causally disconnected boundaries first. We choose to foliate the geometry with spacelike geodesics connecting boundary points of equal proper time $\tau_{-} = \tau_{+}$. Consequently, it is convenient to label each slice with the total proper time
\begin{equation}
    \tau = \tau_{-} + \tau_{+}.
\end{equation}
Taking into account the contributions from both Dirichlet boundaries, the total Hamiltonian generating translations in $\tau$ is given by the average of two terms of the form \eqref{eq:ADM_energy}, since moving along the foliation by $\text{d}\tau$ shifts each boundary time by $\text{d}\tau/2$\footnote{Both contributions here come with $\epsilon = +1$ since the outward normal points in the direction of increasing dilaton.}
\begin{equation}\label{eq:hamiltonian_cutoffcutoff_A}
    H = \frac{H_{-} + H_{+}}{2} = \frac{1}{2} \left( \Phi_b^- - \sqrt{(\Phi_b^-)^2 - \Phi_h^2} + \Phi_b^+ - \sqrt{(\Phi_b^+)^2 - \Phi_h^2} \right).
\end{equation}
Moreover, following \cite{Harlow:2018tqv}, it is natural to choose the length of each slice as the first phase space variable: the computation can be performed in embedding coordinates and yields
\begin{equation}\label{eq:length_cutoff_cutoff}
    \begin{split}
        L(\tau, \Phi_h) &= \text{arccosh}\!\Bigg[ \frac{\Phi_b^- \, \Phi_b^+}{\Phi_h^2} + \\
        &\quad + \frac{\sqrt{((\Phi_b^-)^2 - \Phi_h^2) ((\Phi_b^+)^2 - \Phi_h^2)}}{\Phi_h^2} \cosh\!\left( \left( \frac{\Phi_h}{\sqrt{(\Phi_b^-)^2 - \Phi_h^2}} + \frac{\Phi_h}{\sqrt{(\Phi_b^+)^2 - \Phi_h^2}} \right) \frac{\tau}{2} \right) \Bigg].
    \end{split}
\end{equation}
Finally, inverting Hamilton's equation
\begin{equation}
    \frac{\text{d} L}{\text{d} \tau} = \frac{\partial H}{\partial P},
\end{equation}
we obtain the following expression for the conjugate momentum $P$:
\begin{equation}
    P^2(\tau,\Phi_h) = \frac{\Phi_h^2 \sinh^2\!{L(\tau,\Phi_h)} + (\Phi_b^-)^2 + (\Phi_b^+)^2 - 2 \Phi_b^- \Phi_b^+ \cosh\!{L(\tau,\Phi_h)}}{\sinh^2\!{L(\tau,\Phi_h)}}.
\end{equation}
As a result, the full Hamiltonian \eqref{eq:hamiltonian_cutoffcutoff_A} can be expressed in terms of the canonical variables $(L,P)$ as
\begin{equation}\label{eq:hamiltonian_cutoff_cutoff_LP}
    H(L,P) = \frac{1}{2} \left( \Phi_b^- - \sqrt{(\Phi_b^-)^2 - \Phi_h^2(L,P)} + \Phi_b^+ - \sqrt{(\Phi_b^+)^2 - \Phi_h^2(L,P)} \right),
\end{equation}
where $\Phi_h^2(L,P)$ is given, in turn, by the generalized P\"oschl--Teller Hamiltonian
\begin{equation}\label{eq:phi_h^2_LP}
    \Phi_h^2(L,P) = P^2 + \frac{\Phi_b^- \Phi_b^+}{\cosh^2(L/2)} - \frac{(\Phi_b^- - \Phi_b^+)^2}{\sinh^2{L}}.
\end{equation}
We recover the results by \cite{Griguolo:2026eoi} once we specialize to $\Phi_b^- = \Phi_b^+ = \Phi_b$.

Next, we consider the case represented on the right panel of Figure \ref{fig:penrose_diagram_cutoff_cutoff}, where the two boundaries both lie on the right of the black hole horizon.
An important difference with respect to the previous result is that, on the innermost boundary with smaller dilaton value, the outward normal points in the direction of decreasing dilaton. As a result, adding contributions from both boundaries yields the total Hamiltonian
\begin{equation}\label{eq:hamiltonian_cutoffcutoff_B}
    H = \frac{H_{-} + H_{+}}{2} = \frac{1}{2} \left( \Phi_b^- + \sqrt{(\Phi_b^-)^2 - \Phi_h^2} + \Phi_b^+ - \sqrt{(\Phi_b^+)^2 - \Phi_h^2} \right),
\end{equation}
which differs by a relative sign from \eqref{eq:hamiltonian_cutoffcutoff_A}.
Once again, we foliate the region of interest with geodesics connecting boundary points of equal proper time $\tau_{-} = \tau_{+}$. Notice that such points are spacelike separated only for a limited range of proper time $\tau = \tau_{-} + \tau_{+}$, hence we will only consider those cases. Namely, we will only consider slices with $L \geq 0$.

This time, the computation of their geodesic distance gives
\begin{equation}
    \begin{split}
        L(\tau, \Phi_h) &= \text{arccosh}\!\Bigg[ \frac{\Phi_b^- \, \Phi_b^+}{\Phi_h^2} - \\
        &\quad - \frac{\sqrt{((\Phi_b^-)^2 - \Phi_h^2) ((\Phi_b^+)^2 - \Phi_h^2)}}{\Phi_h^2} \cosh\!\left( \left( \frac{\Phi_h}{\sqrt{(\Phi_b^-)^2 - \Phi_h^2}} - \frac{\Phi_h}{\sqrt{(\Phi_b^+)^2 - \Phi_h^2}} \right) \frac{\tau}{2} \right) \Bigg].
    \end{split}
\end{equation}
This differs from the result \eqref{eq:length_cutoff_cutoff} by two relative minus signs both in the argument of $\text{arccosh}$ and in the frequency. Moreover, we notice that the configuration $L = 0$, at which the two boundaries become null separated, is reached in finite proper time.
One can check that $L(\tau,\Phi_h) < L_\text{max}$ for all $\tau$, hence these trajectories are precisely the bound states discussed in subsection \ref{s:claspot} in the main text.

Inverting Hamilton's equations in this case gives
\begin{equation}
    P^2(\tau,\Phi_h) = \frac{\Phi_h^2 \sinh^2\!{L(\tau,\Phi_h)} + (\Phi_b^-)^2 + (\Phi_b^+)^2 - 2 \Phi_b^- \Phi_b^+ \cosh\!{L(\tau,\Phi_h)}}{\sinh^2\!{L(\tau,\Phi_h)}},
\end{equation}
which allows us to write the Hamiltonian as
\begin{equation}\label{eq:hamiltonian_cutoffcutoff_B_LP}
    H(L,P) = \frac{1}{2} \left( \Phi_b^- + \sqrt{(\Phi_b^-)^2 - \Phi_h^2(L,P)} + \Phi_b^+ - \sqrt{(\Phi_b^+)^2 - \Phi_h^2(L,P)} \right).
\end{equation}
Here $\Phi_h^2(L,P)$ is given by the same expression we found in the case of two boundaries on opposite sides.

From the shape of the potential of Fig.~\ref{fig:Vfcplot} in subsection \ref{s:claspot}, it is clear that there is one further class of classical trajectories, with $\Phi_h^2(L,P)<0$, trapped in the well near $L=0$. These are not described by any of the solutions $L(\tau)$ above, which assume $\Phi_h^2>0$. Nevertheless, they do have a gravitational interpretation as they correspond to the horizonless solution of \eqref{eq:secondordereom}:
\begin{equation}\label{eq:horizonless}
ds^2 = \frac{-\mathrm{d}\eta^2+\mathrm{d}\theta^2}{\cos^2\theta},
\qquad
\Phi = \sqrt{-\Phi_h^2} \, \tan\theta,
\qquad -\infty<\eta<+\infty,
\qquad -\pi/2 < \theta < \pi/2,
\end{equation}
namely empty global AdS$_2$. Both boundaries are then curves $\tan\theta_b^\pm = \Phi_b^{\pm}/\sqrt{-\Phi_h^2}$ in this region, and the geodesic distance between boundary points of equal proper time reads
\begin{equation}\label{eq:Lhorizonless}
    \begin{split}
        L(\tau,\Phi_h)&=\text{arccosh}\!\Bigg[\frac{\Phi_b^- \,\Phi_b^+}{\Phi_h^2} -\\
        &\quad - \frac{\sqrt{((\Phi_b^-)^2-\Phi_h^2)((\Phi_b^+)^2-\Phi_h^2)}}{\Phi_h^2}\cos\!\left(\left(\frac{\sqrt{-\Phi_h^2}}{\sqrt{(\Phi_b^-)^2-\Phi_h^2}}-\frac{\sqrt{-\Phi_h^2}}{\sqrt{(\Phi_b^+)^2-\Phi_h^2}}\right)\frac{\tau}{2}\right)\Bigg].
    \end{split}
\end{equation}
Repeating the same construction as before, we find that the corresponding Hamiltonian as a function of $L$ and $P$ is again given by \eqref{eq:hamiltonian_cutoffcutoff_B_LP}.

\subsection{Cutoff-EOW boundary conditions}
\label{app:cebc}
In this Section, we consider JT gravity with cutoff conditions \eqref{eq: boundary conditions} on the right boundary and EOW conditions \eqref{eq:eowboundaryconditions} on the left. This case generalizes the computation performed in \cite{Gao:2021uro} for ordinary JT gravity with an EOW brane.\footnote{While this work was being finished, an analogous treatment of these configurations was presented in \cite{Aguilar-Gutierrez:2026qqe}.}

Importantly, depending on the value of the solution parameter $\Phi_h$ relative to the boundary data $\Phi_b$, $\mu$, we have two qualitatively different situations, see Figure \ref{fig:penrose_diagram_EOW}. When $\Phi_h^2 < \Phi_b^2 - \mu^2$, the EOW boundary sits on the left of the Penrose diagram, whereas the cutoff boundary sits on the right and the two are separated by a black hole horizon:
\begin{equation}
    \Phi_h^2 < \Phi_b^2 - \mu^2 : \qquad 
    \sigma_\text{cutoff}(\nu) = \pi - \text{arcsin}\!\left(  \frac{\Phi_h}{\Phi_b} \cos\nu\right),
    \qquad
    \sigma_\text{EOW}(\nu) = \text{arccos}\!\left( \frac{\cos\nu}{\sqrt{1 + \Phi_h^2/\mu^2}} \right).
\end{equation}
Instead, when $\Phi_h^2 > \Phi_b^2 - \mu^2$, both boundaries sit on the left of the diagram, and thus there is no black hole horizon separating them. In this case, we have a naked brane.
\begin{equation}
    \Phi_h^2 > \Phi_b^2 - \mu^2 : \qquad 
    \sigma_\text{cutoff}(\nu) = \text{arcsin}\!\left(  \frac{\Phi_h}{\Phi_b} \cos\nu\right),
    \qquad
    \sigma_\text{EOW}(\nu) = \text{arccos}\!\left( \frac{\cos\nu}{\sqrt{1 + \Phi_h^2/\mu^2}} \right).
\end{equation}
In the first case, the cutoff boundary is separated from the EOW boundary by a black hole horizon, whereas in the second case they both lie on the same spacetime wedge and intersect at two points. In this case, the spacetime is the region between the two curves with $\Phi > \Phi_b$.

We now proceed by computing the distance between a point on the cutoff boundary and the brane. Conveniently, this can be done for both configurations at once.
Specifically, we are interested in spacelike geodesics with one endpoint corresponding to the point labeled by $\tau$ on $\partial\mathcal{M}_\text{cutoff}$ that arrive orthogonally to $\partial\mathcal{M}_\text{EOW}$. 
In embedding coordinates, the points on this curve are given by
\begin{equation}
    Y_\text{EOW} = \begin{pmatrix}
        -\frac{k \cos\nu}{\sqrt{1 - k^2 \cos^2\nu}}\\
        \frac{\cos\nu}{\sqrt{1 - k^2 \cos^2\nu}}\\
        \frac{\sin\nu}{\sqrt{1 - k^2 \cos^2\nu}}
    \end{pmatrix},
\end{equation}
where we introduced the parameter
\begin{equation}
    k \equiv \frac{\mu}{\sqrt{\mu^2 + \Phi_h^2}} < 1.
\end{equation}

Following the strategy presented in \cite{Gao:2021uro}, we want to act on the classical geometry \eqref{eq:geometry_on_shell} with an isometry that maps the EOW boundary $\partial\mathcal{M}_\text{EOW}$ to a vertical line
\begin{equation}\label{eq:EOW_vertical}
    \partial\mathcal{M}_\text{EOW}' = \big( \pi/2, \nu' \big) \qquad \to \qquad Y_\text{EOW}' = \begin{pmatrix}
        0\\
        \cos\nu'\\
        \sin\nu'
    \end{pmatrix}.
\end{equation}
Since isometries preserve both angles and the length of curves, it is convenient to compute the geodesic distance in this new coordinate system $(\sigma',\nu')$.
Indeed, in the new coordinates, we compute the length of the horizontal geodesic ($\nu' = \text{const}$) that connects the point labeled by $\tau$ on $\partial\mathcal{M}_\text{cutoff}'$ to the vertical curve \eqref{eq:EOW_vertical}.
Since $\nu' = \text{const}$ slices are geodesics orthogonal to $\sigma' = \pi/2$, their length is simply given by $\sinh{L} = \abs{1/\tan{\sigma'}}$.

The isometry we are looking for is explicitly realized by some matrix $M(\xi)$ such that
\begin{equation}\label{eq:isometry}
    \begin{pmatrix}
        0\\ \cos\nu' \\ \sin\nu'
    \end{pmatrix}
    =
    \underbrace{
    \begin{pmatrix}
        \cosh\xi &\sinh\xi \\
        \sinh\xi &\cosh\xi \\
         & & 1
    \end{pmatrix}
    }_{\equiv \displaystyle{M(\xi)}}
    \begin{pmatrix}
        -\frac{k \cos\nu}{\sqrt{1 - k^2 \cos^2\nu}}\\
        \frac{\cos\nu}{\sqrt{1 - k^2 \cos^2\nu}}\\
        \frac{\sin\nu}{\sqrt{1 - k^2 \cos^2\nu}}
    \end{pmatrix}.
\end{equation}
One can verify that the desired value of the parameter $\xi$ is the one for which
\begin{equation}
    \cosh\xi = \frac{1}{\sqrt{1 - k^2}}, \qquad \sinh\xi = \frac{k}{\sqrt{1 - k^2}}.
\end{equation}
Finally, we act with the same isometry \eqref{eq:isometry} on the cutoff boundary. In terms of proper time, this is parameterized as
\begin{equation}
    Y_\text{cutoff}(\tau) =
    \begin{pmatrix}
        \epsilon \frac{\sqrt{\Phi_b^2 - \Phi_h^2}}{\Phi_h} \cosh\!\left( \frac{\Phi_h}{\sqrt{\Phi_b^2 - \Phi_h^2}} \tau \right),
        &
        \frac{\Phi_b}{\Phi_h},
        &
        \frac{\sqrt{\Phi_b^2 - \Phi_h^2}}{\Phi_h} \sinh\!\left( \frac{\Phi_h}{\sqrt{\Phi_b^2 - \Phi_h^2}} \tau \right)
    \end{pmatrix},
\end{equation}
where $\epsilon = +1$ ($\epsilon = -1$) corresponds to a cutoff boundary at $\sigma > \pi/2$ ($\sigma < \pi/2$). As a result, the coordinate $\sigma$ is related to the boundary proper time $\tau$ as
\begin{equation}
    \tan\sigma = - \epsilon \frac{\Phi_h}{\sqrt{\Phi_b^2 - \Phi_h^2}} \frac{1}{\cosh\left( \frac{\Phi_h}{\sqrt{\Phi_b^2 - \Phi_h^2}} \tau \right)}.
\end{equation}
Finally, one obtains the following expression for the geodesic length
\begin{equation}\label{eq:length}
    L(\tau,\Phi_h) = \text{arcsinh}\!\left[ \frac{\mu \Phi_b}{\Phi_h^2} + \epsilon \frac{\sqrt{\Phi_h^2 + \mu^2} \sqrt{\Phi_b^2 - \Phi_h^2}}{\Phi_h^2} \cosh\!\left( \frac{\Phi_h}{\sqrt{\Phi_b^2 - \Phi_h^2}} \tau \right) \right].
\end{equation}
For $\epsilon = -1$, the argument of $\text{arcsinh}$ vanishes in finite $\tau$, suggesting that the point on the cutoff boundary reaches the two intersections in finite proper time.

Having obtained an expression for the geodesic distance as a function of $\tau$ in both cases, we recall that the boundary Hamiltonian is given by
\begin{equation}\label{eq:energy_EOW}
    H = \Phi_b - \epsilon \sqrt{\Phi_b^2 - \Phi_h^2}.
\end{equation}

Next, we want to find an expression for the ADM energy \eqref{eq:energy_EOW} in terms of the conjugate pair $(L,P)$, where $P$ is the momentum conjugate to $L$. Inverting Hamilton's equation results in the following identity between $H$, $L$ and $P$:
\begin{equation}\label{eq:hamiltonian_EOW_LP}
    P^2 = \Phi_b^2 - (\Phi_b - H)^2 - \frac{\Phi_b^2 - \mu^2 + 2 \mu \Phi_b \sinh L}{\cosh^2\!L},
\end{equation}
which is valid in both the regimes of $\Phi_h$ relative to $\Phi_b$ and $\mu$.
Finally, we can express the ADM energy in terms of the conjugate pair $(L,P)$, yielding
\begin{equation}\label{eq:hamiltonian}
    H(L,P) = \Phi_b - \epsilon \sqrt{\Phi_b^2 - \Phi_h^2(L,P)},
\end{equation}
where
\begin{equation}\label{eq:phi_h^2_EOW_LP}
    \Phi_h^2(L,P) = P^2 + \frac{\Phi_b^2 - \mu^2 + 2 \mu \Phi_b \sinh L}{\cosh^2\!L}.
\end{equation}

Finally, we comment on the global AdS case which is realized when $\Phi_h^2 < 0$. In the coordinates \eqref{eq:horizonless}, the brane and the cutoff boundary are given, respectively, by
\begin{equation}
    \sin \theta_\text{EOW} = \sqrt{1 + \frac{\Phi_h^2}{\mu^2}} \cos\eta,
    \qquad
    \tan \theta_b = \frac{\Phi_b}{\sqrt{- \Phi_h^2}}.
\end{equation}
Importantly, the brane can exist only for $-\Phi_h^2 < \mu^2$ and it intersects the cutoff boundary when the boundary data obey $\mu > \Phi_b$.
Computing the distance between a boundary point of proper time $\tau$ and the brane gives
\begin{equation}
    L(\tau,\Phi_h) = \text{arcsinh}\!\left[ \frac{\mu \Phi_b}{\Phi_h^2} - \frac{\sqrt{(\mu^2 + \Phi_h^2) (\Phi_b^2 - \Phi_h^2)}}{\Phi_h^2} \cos\!\left( \frac{\sqrt{-\Phi_h^2}}{\sqrt{\Phi_b^2 - \Phi_h^2}} \tau \right) \right],
\end{equation}
and an analogous computation for the corresponding conjugate momentum $P$ yields the same expression for the Hamiltonian \eqref{eq:hamiltonian_EOW_LP}.

Some comments on the Hamiltonian \eqref{eq:hamiltonian} are in order.
First, we notice that it reduces to the Morse Hamiltonian found in \cite{Gao:2021uro} for JT gravity in the presence of an EOW and an asymptotic boundary once we take its large-cutoff limit. The latter is obtained by sending $\Phi_b \to \infty$ while keeping the renormalized length
\begin{equation}
    L_\text{ren} \equiv L - \log \Phi_b
\end{equation}
finite. Indeed, taking this limit at the level of the Hamiltonian \eqref{eq:hamiltonian} (with $\epsilon = +1$) results in
\begin{equation}
    H(L_\text{ren},P) \overset{\Phi_b \to \infty}{\sim} \frac{1}{2 \Phi_b} \left( P^2 + 4 e^{-2 L_\text{ren}} + 4 \mu e^{-L_\text{ren}} \right).
\end{equation}

The potential term in \eqref{eq:phi_h^2_EOW_LP} is non-singular for all values of the boundary data and must be supplemented by an infinite barrier at the origin preventing unphysical configurations with $L < 0$.
Depending on the hierarchy between $\Phi_b$ and $\mu$, we can have two qualitatively different situations. As depicted in Figure \ref{fig:VEOW}, $V(L)$ has a maximum at 
\begin{equation}
    L_\text{max} = \text{arcsinh}\!\left( \frac{\mu}{\Phi_b} \right)
\end{equation}
at which $V(L_\text{max}) = \Phi_b^2$. Furthermore, at $L=0$ the value of the potential is $V(L=0)=\Phi_b^2 - \mu^2$.
From this analysis, we understand that the $(\epsilon = +)$-trajectories given in \eqref{eq:length} (and depicted on the left of Figure \ref{fig:penrose_diagram_EOW}) describe the motion of a particle that comes from $L \to \infty$ at large negative times, scatters against $V(L)$ and then goes back to $L \to \infty$ at large positive times.
Instead, the $(\epsilon = -)$-trajectories in \eqref{eq:length} (and depicted on the right of Figure \ref{fig:penrose_diagram_EOW}) describe the motion of a particle trapped in the region $L < L_\text{max}$, which reaches $L = 0$ in finite time, when the cutoff boundary meets the brane.
All these trajectories are classically bounded, whereas quantum mechanically they can tunnel through the barrier towards $L \to \infty$ and are therefore metastable.

\bibliographystyle{ourbst}
\bibliography{boundaries.bib}
\end{document}

%% file: images/penrose_diagram_EOW.tikz
\begin{tikzpicture}[x=0.75pt,y=0.75pt,yscale=-1,xscale=1]

\draw  [draw opacity=0][fill={rgb, 255:red, 230; green, 230; blue, 255 }  ,fill opacity=1 ] (337,99) .. controls (369.8,130.2) and (367.8,152.2) .. (338,181) .. controls (318.67,157.8) and (318.67,123.8) .. (337,99) -- cycle ;
\draw [color={rgb, 255:red, 208; green, 2; blue, 27 }  ,draw opacity=1 ][line width=1.5]    (290,50) .. controls (385.73,149.14) and (384.73,131.14) .. (290,230) ;
\draw [color={rgb, 255:red, 37; green, 170; blue, 80 }  ,draw opacity=1 ][line width=1.5]    (380,230) .. controls (303.67,149.8) and (303.67,129.8) .. (380,50) ;
\draw [color={rgb, 255:red, 0; green, 0; blue, 0 }  ,draw opacity=1 ] [dash pattern={on 4.5pt off 4.5pt}]  (470,50) -- (290,230) ;
\draw [color={rgb, 255:red, 0; green, 0; blue, 0 }  ,draw opacity=1 ] [dash pattern={on 4.5pt off 4.5pt}]  (290,50) -- (470,230) ;
\draw  [color={rgb, 255:red, 0; green, 0; blue, 255 }  ,draw opacity=1 ][fill={rgb, 255:red, 0; green, 0; blue, 255 }  ,fill opacity=1 ] (323.37,121) .. controls (323.37,119.55) and (324.55,118.37) .. (326,118.37) .. controls (327.45,118.37) and (328.63,119.55) .. (328.63,121) .. controls (328.63,122.46) and (327.45,123.64) .. (326,123.64) .. controls (324.55,123.64) and (323.37,122.46) .. (323.37,121) -- cycle ;
\draw [color={rgb, 255:red, 208; green, 2; blue, 27 }  ,draw opacity=1 ][line width=0.75]  [dash pattern={on 4.5pt off 4.5pt}]  (470,230) .. controls (374.27,130.87) and (375.27,148.87) .. (470,50) ;
\draw  [color={rgb, 255:red, 0; green, 0; blue, 255 }  ,draw opacity=1 ][fill={rgb, 255:red, 0; green, 0; blue, 255 }  ,fill opacity=1 ] (356.37,130) .. controls (356.37,128.55) and (357.55,127.37) .. (359,127.37) .. controls (360.45,127.37) and (361.63,128.55) .. (361.63,130) .. controls (361.63,131.46) and (360.45,132.64) .. (359,132.64) .. controls (357.55,132.64) and (356.37,131.46) .. (356.37,130) -- cycle ;
\draw [color={rgb, 255:red, 0; green, 0; blue, 255 }  ,draw opacity=1 ][line width=1.5]    (326,121) .. controls (333.27,124.6) and (348.2,129.31) .. (359,130) ;
\draw  [draw opacity=0][fill={rgb, 255:red, 230; green, 230; blue, 255 }  ,fill opacity=1 ] (140,50.07) .. controls (181.93,50.78) and (188.28,49.72) .. (230,50.07) .. controls (163.2,130.67) and (163.2,151.67) .. (230,230.07) .. controls (187.22,229.65) and (182.99,229.65) .. (140,230.07) .. controls (98.6,157.94) and (97.6,115.94) .. (140,50.07) -- cycle ;
\draw [color={rgb, 255:red, 208; green, 2; blue, 27 }  ,draw opacity=1 ][line width=1.5]    (230,230.07) .. controls (162.33,145.81) and (162.33,133.81) .. (230,50.07) ;
\draw [color={rgb, 255:red, 37; green, 170; blue, 80 }  ,draw opacity=1 ][line width=1.5]    (140,230.07) .. controls (98.6,159.87) and (97.6,120.87) .. (140,50.07) ;
\draw [color={rgb, 255:red, 0; green, 0; blue, 0 }  ,draw opacity=1 ] [dash pattern={on 4.5pt off 4.5pt}]  (230,50.07) -- (50,230.07) ;
\draw [color={rgb, 255:red, 0; green, 0; blue, 0 }  ,draw opacity=1 ] [dash pattern={on 4.5pt off 4.5pt}]  (50,50.07) -- (230,230.07) ;
\draw  [color={rgb, 255:red, 0; green, 0; blue, 255 }  ,draw opacity=1 ][fill={rgb, 255:red, 0; green, 0; blue, 255 }  ,fill opacity=1 ] (108.37,116.07) .. controls (108.37,114.62) and (109.55,113.44) .. (111,113.44) .. controls (112.45,113.44) and (113.63,114.62) .. (113.63,116.07) .. controls (113.63,117.52) and (112.45,118.7) .. (111,118.7) .. controls (109.55,118.7) and (108.37,117.52) .. (108.37,116.07) -- cycle ;
\draw  [color={rgb, 255:red, 0; green, 0; blue, 255 }  ,draw opacity=1 ][fill={rgb, 255:red, 0; green, 0; blue, 255 }  ,fill opacity=1 ] (177.37,132.7) .. controls (177.37,131.25) and (178.55,130.07) .. (180,130.07) .. controls (181.45,130.07) and (182.63,131.25) .. (182.63,132.7) .. controls (182.63,134.16) and (181.45,135.34) .. (180,135.34) .. controls (178.55,135.34) and (177.37,134.16) .. (177.37,132.7) -- cycle ;
\draw [color={rgb, 255:red, 208; green, 2; blue, 27 }  ,draw opacity=1 ][line width=0.75]  [dash pattern={on 4.5pt off 4.5pt}]  (50,50.07) .. controls (117.67,134.34) and (117.67,146.34) .. (50,230.07) ;
\draw [color={rgb, 255:red, 0; green, 0; blue, 255 }  ,draw opacity=1 ][line width=1.5]    (111,116.07) .. controls (134.2,122.31) and (152.67,130.47) .. (180,132.7) ;
\draw  [draw opacity=0][fill={rgb, 255:red, 255; green, 255; blue, 255 }  ,fill opacity=1 ] (39.93,39.93) -- (239.93,39.93) -- (239.93,239.93) -- (39.93,239.93) -- cycle(230,49.86) -- (49.87,49.86) -- (49.87,230) -- (230,230) -- cycle ;
\draw  [line width=0.75]  (50,50) -- (230,50) -- (230,230) -- (50,230) -- cycle ;
\draw [line width=0.75]    (40,50) -- (230,50) ;
\draw [line width=0.75]    (40,230) -- (230,230) ;
\draw [line width=0.75]    (49.87,239.86) -- (49.87,49.86) ;
\draw [line width=0.75]    (230,240) -- (230,50) ;
\draw  [draw opacity=0][fill={rgb, 255:red, 255; green, 255; blue, 255 }  ,fill opacity=1 ] (280,40) -- (480,40) -- (480,240) -- (280,240) -- cycle(470.07,49.93) -- (289.93,49.93) -- (289.93,230.07) -- (470.07,230.07) -- cycle ;
\draw  [line width=0.75]  (290.07,50.07) -- (470.07,50.07) -- (470.07,230.07) -- (290.07,230.07) -- cycle ;
\draw [line width=0.75]    (280.07,50.07) -- (470.07,50.07) ;
\draw [line width=0.75]    (280.07,230.07) -- (470.07,230.07) ;
\draw [line width=0.75]    (289.93,239.93) -- (289.93,49.93) ;
\draw [line width=0.75]    (470.07,240.07) -- (470.07,50.07) ;

\draw (140,243.4) node [anchor=north] [inner sep=0.75pt]    {$\sigma $};
\draw (38,140) node [anchor=east] [inner sep=0.75pt]    {$\nu $};
\draw (38,50) node [anchor=east] [inner sep=0.75pt]    {$+\frac{\pi }{2}$};
\draw (38,230) node [anchor=east] [inner sep=0.75pt]    {$-\frac{\pi }{2}$};
\draw (50,243.4) node [anchor=north] [inner sep=0.75pt]    {$0$};
\draw (230,243.4) node [anchor=north] [inner sep=0.75pt]    {$\pi $};
\draw (380,243.4) node [anchor=north] [inner sep=0.75pt]    {$\sigma $};
\draw (289.93,243.33) node [anchor=north] [inner sep=0.75pt]    {$0$};
\draw (470.07,243.47) node [anchor=north] [inner sep=0.75pt]    {$\pi $};
\draw (278,140) node [anchor=east] [inner sep=0.75pt]    {$\nu $};
\draw (278.07,50.07) node [anchor=east] [inner sep=0.75pt]    {$+\frac{\pi }{2}$};
\draw (278.07,230.07) node [anchor=east] [inner sep=0.75pt]    {$-\frac{\pi }{2}$};
\draw (146,121.67) node [anchor=south] [inner sep=0.75pt]  [color={rgb, 255:red, 0; green, 0; blue, 255 }  ,opacity=1 ]  {$\textcolor[rgb]{0,0,1}{L}$};
\draw (341,130.4) node [anchor=north] [inner sep=0.75pt]  [color={rgb, 255:red, 0; green, 0; blue, 255 }  ,opacity=1 ]  {$\textcolor[rgb]{0,0,1}{L}$};
\draw (182,132.7) node [anchor=west] [inner sep=0.75pt]  [color={rgb, 255:red, 0; green, 0; blue, 0 }  ,opacity=1 ]  {$\textcolor[rgb]{0.82,0.01,0.11}{\tau }$};
\draw (361,130) node [anchor=west] [inner sep=0.75pt]  [color={rgb, 255:red, 0; green, 0; blue, 0 }  ,opacity=1 ]  {$\textcolor[rgb]{0.82,0.01,0.11}{\tau }$};
\draw (135,270.4) node [anchor=north] [inner sep=0.75pt]    {$\Phi _{h}^{2} < \Phi _{b}^{2} -\mu ^{2}$};
\draw (380.07,270.47) node [anchor=north] [inner sep=0.75pt]    {$\Phi _{h}^{2}  >\Phi _{b}^{2} -\mu ^{2}$};

\end{tikzpicture}

%% file: images/boundary_frame.tikz
\begin{tikzpicture}[x=0.75pt,y=0.75pt,yscale=-1,xscale=1]

\draw  [draw opacity=0][fill={rgb, 255:red, 230; green, 230; blue, 255 }  ,fill opacity=1 ] (140,50.07) .. controls (181.93,50.78) and (188.28,49.72) .. (230,50.07) .. controls (163.2,130.67) and (163.2,151.67) .. (230,230.07) .. controls (187.22,229.65) and (182.99,229.65) .. (140,230.07) .. controls (58.2,157.67) and (63.2,117.67) .. (140,50.07) -- cycle ;
\draw [color={rgb, 255:red, 208; green, 2; blue, 27 }  ,draw opacity=1 ][line width=1.5]    (230,230.07) .. controls (162.33,145.81) and (162.33,133.81) .. (230,50.07) ;
\draw [color={rgb, 255:red, 37; green, 170; blue, 80 }  ,draw opacity=1 ][line width=1.5]    (140,230.07) .. controls (60.96,159.87) and (59.05,120.87) .. (140,50.07) ;
\draw  [draw opacity=0][fill={rgb, 255:red, 255; green, 255; blue, 255 }  ,fill opacity=1 ] (39.93,39.93) -- (239.93,39.93) -- (239.93,239.93) -- (39.93,239.93) -- cycle(230,49.86) -- (49.87,49.86) -- (49.87,230) -- (230,230) -- cycle ;
\draw  [line width=0.75]  (50,50) -- (230,50) -- (230,230) -- (50,230) -- cycle ;
\draw [line width=0.75]    (50,50) -- (230,50) ;
\draw [line width=0.75]    (49.87,230) -- (230,230) ;
\draw [line width=0.75]    (49.87,230) -- (49.87,49.86) ;
\draw [line width=0.75]    (230,230.07) -- (230,50) ;
\draw    (180,128.06) -- (225.17,141.03) ;
\draw [shift={(228.06,141.86)}, rotate = 196.03] [fill={rgb, 255:red, 0; green, 0; blue, 0 }  ][line width=0.08]  [draw opacity=0] (8.93,-4.29) -- (0,0) -- (8.93,4.29) -- cycle    ;
\draw    (180,128.06) -- (192.98,82.88) ;
\draw [shift={(193.81,80)}, rotate = 106.03] [fill={rgb, 255:red, 0; green, 0; blue, 0 }  ][line width=0.08]  [draw opacity=0] (8.93,-4.29) -- (0,0) -- (8.93,4.29) -- cycle    ;

\draw [line width=0.75]    (81.25,150) -- (128.01,145.34) ;
\draw [shift={(131,145.04)}, rotate = 174.31] [fill={rgb, 255:red, 0; green, 0; blue, 0 }  ][line width=0.08]  [draw opacity=0] (8.93,-4.29) -- (0,0) -- (8.93,4.29) -- cycle    ;
\draw [line width=0.75]    (81.25,150) -- (76.59,103.23) ;
\draw [shift={(76.29,100.25)}, rotate = 84.31] [fill={rgb, 255:red, 0; green, 0; blue, 0 }  ][line width=0.08]  [draw opacity=0] (8.93,-4.29) -- (0,0) -- (8.93,4.29) -- cycle    ;
\draw [color={rgb, 255:red, 0; green, 0; blue, 0 }  ,draw opacity=1 ] [dash pattern={on 4.5pt off 4.5pt}]  (230,50) -- (50,230) ;
\draw [color={rgb, 255:red, 0; green, 0; blue, 0 }  ,draw opacity=1 ] [dash pattern={on 4.5pt off 4.5pt}]  (50,50) -- (230,230) ;

\draw (191.81,80) node [anchor=east] [inner sep=0.75pt]    {$e^{/\!/}$};
\draw (226.06,145.26) node [anchor=north east] [inner sep=0.75pt]    {$e^{\perp }$};
\draw (74.29,100.25) node [anchor=east] [inner sep=0.75pt]    {$e^{/\!/}$};
\draw (119.25,148.44) node [anchor=north] [inner sep=0.75pt]    {$e^{\perp }$};

\end{tikzpicture}

%% file: images/local_frame.tikz
\begin{tikzpicture}[x=0.75pt,y=0.75pt,yscale=-1,xscale=1]

\draw  [draw opacity=0][fill={rgb, 255:red, 230; green, 230; blue, 255 }  ,fill opacity=1 ] (50,50) .. controls (78.82,50.49) and (133.33,50.14) .. (178.16,49.99) .. controls (198.56,49.92) and (216.95,49.89) .. (230,50) .. controls (217.2,91) and (216.2,117) .. (230,160) .. controls (187.22,159.58) and (92.99,159.58) .. (50,160) .. controls (66.2,117) and (70.2,97) .. (50,50) -- cycle ;
\draw [color={rgb, 255:red, 208; green, 2; blue, 27 }  ,draw opacity=1 ][line width=1.5]    (230,160) .. controls (215.2,124) and (215.2,87) .. (230,50) ;
\draw [color={rgb, 255:red, 208; green, 2; blue, 27 }  ,draw opacity=1 ][line width=1.5]    (50,50) .. controls (70.2,87) and (69.2,124) .. (50,160) ;
\draw  [color={rgb, 255:red, 0; green, 0; blue, 255 }  ,draw opacity=1 ][fill={rgb, 255:red, 0; green, 0; blue, 255 }  ,fill opacity=1 ] (55,66.6) .. controls (55,68.05) and (56.18,69.23) .. (57.63,69.23) .. controls (59.09,69.23) and (60.27,68.05) .. (60.27,66.6) .. controls (60.27,65.15) and (59.09,63.97) .. (57.63,63.97) .. controls (56.18,63.97) and (55,65.15) .. (55,66.6) -- cycle ;
\draw  [color={rgb, 255:red, 0; green, 0; blue, 255 }  ,draw opacity=1 ][fill={rgb, 255:red, 0; green, 0; blue, 255 }  ,fill opacity=1 ] (219.37,135.63) .. controls (219.37,137.09) and (220.55,138.27) .. (222,138.27) .. controls (223.45,138.27) and (224.63,137.09) .. (224.63,135.63) .. controls (224.63,134.18) and (223.45,133) .. (222,133) .. controls (220.55,133) and (219.37,134.18) .. (219.37,135.63) -- cycle ;
\draw [color={rgb, 255:red, 0; green, 0; blue, 255 }  ,draw opacity=1 ][line width=1.5]    (57.63,66.6) .. controls (130.2,70.8) and (127.2,127.8) .. (222,135.63) ;
\draw    (140,107.13) -- (178.66,133.86) ;
\draw [shift={(181.13,135.56)}, rotate = 214.66] [fill={rgb, 255:red, 0; green, 0; blue, 0 }  ][line width=0.08]  [draw opacity=0] (8.93,-4.29) -- (0,0) -- (8.93,4.29) -- cycle    ;
\draw    (140,107.13) -- (166.73,68.47) ;
\draw [shift={(168.44,66)}, rotate = 124.66] [fill={rgb, 255:red, 0; green, 0; blue, 0 }  ][line width=0.08]  [draw opacity=0] (8.93,-4.29) -- (0,0) -- (8.93,4.29) -- cycle    ;

\draw (166.44,60) node [anchor=east] [inner sep=0.75pt]    {$e^{0}$};
\draw (181.13,138.96) node [anchor=north] [inner sep=0.75pt]    {$e^{1}$};
\draw (95.45,79.4) node [anchor=north east] [inner sep=0.75pt]    {$\textcolor[rgb]{0,0,1}{\gamma }$};

\end{tikzpicture}

%% file: images/bulklocality.tikz





\begin{tikzpicture}[x=0.75pt,y=0.75pt,yscale=-1,xscale=1]

\draw  [draw opacity=0][fill={rgb, 255:red, 230; green, 230; blue, 255 }  ,fill opacity=1 ] (135.55,50) -- (166.66,50) -- (166.66,190) -- (135.55,190) -- cycle ;
\draw [color={rgb, 255:red, 208; green, 2; blue, 27 }  ,draw opacity=1 ][line width=1.5]    (135.55,50.01) -- (135.55,190) ;
\draw [color={rgb, 255:red, 208; green, 2; blue, 27 }  ,draw opacity=1 ][line width=1.5]    (166.66,50) -- (166.66,189.99) ;
\draw [color={rgb, 255:red, 0; green, 0; blue, 0 }  ,draw opacity=1 ][line width=1.5]    (50,50) -- (50,190) ;
\draw [color={rgb, 255:red, 0; green, 0; blue, 0 }  ,draw opacity=1 ][line width=1.5]    (190,50) -- (190,190) ;
\draw [color={rgb, 255:red, 0; green, 0; blue, 255 }  ,draw opacity=1 ][line width=1.5]    (135.55,135.55) -- (166.66,135.55) ;
\draw  [color={rgb, 255:red, 0; green, 0; blue, 255 }  ,draw opacity=1 ][fill={rgb, 255:red, 0; green, 0; blue, 255 }  ,fill opacity=1 ] (133.51,135.55) .. controls (133.51,134.42) and (134.42,133.51) .. (135.55,133.51) .. controls (136.68,133.51) and (137.6,134.42) .. (137.6,135.55) .. controls (137.6,136.68) and (136.68,137.6) .. (135.55,137.6) .. controls (134.42,137.6) and (133.51,136.68) .. (133.51,135.55) -- cycle ;
\draw  [color={rgb, 255:red, 0; green, 0; blue, 255 }  ,draw opacity=1 ][fill={rgb, 255:red, 0; green, 0; blue, 255 }  ,fill opacity=1 ] (164.62,135.55) .. controls (164.62,134.42) and (165.53,133.51) .. (166.66,133.51) .. controls (167.8,133.51) and (168.71,134.42) .. (168.71,135.55) .. controls (168.71,136.68) and (167.8,137.6) .. (166.66,137.6) .. controls (165.53,137.6) and (164.62,136.68) .. (164.62,135.55) -- cycle ;

\draw (133.55,135.55) node [anchor=east] [inner sep=0.75pt]  [color={rgb, 255:red, 0; green, 0; blue, 255 }  ,opacity=1 ]  {$\textcolor[rgb]{0.82,0.01,0.11}{\Phi _{b}^{-}}$};
\draw (168.66,135.55) node [anchor=west] [inner sep=0.75pt]  [color={rgb, 255:red, 0; green, 0; blue, 255 }  ,opacity=1 ]  {$\textcolor[rgb]{0.82,0.01,0.11}{\Phi _{b}^{+}}$};
\draw (151.11,132.15) node [anchor=south] [inner sep=0.75pt]    {$\textcolor[rgb]{0,0,1}{L}$};

\end{tikzpicture}

%% file: images/raddress.tikz
\begin{tikzpicture}[x=0.75pt,y=0.75pt,yscale=-1,xscale=1]

\draw  [draw opacity=0][fill={rgb, 255:red, 230; green, 230; blue, 255 }  ,fill opacity=1 ] (50,50) .. controls (78.82,50.49) and (133.33,50.14) .. (178.16,49.99) .. controls (198.56,49.92) and (216.95,49.89) .. (230,50) .. controls (217.2,91) and (216.2,117) .. (230,160) .. controls (187.22,159.58) and (92.99,159.58) .. (50,160) .. controls (66.2,117) and (70.2,97) .. (50,50) -- cycle ;
\draw [color={rgb, 255:red, 208; green, 2; blue, 27 }  ,draw opacity=1 ][line width=1.5]    (230,160) .. controls (215.2,124) and (215.2,87) .. (230,50) ;
\draw [color={rgb, 255:red, 208; green, 2; blue, 27 }  ,draw opacity=1 ][line width=1.5]    (50,50) .. controls (70.2,87) and (69.2,124) .. (50,160) ;
\draw  [color={rgb, 255:red, 0; green, 0; blue, 255 }  ,draw opacity=1 ][fill={rgb, 255:red, 0; green, 0; blue, 255 }  ,fill opacity=1 ] (55,66.6) .. controls (55,68.05) and (56.18,69.23) .. (57.63,69.23) .. controls (59.09,69.23) and (60.27,68.05) .. (60.27,66.6) .. controls (60.27,65.15) and (59.09,63.97) .. (57.63,63.97) .. controls (56.18,63.97) and (55,65.15) .. (55,66.6) -- cycle ;
\draw  [color={rgb, 255:red, 0; green, 0; blue, 255 }  ,draw opacity=1 ][fill={rgb, 255:red, 0; green, 0; blue, 255 }  ,fill opacity=1 ] (219.37,135.63) .. controls (219.37,137.09) and (220.55,138.27) .. (222,138.27) .. controls (223.45,138.27) and (224.63,137.09) .. (224.63,135.63) .. controls (224.63,134.18) and (223.45,133) .. (222,133) .. controls (220.55,133) and (219.37,134.18) .. (219.37,135.63) -- cycle ;
\draw [color={rgb, 255:red, 0; green, 0; blue, 255 }  ,draw opacity=1 ][line width=1.5]    (57.63,66.6) .. controls (130.2,70.8) and (127.2,127.8) .. (222,135.63) ;
\pgfpathcurvebetweentime{0}{0.47}{\pgfpointxy{57.63}{66.6}}{\pgfpointxy{130.2}{70.8}}{\pgfpointxy{127.2}{127.8}}{\pgfpointxy{222}{135.63}}
  \pgfsetstrokecolor{black}
  \pgfsetlinewidth{2pt}
  \pgfusepath{stroke}
\fill[black] (130,98) circle (4);

\draw (110,70) node [anchor=east] [inner sep=0.75pt]    {$s$};
\draw (170,79.4) node [anchor=north east] [inner sep=0.75pt]    {$\textcolor[rgb]{0,0,1}{L}$};
\draw (135,110) node [anchor=north east] [inner sep=0.75pt]    {$\textcolor[rgb]{0,0,0}{\Phi(L,s)}$};

\end{tikzpicture}

%% file: images/penrose_diagram_cutoff_cutoff.tikz
\begin{tikzpicture}[x=0.75pt,y=0.75pt,yscale=-1,xscale=1]

\draw  [draw opacity=0][fill={rgb, 255:red, 230; green, 230; blue, 255 }  ,fill opacity=1 ] (470.07,50.07) .. controls (512,50.78) and (428.35,49.72) .. (470.07,50.07) .. controls (437.47,130.71) and (437.47,147.71) .. (470.07,230.07) .. controls (427.29,229.64) and (513.05,229.51) .. (470.07,229.93) .. controls (387.8,150.24) and (388.8,128.24) .. (470.07,50.07) -- cycle ;
\draw [color={rgb, 255:red, 208; green, 2; blue, 27 }  ,draw opacity=1 ][line width=1.5]    (470.07,230.07) .. controls (437.47,146.71) and (437.47,133.71) .. (470.07,50.07) ;
\draw [color={rgb, 255:red, 208; green, 2; blue, 27 }  ,draw opacity=1 ][line width=1.5]    (470.07,229.93) .. controls (388.67,145.89) and (388.67,133.89) .. (470.07,49.93) ;
\draw  [draw opacity=0][fill={rgb, 255:red, 230; green, 230; blue, 255 }  ,fill opacity=1 ] (50,50) .. controls (91.93,50.71) and (188.28,49.65) .. (230,50) .. controls (197.4,130.64) and (197.4,147.64) .. (230,230) .. controls (187.22,229.57) and (92.99,229.57) .. (50,230) .. controls (130.4,149.04) and (130.4,131.04) .. (50,50) -- cycle ;
\draw [color={rgb, 255:red, 208; green, 2; blue, 27 }  ,draw opacity=1 ][line width=1.5]    (230,230) .. controls (197.4,146.64) and (197.4,133.64) .. (230,50) ;
\draw [color={rgb, 255:red, 208; green, 2; blue, 27 }  ,draw opacity=1 ][line width=1.5]    (50,50) .. controls (131.4,134.04) and (131.4,146.04) .. (50,230) ;
\draw [color={rgb, 255:red, 0; green, 0; blue, 0 }  ,draw opacity=1 ] [dash pattern={on 4.5pt off 4.5pt}]  (230,50) -- (50,230) ;
\draw [color={rgb, 255:red, 0; green, 0; blue, 0 }  ,draw opacity=1 ] [dash pattern={on 4.5pt off 4.5pt}]  (50,50) -- (230,230) ;
\draw  [draw opacity=0][fill={rgb, 255:red, 255; green, 255; blue, 255 }  ,fill opacity=1 ] (39.93,39.93) -- (239.93,39.93) -- (239.93,239.93) -- (39.93,239.93) -- cycle(230,49.86) -- (49.87,49.86) -- (49.87,230) -- (230,230) -- cycle ;
\draw  [line width=0.75]  (50,50) -- (230,50) -- (230,230) -- (50,230) -- cycle ;
\draw [line width=0.75]    (40,50) -- (230,50) ;
\draw [line width=0.75]    (40,230) -- (230,230) ;
\draw [line width=0.75]    (49.87,239.86) -- (49.87,49.86) ;
\draw [line width=0.75]    (230,240) -- (230,50) ;
\draw  [color={rgb, 255:red, 0; green, 0; blue, 255 }  ,draw opacity=1 ][fill={rgb, 255:red, 0; green, 0; blue, 255 }  ,fill opacity=1 ] (91.37,100) .. controls (91.37,98.55) and (92.55,97.37) .. (94,97.37) .. controls (95.45,97.37) and (96.63,98.55) .. (96.63,100) .. controls (96.63,101.45) and (95.45,102.63) .. (94,102.63) .. controls (92.55,102.63) and (91.37,101.45) .. (91.37,100) -- cycle ;
\draw  [color={rgb, 255:red, 0; green, 0; blue, 255 }  ,draw opacity=1 ][fill={rgb, 255:red, 0; green, 0; blue, 255 }  ,fill opacity=1 ] (204.37,121) .. controls (204.37,119.55) and (205.55,118.37) .. (207,118.37) .. controls (208.45,118.37) and (209.63,119.55) .. (209.63,121) .. controls (209.63,122.45) and (208.45,123.63) .. (207,123.63) .. controls (205.55,123.63) and (204.37,122.45) .. (204.37,121) -- cycle ;
\draw [color={rgb, 255:red, 0; green, 0; blue, 0 }  ,draw opacity=1 ] [dash pattern={on 4.5pt off 4.5pt}]  (470.07,50.07) -- (290.07,230.07) ;
\draw [color={rgb, 255:red, 0; green, 0; blue, 0 }  ,draw opacity=1 ] [dash pattern={on 4.5pt off 4.5pt}]  (290.07,50.07) -- (470.07,230.07) ;
\draw  [draw opacity=0][fill={rgb, 255:red, 255; green, 255; blue, 255 }  ,fill opacity=1 ] (280,40) -- (480,40) -- (480,240) -- (280,240) -- cycle(470.07,49.93) -- (289.93,49.93) -- (289.93,230.07) -- (470.07,230.07) -- cycle ;
\draw  [line width=0.75]  (290.07,50.07) -- (470.07,50.07) -- (470.07,230.07) -- (290.07,230.07) -- cycle ;
\draw [line width=0.75]    (280.07,50.07) -- (470.07,50.07) ;
\draw [line width=0.75]    (280.07,230.07) -- (470.07,230.07) ;
\draw [line width=0.75]    (289.93,239.93) -- (289.93,49.93) ;
\draw [line width=0.75]    (470.07,240.07) -- (470.07,50.07) ;
\draw  [color={rgb, 255:red, 0; green, 0; blue, 255 }  ,draw opacity=1 ][fill={rgb, 255:red, 0; green, 0; blue, 255 }  ,fill opacity=1 ] (414.73,112.63) .. controls (414.73,111.18) and (415.91,110) .. (417.37,110) .. controls (418.82,110) and (420,111.18) .. (420,112.63) .. controls (420,114.09) and (418.82,115.27) .. (417.37,115.27) .. controls (415.91,115.27) and (414.73,114.09) .. (414.73,112.63) -- cycle ;
\draw  [color={rgb, 255:red, 0; green, 0; blue, 255 }  ,draw opacity=1 ][fill={rgb, 255:red, 0; green, 0; blue, 255 }  ,fill opacity=1 ] (444,127.63) .. controls (444,126.18) and (445.18,125) .. (446.63,125) .. controls (448.09,125) and (449.27,126.18) .. (449.27,127.63) .. controls (449.27,129.09) and (448.09,130.27) .. (446.63,130.27) .. controls (445.18,130.27) and (444,129.09) .. (444,127.63) -- cycle ;
\draw [color={rgb, 255:red, 0; green, 0; blue, 255 }  ,draw opacity=1 ][line width=1.5]    (94,100) .. controls (133.2,100.84) and (158.2,118.84) .. (207,121) ;
\draw [color={rgb, 255:red, 0; green, 0; blue, 255 }  ,draw opacity=1 ][line width=1.5]    (417.37,112.63) .. controls (420.4,116.84) and (438.4,126.84) .. (446.63,127.63) ;

\draw (140,243.4) node [anchor=north] [inner sep=0.75pt]    {$\sigma $};
\draw (38,140) node [anchor=east] [inner sep=0.75pt]    {$\nu $};
\draw (38,50) node [anchor=east] [inner sep=0.75pt]    {$+\frac{\pi }{2}$};
\draw (38,230) node [anchor=east] [inner sep=0.75pt]    {$-\frac{\pi }{2}$};
\draw (50,243.4) node [anchor=north] [inner sep=0.75pt]    {$0$};
\draw (230,243.4) node [anchor=north] [inner sep=0.75pt]    {$\pi $};
\draw (92,100) node [anchor=east] [inner sep=0.75pt]  [color={rgb, 255:red, 37; green, 170; blue, 80 }  ,opacity=1 ]  {$\textcolor[rgb]{0.82,0.01,0.11}{\tau }\textcolor[rgb]{0.82,0.01,0.11}{_{-}}$};
\draw (209,121) node [anchor=west] [inner sep=0.75pt]  [color={rgb, 255:red, 0; green, 0; blue, 0 }  ,opacity=1 ]  {$\textcolor[rgb]{0.82,0.01,0.11}{\tau }\textcolor[rgb]{0.82,0.01,0.11}{_{+}}$};
\draw (415.37,112.63) node [anchor=east] [inner sep=0.75pt]  [color={rgb, 255:red, 0; green, 0; blue, 0 }  ,opacity=1 ]  {$\textcolor[rgb]{0.82,0.01,0.11}{\tau }\textcolor[rgb]{0.82,0.01,0.11}{_{-}}$};
\draw (448.63,127.63) node [anchor=west] [inner sep=0.75pt]  [color={rgb, 255:red, 0; green, 0; blue, 0 }  ,opacity=1 ]  {$\textcolor[rgb]{0.82,0.01,0.11}{\tau }\textcolor[rgb]{0.82,0.01,0.11}{_{+}}$};
\draw (380,243.4) node [anchor=north] [inner sep=0.75pt]    {$\sigma $};
\draw (289.93,243.33) node [anchor=north] [inner sep=0.75pt]    {$0$};
\draw (470.07,243.47) node [anchor=north] [inner sep=0.75pt]    {$\pi $};
\draw (278,140) node [anchor=east] [inner sep=0.75pt]    {$\nu $};
\draw (278.07,50.07) node [anchor=east] [inner sep=0.75pt]    {$+\frac{\pi }{2}$};
\draw (278.07,230.07) node [anchor=east] [inner sep=0.75pt]    {$-\frac{\pi }{2}$};
\draw (152,104.6) node [anchor=south] [inner sep=0.75pt]  [color={rgb, 255:red, 0; green, 0; blue, 255 }  ,opacity=1 ]  {$L$};
\draw (436,116.6) node [anchor=south] [inner sep=0.75pt]  [color={rgb, 255:red, 0; green, 0; blue, 255 }  ,opacity=1 ]  {$L$};

\end{tikzpicture}